\documentclass[acmsmall]{acmart}
\AtBeginDocument{%
  \providecommand\BibTeX{{%
    \normalfont B\kern-0.5em{\scshape i\kern-0.25em b}\kern-0.8em\TeX}}}

\acmDOI{10.1145/1122445.1122456}
\usepackage{booktabs}% http://ctan.org/pkg/booktabs

\usepackage{afterpage}  % load the afterpage package
\usepackage{adjustbox}
\usepackage{blindtext}
\usepackage{graphicx,array}
\usepackage{makecell}
\usepackage{tikz}
\usepackage{forest}
\usepackage{color, colortbl}
\usepackage{verbatim}
\usepackage[skins]{tcolorbox}
\usepackage{tabularx}
\usepackage{longtable}
\usepackage{enumerate}
\usepackage[shortlabels]{enumitem}

\usepackage{float}
\newcommand{\cmmnt}[1]{}
\usepackage[T1]{fontenc}
\usepackage{lmodern}
\definecolor{Gray}{gray}{0.9}

\setcopyright{none}
\renewcommand\footnotetextcopyrightpermission[1]{}
\begin{document}

\title{SOK: Fake News Outbreak 2021: Can We Stop the Viral Spread?}

\author{Tanveer Khan}
\affiliation{\institution{Tampere University}}
\email{tanveer.khan@tuni.fi}
\orcid{0000-0001-7296-2178}

\author{Antonis Michalas}\tikzstyle{selected}=[draw=red,fill=red!30]
\tikzstyle{optional}=[dashed,fill=gray!50]
\affiliation{\institution{Tampere University}}
\email{antonios.michalast@tuni.fi}
\orcid{0000-0002-0189-3520}

\author{Adnan Akhunzada}
\affiliation{\institution{Technical University of Denmark}}
\email{adnak@dtu.dk}

\begin{abstract}
Social Networks' omnipresence and ease of use has revolutionized the generation and distribution of information in today's world. However, easy access to information does not equal an increased level of public knowledge. Unlike traditional media channels, social networks also facilitate faster and wider spread of disinformation and misinformation. Viral spread of false information has serious implications on the behaviours, attitudes and beliefs of the public, and ultimately can seriously endanger the democratic processes. Limiting false information's negative impact through early detection and control of extensive spread presents the main challenge facing researchers today. In this survey paper, we extensively analyse a wide range of different solutions for the early detection of fake news in the existing literature. More precisely, we examine Machine Learning (ML) models for the identification and classification of fake news, online fake news detection competitions, statistical outputs as well as the advantages and disadvantages of some of the available data sets. Finally, we evaluate the online web browsing tools available for detecting and mitigating fake news and present some open research challenges.

%The ease of use of Online Social Networks has revolutionized the generation and distribution of information. In every category of information, false information--disinformation and misinformation, can spread faster, deeper and more widely on social media than in traditional media channels. If false information shared on social media goes viral, it has a dramatic effect on the opinions and decisions of an individual. The big challenge for researchers regarding fake news is its early detection, controlling its extensive spread and limiting its impact on society.  In this survey paper, we extensively analyze a wide range of different solutions that are used for the early detection of fake news in the existing literature. More precisely, we look at how researchers formulate Machine Learning~(ML) models for the automatic identification and classification of fake news. We discuss online competitions for fake news detection followed by the statistical outputs, as well as the advantages and disadvantages of some of the available data sets and evaluate the online web browsing tools available for detecting and mitigating fake news. Finally, various open research challenges that can direct future research are identified and addressed. 
\end{abstract}

\keywords{}

\maketitle

\section{Introduction}
\label{sec:intro}

The popularity of Online Social Networks (OSNs)  has rapidly increased in recent years. Social media has shaped the digital world to an extent it is now an indispensable part of life for most of us~\cite{gazi2017research}. Rapid and extensive adoption of online services is influencing and changing how we access information, how we organize to demand political change and how we find partners. One of the main advantages and attractions of social media is the fact that it is fast and free. This technology has dramatically reshaped the news and media industries since becoming a dominant and growing source of news and information for hundreds of millions of people~\cite{kaplan2015social}. In the United States today more people are using social media as a news source than ever before~\cite{Peter2019}. Social media has progressively changed the way we consume and create news. The ease of producing and distributing news through OSNs has also simultaneously sharply increased the spread of fake news.

Fake news is not a new phenomenon; it existed long before the arrival of social media. However, following the~2016 US presidential election it has become a buzzword~\cite{Jonathan}. There are numerous examples of fake news trough history. A notable one from  the antiquity is the Mark Anthony smear campaign circa~44 BC~\cite{posetti2018short}. In more recent times, examples include the anti-German campaign, German corpse factory in~1917~\cite{neander2010media} and the Reich Ministry of Public Enlightenment and Propaganda established in~1933 by the Nazis to spread Nazi ideology and incite violence against Jews~\cite{bytwerk2010grassroots}.

Although propaganda campaigns and spread of fabricated news may have been around for centuries, their fast and effective dissemination only became possible by means of a modern technology such as the internet. The internet revolutionized fake news, regardless of how the misinformation is manifested: whether we are talking about a rumor, disinformation, or biased, sloppy, erroneous reporting. In a recent study~\cite{All:Fake:News:Traffic:Facebook}, it was found that almost~50 percent of traffic taken from Facebook is fake and hyperpartisan, while at the same time, news publishers relied on Facebook for~20 percent of their traffic. In another study, it was found that~8 percent of~25 million Universal Resource Locator (URLs) posted on social media were indicative of malware, phishing and scams~\cite{thomas2013role}. %Also, in~2017, a committee in Egypt identified the distribution of~53,000 false rumors within a two-month period~\cite{Egypt:Fake:News:2017}.

Researchers in Germany conducted a study regarding fake news distribution in the country and people's attitudes and reactions towards it~\cite{Fake:News:Perception:Germany}. Based on the published results,~59 percent of participants stated that they had encountered fake news; in some regions, this number increased to almost~80 percent~\cite{TheLawLibrary}. Furthermore, more than~80 percent of participants agreed fake news poses a threat and~78 percent strongly believed it directly harms democracy. 
%Fake news is now becoming a significant threat to democracy. 
Government institutions and powerful individuals use it as a weapon against their opponents~\cite{Thetelegraph}. In the~2016 US presidential election, a significant shift in how social media was used to reinforce and popularize narrow opinions was observed. In November of the same year,~159 million visits to fake news websites were recorded~\cite{allcott2017social}, while the most widely shared stories were considered to be fake~\cite{Buzzfeednews}. Similarly, it is believed that the distribution of fake news influenced the UK European Union membership referendum~\cite{Independent}. 

However, fake news is not only about politics. During the recent fires in Australia, several maps and pictures of Australia's unprecedented bushfires spread widely on social media. While users posted them to raise awareness, the result was exactly the opposite since some of the viral maps were misleading, spreading disinformation that could even cost human lives~\cite{rannard2020australia}. The recent COVID-19 pandemic accelerated the rise of conspiracy theories in social media. Some were alleging that the novel coronavirus is a bio-weapon funded by Bill Gates to increase the selling of vaccines~\cite{Madeforminds}. Undoubtedly fake news threaten multiple spheres of life and can bring devastation not only to economic and political aspects but peoples' wellbeing and lives.

\subsection{An overview of this Survey}
\label{subsec:overviewofsurvey}
%Since fake news affects different aspect of our life ranging from democratic processes, general well-being, social conflict etc, this has warranted numerous research into ways and means to eradicate this problem. A keyword search in scopus using the keyword `fake news' generated numerous articles that have discussed this problem, meaning that there is tremendous interest from researchers all over the world pertaining to this subject. 
The main motivation behind our study was to provide a comprehensive overview of the methods already used in fake news  detection as well as bridge the knowledge gap in the field, thereby helping boost interdisciplinary research collaboration. This work's main aim is to provide a general introduction to the current state of research in the field. 

We performed an extensive search of a wide range of existing solutions designed primarily to detect fake news. The studies used deal with identification of fake news based on ML models, network propagation models, fact-checking methods etc.  More precisely, we start by examining how researchers formulate ML models for the identification and classification of fake news, which tools are used for detecting fake news and conclude by identifying open research challenges in this domain. 

\paragraph{Comparison to Related Surveys} 
In a related work by Vitaly Klyuev~\cite{klyuev2018fake}, an overview of the different semantic methods by concentrating on Natural Language Processing (NLP) and text mining techniques was provided. In addition, the author also discussed automatic fact-checking as well as the detection of social bots. In another study, Oshikawa \textit{et al.}~\cite{oshikawa2018survey} focused on the automatic detection of fake news using NLP techniques only. Two studies can be singled out as being the closest to our work. First, Study by Collins \textit{et al.}~\cite{collins2020trends} which examined fake news detection models by studying the various variants of fake news and provided a review of recent trends in combating malicious contents on social media. Second, a study by Shu \textit{et al.}~\cite{shu2020combating} which mostly focused on various forms of disinformation, factors influencing it and mitigating approaches.
 
Although some similarities are inevitable, our work varies from the aforementioned ones. We provide a more detailed description of some of the approaches used and highlight the advantages and limitations of some of the methods. Additionally, our work is not limited to NLP techniques, but also examines types of detection models available, such as, knowledge-based approaches, fact-checking (manual and automatic) and hybrid approaches. Furthermore, our approach considers how the NLP techniques are used for the detection of other variants of fake news such as rumors, clickbaits, misinformation and disinformation. Finally, it also examines the governmental approaches taken to combat fake news and its variants. 
%For the same purpose, we have also explored the tools and data resources that are available.
 
\subsection{Organization}
\label{subsec:organization}
The rest of this paper is organized as follows: Section~\ref{sec:fakenewsanalysis}  discusses the most important methods for detecting fake news, in Section~\ref{sec:factchecking}, we detailed both the automatic and manual assessment of news and analyzed different ways of measuring the relevance, credibility and quality of sources. To automate the process of fake news detection, the analysis of comprehensive data sets is of paramount importance. To this end, in Section~\ref{sec:toolsanddataset}, we first discuss the characteristics of online tools used for identifying fake news and then compare and discuss different data sets used to train ML algorithms to effectively identify fake news. The classification of existing literature, identified challenges, future directions and existing governmental strategies to tackle the problem of fake news detection are discussed in Section~\ref{sec:discussionandchallenges}. Finally, the concluding remarks are given in Section~\ref{sec:Conclusion}. 

%\section{Governmental Strategies and Fake News}
%\label{sec:governmentstrategies}
%
%%\subsection{Types of Fake News}
%%\label{subsec:Types of Fake News}
%
%As not all fake news is the same nor easily spotted, this section starts by reviewing how the fake news is presented in current research. Afterwards, the governmental initiatives taken to tackle the problem of fake news are discussed.

%\textcolor{red}{Following this, we will tackle the problem of spotting fake news, describe a set of possible solutions for successfully detecting disinformation of this nature, present the most important methods for detecting fake news and analyze different ways of measuring the relevance, credibility and quality of sources. Fake news is not a new term, it has a long legacy reaching back centuries.However, following the 2016 US presidential elections, it has become a buzzword~\cite{Jonathan}.}  \textcolor{red}{In this review we have conducted an extensive literature search of article concerning fake news, many of the articles reviewed classified fake news into the following categories~\cite{tandoc2018defining, rubin2015deception}:}

%Based on the current literature review~\cite{tandoc2018defining, rubin2015deception}, fake news can be classified into following types:\textcolor{red}{Certain forms refer to human prejudices or human errors~(misinformation) in the categories above, and others are true fake news~(disinformation). They are collectively referred to as fake news because both have a loose link to real news while the intention is the same~(to deceive).}

\section{Fake News Analysis}
\label{sec:fakenewsanalysis}
People are heavily dependent on social media for getting information and spend a substantial amount of time interacting on it. In~2018, the \href{https://www.journalism.org/2018/09/10/news-use-across-social-media-platforms-2018/}{Pew Research Center} revealed that~68 percent of Americans~\cite{ElisaShearer} used social media to obtain information. On average,~45 percent of the world's population spend~2 hours and~23 minutes per day on social media and this figure is constantly increasing~\cite{Evanasano}. The biggest problem with information available on social media is its low quality. Unlike the traditional media, at the moment, there is no regulatory authority checking the quality of information shared on social media. The negative potential of such unchecked information became evident during the the~2016 US presidential election\footnote{\url{https://www.independent.co.uk/life-style/gadgets-and-tech/news/tumblr-russian-hacking-us-presidential-election-fake-news-internet-research-agency-propaganda-bots-a8274321.html}}. In short, it is of paramount importance to start considering fake news as a critical issue that needs to be solved.

In spite of the overwhelming evidence supporting the need to detect fake news, there is, as yet,  no universally accepted definition of fake news. According to~\cite{lazer2018science}, ``fake news is fabricated information that mimics news media content in form but not in organizational process or intent''. In a similar way, fake news is defined as ``a news article that is intentionally and verifiable false''~\cite{shu2017fake}. Some articles also associate fake news with terms such as deceptive news~\cite{allcott2017social}, satire news~\cite{rubin2015deception}, clickbait~\cite{chen2015misleading}, rumors~\cite{zubiaga2018detection}, misinformation~\cite{kucharski2016study}, and disinformation~\cite{kshetri2017economics}. Hence, these terms are used interchangeably in this survey. 

The following forms of misuse of information have been considered as variants of  fake news in the existing literature~\cite{tandoc2018defining, rubin2015deception}: 
\begin{itemize}
	\item \textbf{Clickbait:} Snappy headlines that easily capture user attention without fulfilling user expectations since they are often tenuously related to the actual story. Their main aim is to increase revenue by increasing the number of visitors to a website.
	\item \textbf{Propaganda:} Deliberately biased information designed to mislead the audience. Recently, an increased interest has been observed in propaganda due to its relevance to the political events~\cite{rubin2015deception}.
	\item \textbf{Satire or Parody:} Fake information published by several websites for the entertainment of users such as \href{https://www.thedailymash.co.uk/}{``The Daily Mash'' website.} This type of fake news typically use exaggeration or humor to present audiences with news updates.  
	\item \textbf{Sloppy Journalism:} Unreliable and unverified information shared by journalists that can mislead readers.
	\item \textbf{Misleading Headings:} Stories that are not completely false, but feature sensationalist or misleading headlines.
	\item \textbf{Slanted or Biased News:} Information that describes one side of a story by suppressing evidence that supports the other side or argument.
\end{itemize} 
 
For years, researchers have been working to develop algorithms to analyze the content and evaluate the context of information published by users. 
%For instance, in~\cite{shu2017fake}, the problem of fake news detection in social media is thoroughly discussed. The study identifies and classifies the process in two distinct phases: the \textit{characterization} phase and the \textit{detection} phase. To analyze the characterization phase, they focused on the basic concepts and principles of fake news in both traditional and social media, while the detection phase consisted of a review of fake news detection using a data mining approach.
%\textcolor{red}{In the same way, the following subsections look at how researchers use various social media features and suggest solutions to the problem of fake news.} 
Our review of the existing literature is organised in the following way: subsection~\ref{subsec:detectingmaliciousbots}, examines approaches to identifying different types of user accounts such as bots, spammers and cyborgs. It is followed by subsection~\ref{subsec:rumorsandclickbaits}, where different methods used for identifying rumors and clickbaits are discussed. In subsection~\ref{subsec:contentandcontextanalysis}, the users' content and context features are considered while in subsection~\ref{subsec:networkpropagation}, different approaches for the early detection of fake news by considering its propagation are discussed.

%, whereas different computation approaches in fact-checking for claims are discussed in subsection~\ref{subsec:FC}. Finally, by taking into account the credibility, relevance, and quality of the source, we address the state of the art approaches adopted by the researcher to mitigate the propagation of fake news in subsection~\ref{sec:TAC}. 
 
%\subsection{Account Analysis}
%\textcolor{blue}{\subsection{Account Classification into Users and Automated Agents}}
\subsection{User Account Analysis}
\label{subsec:detectingmaliciousbots}

According to a report published in~2021 Twitter alone has~340 million users,~11.7 million registered apps, delivers~500 million tweets a day and~200 billion tweets a year~\cite{Aslam2021}. It's popularity has made it an ideal target for bots, or automated programs~\cite{kaur2018rise}. Recently, it was reported that around~5-10 percent of Twitter accounts are bots and responsible for the generation of~20-25 percent of all tweets~\cite{UEFL}. Some of the bots are legitimate, comply with Twitter objectives, and can generate a substantial volume of benign tweets like blogs and news updates. Other bots, however, can be used for malicious purposes such as a malware that gathers passwords or a spam that adds random users as friends and expects to be followed back~\cite{Digital}. Such bots have a more detrimental effect particularly when spreading fake news. The significance of differentiating the legitimate bots from the malicious ones emerged from the fact that malicious bots can also be used to mimic human behaviour in a negative way. 

Researchers examined bots, in a number of existing publications~\cite{cresci2017paradigm, gibert2020rise, edwards2014bot, lee2011seven, wu2013detecting, stone2009your}. Gilani \textit{et al.}~\cite{gilani2017classification} focused on classifying Twitter accounts into ``human'' and ``bots'' and analyzing the impact each has on Twitter. The proposed technique was based on previous work by ``Stweeler''~\cite{STCSZafar} for the collection, processing, analysis, and annotation of data. For the identification of bots, human annotation was used, where participants differentiated bots from humans and generated a reliable data set for classification. The process provided an in-depth characterization of bots and humans by observing differences and similarities. The finding stated that the bots' removal from Twitter causes serious repercussions for content production and information dissemination and also indicated that bots count on re-tweeting, redirecting users via URLs, and uploading media. However, the imprecision in the existing algorithm revealed by the authors and the manual collection of data limited the ability to analyse accounts. 

Similarly, Giachanou \textit{et al.}~\cite{giachanou2019bot} investigated whether the Twitter account author is human or a bot and further determined the gender of a human account. For this purpose, a linear Support Vector Machines (SVM) classifier was trained to analyse words, character grams, and stylistic features. For the identification of human gender, a stochastic gradient descent classifier was used to assess the sentiment of tweets, words, and character grams and point wise mutual information features -- the importance of terms per gender. The data set used consisted of tweets in English and Spanish. The experiments illustrated the accuracy of bot detection, i.e~0.906 for bots in English and~0.856 for Spanish. Similarly, for the identification of gender, the accuracy for English tweets amounted to~0.773 and~0.648 for Spanish tweets. In the long run, the bot detection model outperformed the gender detection model. 

Another account type that can be generated on Twitter is a Cyborg. Cyborg refers to a human-assisted bot or bot-assisted human~\cite{chu2010tweeting}. Cyborgs have characteristics of both human-generated and bot-generated accounts and as such require a level of human engagement.  These accounts faciltate posting various information more frequently, rapidly and long-term~\cite{DFRLab}. Differentiating a cyborg from a human can be a challenging task. The automated turing test~\cite{von2004telling} used to detect undesirable or bot programs is not capable of differentiating cyborgs from humans. However, Jeff Yan~\cite{yan2006bot} proposed that a cyborg might be differentiated by comparing the characteristics of a machine and human elements of a cyborg. Similarly, Chu \textit{et al.}~\cite{chu2012detecting} differentiate between bot, cyborg and human accounts by taking into account tweet content, tweeting behaviour and features of the account.

OSNs also serve as platforms for the rapid creation and spread of spam. Spammers act similarly to bots and are responsible for posting malicious links, prohibited content and phishing sites~\cite{Sarah, Michalas:17:Pies:CCS}. %Unsolicited information disseminated by social spammers has a negative impact on the social networking system. 
Traditional methods of detecting spammers that utilize network structure are classified into three categories:

\begin{itemize}
	\item Link-based, where the number of links is used as a measure of trust. These links are considered to be built by legitimate users~\cite{lee2011seven}.
	\item Neighbor-based, which treats links as a measure of homophily, the tendency for linked users to share similar beliefs and values~\cite{hu2013social,rayana2015collective, li2016robust}.
	\item Group-based, which recognizes that spammers often work in groups to coordinate attacks~\cite{jindal2007review}. Group-based methods detect spammers by taking advantage of the group structure hidden in the social network. Additionally, spammers behave differently from legitimate users so they can be treated as outliers~\cite{gao2010community, akoglu2015graph}.
\end{itemize}    

Current efforts for detection of social spammers utilize the structure and behavioural patterns of social spammers in an attempt to discover how their behaviour can be differentiated from legitimate users~\cite{lim2010detecting, chu2010tweeting, ye2016temporal,li2015analyzing, xue2013votetrust, yang2014uncovering}. However, spammers often find ways to create a link with legitimate users, making it more difficult to detect specific spamming patterns. Wu \textit{et al.}~\cite{wu2017adaptive} tackled this problem by taking into account both content and network structure. They proposed ``Sparse Group Modelling for Adaptive Spammer Detection (SGASD)'' that can detect both types of spammers -- those within a group and individuals.

Another challenging task is detection of camouflaged content polluters on OSNs. Content polluters -- spammers, scanners and fraudsters -- first establish links with a legitimate user and then merge the malicious with real content. Due to insufficient label information available for camouflaged posts in online media, the use of these manipulated links and contents as camouflage makes detecting polluters very difficult. In order to tackle this challenge, Wu \textit{et al.}~\cite{wu2017detecting} studied how camouflaged content polluters can be detected and proposed a method called ``Camouflaged Content Polluters using Discriminate Analysis~(CCPDA)'' which can detect content polluters using the patterns of camouflaged pollution.
 
Chris \textit{et al.}~\cite{grier2010spam} spam detection analysis juxtaposed two different types of Twitter accounts -- a ``professional spamming account'' whose sole purpose is to distribute spam, versus ``accounts compromised by spammers''. The authors found that accounts currently sending spam had been compromised by spammers; once legitimate, they became controlled by spammers. Furthermore, to detect spam activity on Twitter, a directed social graph model~\cite{wang2010don} based on friend and follower relationships was proposed. Different classifier techniques were used to distinguish between the spammer and normal behaviour and determined that the Naive Bayes classifier performs better with respect to F-measure. 

Huge momentum has been observed where user-generated content is exploited in micro-blogs for predicting real-world phenomena such as prices and traded stock volume on financial markets~\cite{cresci2016dna}. Research efforts in this domain targeted sentiment metrics as a predictor for stock prices~\cite{bollen2011modeling, chen2014wisdom, gabrovvsek2017twitter}, company tweets and the topology of the stock network~\cite{mao2012correlating, ruiz2012correlating} and used weblogs pointing to the relationship between companies~\cite{kharratzadeh2012weblog}. Cresci \textit{et al.}~\cite{cresci2019cashtag} demonstrated the use of twitter stock micro-blogs as a platform for bots and spammers to practice cash-tag piggybacking -- an activity for promoting low-value stocks by exploiting the popularity of real high-value stocks. They employed a spambot detection algorithm to detect accounts that issue suspicious financial tweets. Nine million tweets from five main US financial markets, which presented stocks with unusually high social significance compared to their low financial relevance, were investigated with respect to their social and financial significance. These tweets were compared with financial data from Google finance. The results indicated that~71 percent of users were classified as bots and that high discussion of low-value financial stocks was due to a massive number of synchronized tweets.
%Twitter currently has no defined policy for addressing automated malicious programs. In the near future, it is expected that these malicious accounts will be deleted~\cite{Oscar}. However, in the existing literature~\cite{ercsahin2017twitter,lee2010uncovering, gilani2017classification, davis2016botornot, antoniadis2015model,weimer2007automatically, adewole2017malicious, wanas2008automatic, weerkamp2008credibility, morris2012tweeting}, the important characteristics that are used for the identification of fake accounts in Twitter are studied which consist of; mostly unprotected accounts with empty description and default account image, few followers as compared to the number of following, have a high-status favorite and listed count, active for a short period and publish duplicate tweets, acts as a phisher and involve in the promotion. 

Twitter currently has no defined policy for addressing automated malicious programs operating on its platform. However, it is expected that these malicious accounts will be deleted in the near future~\cite{Oscar}. A survey of the literature has identified numerous studies~\cite{ercsahin2017twitter,lee2010uncovering, gilani2017classification, davis2016botornot, antoniadis2015model,weimer2007automatically, adewole2017malicious, wanas2008automatic, weerkamp2008credibility, morris2012tweeting} that describe the important characteristics which can be used for the identification of bots on Twitter. Despite these attempts, limitations still exist in employing these characteristics for detecting fake news, especially, early detection of fake news during it's propagation. Other methods, such as network propagation, have to be utilized for this purpose.

\subsection{Identifying Rumors and Clickbaits}
\label{subsec:rumorsandclickbaits} 

Social media is like a blank sheet of paper on which anything can be written~\cite{Niam}, and people easily become dependent on it as a channel for sharing information. This exactly is the reason why social media platforms (e.g. Twitter and Facebook) are highly scrutinized for the information shared on them~\cite{haralabopoulos2015lifespan}. These platforms have undertaken some efforts to combat the spread of fake news but have largely failed to minimize its effect. For instance, in the United States,~60 percent of adults  who depend on social media for news consumption are sharing false information~\cite{Queenie}. In April~2013, two explosions during the \href{https://edition.cnn.com/2013/06/03/us/boston-marathon-terror-attack-fast-facts/index.html}{Boston Marathon} gained tremendous notoriety in the news and on social media, and the tragedy was commented on in millions of tweets. However, many of those tweets were rumors (controversial factual claims) and contained fake information, including conspiracy theories. Similarly, a survey published by \href{https://www.kroll.com/en}{Kroll} -- a business intelligence and investigating firm -- states that~84 percent of companies feel threatened by the rise of rumors and fake news fuelled by social media~\cite{CarolineBinham}. On \href{https://en.wikipedia.org/wiki/Sina_Weibo}{Weibo}, rumors were detected in more than one-third of trending topics~\cite{zhao2015enquiring}. The spread of rumors on social media has also become an important issue for companies worldwide. Still, there is no clear policy defined by social media administrators to verify shared content. Below, we discuss different techniques that have been proposed by researchers to address this problem. 

Traditionally, human observers have been used to identify trending rumors. Currently, research is focused on building an automated rumor identification tool. For this purpose, a rumor detection technique~\cite{zhao2015enquiring} was designed. In this technique, two types of clusters were generated: posts containing words of inquiry such as ``Really'', ``What'', ``Is it true?'' were grouped into one cluster. These inquiries were then used to detect rumor clusters. Similarly, posts without words of inquiry were grouped into another cluster. Similar statements were extracted from both clusters. The clusters were then ranked, based on their likelihood of containing these words. Later, the entire cluster was scanned for disputed claims. These experiments, performed with Twitter data, resulted in earlier and effective detection of rumors (almost~50 rumor clusters were identified). However, there is still considerable space to improve these results~\cite{zhao2015enquiring}. For instance, the manual collection of inquiry words could be improved by training a classifier and the process of ranking could be improved by exploring more features for the rumor cluster algorithm. 

People can share fake information on social media for various reasons. One of those is to increase readership, which is easily achievable by using clickbait. Clickbait is a false advertisement with an attached hyperlink. It is specifically designed to get users to view and read the contents inside the link~\cite{AnnaEscher}. These advertisements attract users with catchy headlines but contain little in the way of meaningful content. A large number of users are lured by clickbait. Monther \textit{et al.}~\cite{aldwairi2018detecting} provided a solution to protect users from clickbait in the form of a tool that filters and detects sites containing fake news. In categorizing a web page as a source of fake news, they considered several factors. The tool navigates the content of a web page, analyzes the syntactical structure of the links and searches for words that might have a misleading effect. The user is then notified before accessing the web page. In addition, the tool searches for the words associated with the title in the links and compares it with a certain threshold. It also monitors punctuation marks such as question and exclamation marks used on the web page, marking it as a potential clickbait. Furthermore, they examined the bounce rate factor--percentage of visitors who leave a website, associated with the web page. Where the bounce rate factor was high, the content was marked as a potential source of misleading information.

A competition was organised with the aim of building a classifier rating the extent to which a media post can be described as clickbait. In the \href{https://www.clickbait-challenge.org/}{clickbait competition}, the data set was generated from Twitter and consisted of~38,517 Twitter posts from~27 US news publishers~\cite{PotthastMartin}. Out of~38,517 tweets,~19,538 were available in the training set and~18,979 were available for testing. For each tweet, a clickbait score was assigned by five annotators from \href{https://www.mturk.com/}{Amazon Mechanical turk}. The clickbait scores assigned by human evaluators were:~1.0 heavily clickbaity,~0.66 considerably clickbaity,~0.33 slightly clickbaity and~0.0 not clickbaity.
% clickbaity~\cite{potthast2018clickbait}
The goal was to propose a regression model that could determine the probable clickbaitiness of a post. The evaluation metric used for the competition was Mean Squared Error (MSE). In this competition, Omidvar \textit{et al.}~\cite{omidvar2018using} proposed a model using the deep learning method and won the challenge. They achieved the lowest MSE \iffalse ~0.0315200660647 \fi for clickbait detection by using a bi-directional Gated Recurrent Unit~(biGRU). Instead of solving the clickbait challenge using a regression model, Yiewi zhou~\cite{zhou2017clickbait} reformulated the problem as a multi-classification. On the hidden state of biGRU, a token level self-attentive mechanism was applied to perform multi-classification. This self attentive Neural Network (NN) was trained without performing any manual feature engineering. They used~5 self-attentive NNs with a~80-20 percent split and obtained the second lowest MSE value. \iffalse of~0.033.\fi Similarly, Alexey Grigorev~\cite{grigorev2017identifying} proposed an ensemble of Linear SVM models to solve the clickbait problem and achieved the third lowest MSE value. \iffalse of~0.036.\fi In addition to the given data set, they gathered more data from multiple Facebook groups that mostly contained clickbait posts by using the approach described in ``identified clickbaits using ML''~\cite{AbhishekThakur}.

\afterpage{
	\setlength\LTleft{-.7cm}
	{\footnotesize
		\begin{longtable}{|p{0.18\textwidth}|p{0.2\textwidth}|p{0.13\textwidth}|p{0.19\textwidth}|p{0.15\textwidth}|p{0.09\textwidth}|}
			\hline
			\rowcolor{Gray}
			\textbf{Approach} & \textbf{Data set and Features} & \textbf{Evaluation Metrics} & \textbf{Finding or Outcomes} & \textbf{Weaknesses} & \textbf{Platform}\\
			\hline
			URL blacklisting for Spam detection~\cite{grier2010spam} & 400 million tweets, 25 million URLs&Click through, measuring delay& 8\% of 25 million URLs indicative of phishing, scams and malware&Inaccurate when used for web services~\cite{thomas2011design}&Twitter\\ \hline
			Bayesian classifier for Spam detection~\cite{wang2010don} & 25 K Users, 500 K tweets, 49 million follower/friend&Precision& Web crawler, directed social graph model, 89\% precision in 
			spam detection&Manual analysis of collected data~\cite{lee2013warningbird}&Twitter\\ \hline
			SVM for Spam detection~\cite{benevenuto2010detecting}&  54 million users, 1.9 billion links, 1.8billion tweets& Precision, recall, Micro F-1, Macro F-1& Correctly classified: 70\% spammers, 96\% non-spammers&Manually labelled data set~\cite{ghosh2012understanding}&Twitter\\ \hline
			Naive bayes for account classification~\cite{ercsahin2017twitter}& 501 fake accounts, 499 real accounts, profile and tweets & ROC curve, F1 score, confusion matrix& Accuracy 90.9\%&Manually labelled data set~\cite{alothali2018detecting}&Twitter\\ \hline
			\hline
			Ranking model for rumor detection~\cite{zhao2015enquiring}&10,240,066 tweets, keyword search&Precision, detection time& Clustering and classification performs effectively, earlier detection of rumors&More features can be explored for rumor clustering&Twitter\\
			\hline
			SVM-rank for account classification~\cite{gilani2017classification}&60 million tweets , tweets frequency&Classification accuracy, precision, recall and $F_{1}$ measure&Develop and evaluate a mechanism to classify automated agents and human users &Cannot check relative frequency of any particular URL~\cite{nasim2018real}&Twitter\\ 
			\hline
			SVM and Stochastic Gradient Descent for bots and gender profiling~\cite{giachanou2019bot}&English and Spanish Tweets, textual, stylistic&Accuracy&Words and char grams are important feature for gender and bot detection&-&Twitter\\ \hline
			SGASD for spam detection~\cite{wu2017adaptive}&TwitterS, TwitterH, Network, content&Precision, recall, $F_{1}$&Present SGASD framework for spammer detection&Network information focuses on user instead of information&Twitter\\ \hline
			Logistic Regression for stance detection~\cite{ferreira2016emergent}&300 rumors, 2,595 news
			articles, headlines&Accuracy, precision, recall&Emergent Dataset used for a variety of NLP tasks&Data set~(cannot learn all nuances of tasks)&Emergent project\\ \hline
			
			\caption{Detailed Summary of the Studies used in Bots, Clickbaits and Rumors}
			\label{tab:botsandclickbaits}
\end{longtable}}}

\subsection{Content and Context Analysis for Fake News}
\label{subsec:contentandcontextanalysis}

The rapid dissemination of fake news is so pernicious that researchers resolved towards trying to automate the process by using ML techniques such as Deep Neural Networks (DNN). However, the Black box problem -- a lack of transparency in decision-making in the NN -- obscures reliability. Nicole \textit{et al.}~\cite{o2018language} addressed the deep learning ``black-box problem'' for fake news detection. A data set composed of~24,000 articles was created, consisting of~12,000 fake and~12,000 genuine articles. The fake articles were collected from \href{https://www.kaggle.com/mrisdal/fake-news}{Kaggle} while genuine ones were sourced from \href{https://open-platform.theguardian.com/access/}{The Guardian} and \href{https://developer.nytimes.com/}{The New York Times}. The study concluded that DNNs can be used to detect the language patterns in fabricated news. Additionally, the algorithm can also be used for detecting fake news in novel topics. 
%However, the study has limitations and can only predict the factual of claims by mainly focusing on only the input claims~(metadata information).
% (metadata information)~\cite{nadeem2019fakta}

Another technique to tackle the deep learning ``black-box problem'' in fake news detection is CSI (capture, score and integrate) -- a three-step system which incorporates the three basic characteristics of fabricated news~\cite{ruchansky2017csi}. These characteristics include text, source, and the response provided by users to articulate missing information. In the first step, a Recurrent Neural Network (RNN) is used to capture the momentary pattern of user activity. The second step estimates the source of suspicions related to user behaviour. The third, hybrid step involves integration of steps one and two and is used to predict fake articles. The experiments were performed on real-world data sets and demonstrated a high level of accuracy in predicting fake articles. Still, a bottleneck in using a computationally intensive model is posed by the lack of a manually labelled fake news data set. William Yang Wang~\cite{wang2017liar} addressed the limited availability of labelled data sets for combating fake news using statistical approaches and chose a contemporary publicly available data set called LIAR. This data set was utilized to investigate fabricated news using linguistic patterns. The results were based on an evaluation of several approaches, including Logistic Regression (LR), the Convolution Neural Network (CNN) model, Long Short-Term Memory (LSTM) networks and SVM. They concluded that combination of meta-data with text significantly improves the detection of fake news. According to the authors, this body of information can also be used to detect rumors, classify stance and carry out topic modeling, argument mining, and political Natural Language Processing (NLP) research. Table~\ref{tab:botsandclickbaits} present a summary of the different approaches proposed for both the account as well as content and context analysis of fake news.

%\textcolor{red}{Twitter, and other social networking websites, have been very reluctant to act on fake news. Buket \textit{et al.}~\cite{ercsahin2017twitter} conducted a study to distinguish between fake and real accounts, using a test group of 1,000~accounts, and determined that 501 were fake and 499 were genuine. The determination (fake or real) is made by examining the profile image and background, user name, number of friends and followers, number of tweets, account description and tweet contents. In total, 16 features are used to generate the data set. The data is pre-processed using ``Entropy Minimization Discretization~(MDE)''. The Naive Bayes learning algorithm results are analyzed with an accuracy rate of 90.9 percent.}

In~2017 a competition named \href{http://www.fakenewschallenge.org/\#fn:4}{Fake News Challenge (FNC)} was held with the aim to use Artificial Intelligence (AI) techniques to combat the problem of fake news. During the initial phase, stance detection was used. It refers to the relative stance to any issue, claim or topic made by two pieces of relevant text (what other organizations say about the topic). A two-level scoring system was applied -- ~25 percent weight was assigned if the text was deemed to be related or unrelated to its headline and ~75 percent weight was assigned on the basis of labelling the related pairs as agrees, disagrees, discusses or unrelated. In this competition, the top team submitted an ensemble model for a Deep Convolution Neural Network (DCNN) and Gradient-Boosted Decision Tree (GBDT) with a weighted average of~50/50~\cite{Sean}. The DCNN and GBDT separately did not achieve perfect accuracy. However, the combination of both approaches correctly detected the stance of each headline with a score of~82.01. Similarly, approach proposed by team Athene~\cite{Andreas} achieved a score of~81.97 and won second place in the competition. They used an ensemble approach involving multi-layer perception and applied MLP and Bag-of-Word (BoW) features to the challenge. The team in third place, Riedel \textit{et al.}~\cite{riedel2017simple}, proposed a stance detection system for FNC Stage~1. For the input text, they used two BoW representations. A MLP classifier was used with one hidden layer having~100 units. For the hidden layer, a Rectified Linear Unit (RELU) activation function was used while the final linear layer utilized a soft-max. They achieved an accuracy of~81.72.

At a different competition named \href{http://www.wsdm-conference.org/2019/}{Web Search and Data Mining (WSDN)~2019}, fake news was detected by classifying the titles of articles. Using a given title for any fake news article `A' and a title for another incoming news article `B', people were asked to classify the incoming article into one of three categories: agrees, disagrees and unrelated~\cite{Kaggle}. The winner of this competition Lam Pham~\cite{pham2019transferring}, who achieved~88.298 percent weighted accuracy on the private leader boards and~88.098 percent weighted accuracy on the public leader boards. This ensemble approach incorporated NNs and gradient boosting trees. In addition, Bidirectional Encoder Representation from Transformer (BERT) was used for encoding news title pairs, transforming and incorporating them into a new representational space. The approach by Liu \textit{et al.} won a second place~\cite{liu2019trust} by proposing a novel ensemble framework based on the Natural Language Interference (NLI) task. Their proposed framework  for fake news classification consisted of three-level architecture with a~25 BERT model along with a blending ensemble strategy in the first level followed by~6 ML models and finally a single LR for the final classification.  Yang \textit{et al.}~\cite{yang2019fake} also considered this problem as a NLI task and considered both the NLI model as well as the BERT. They trained the strongest NLI models, Dense RNN, Dense CNN, ESIM, Gate CNN~\cite{dauphin2017language} and decomposable attention, and achieved an accuracy of~88.063 percent. 

\subsection{Network Propagation and Detection of Fake News}
\label{subsec:networkpropagation}
%\textcolor{red}{So far, researchers have primarily exploited various data features to tackle the problem of fake news detection, which significantly limits early detection of fake news. Additionally, the accuracy of early detection of fake news is very low. } 

One of the OSNs main strong points is facilitating the propagation of information between users. The information of interest to users is further shared with relatives, friends, etc~\cite{kong2019academic}. In order to detect the propagation of fake news at its early stage, it is crucial to  be able to understand and measure the information propagation process. The influence of propagation on OSNs and their impact on network structure was studied in~\cite{saxena2015understanding, hong2011predicting}.  Ye \textit{et al.}~\cite{ye2010measuring} study revealed that more than~45.1 percent of information shared by a user on social media is further propagated by his/her followers. Furthermore, approximately~37.1 percent of the information shared is propagated up to~4 hops from the original publisher. %Three types of heterogeneous networks (knowledge, stance and interaction networks) are discussed~\cite{shu2019studying} in the analysis of fake news. 

Liu and Wu~\cite{liu2018early} used the data network features and introduced a popular network model for the early detection of fake news. They addressed the limitation of low accuracy of early fake news detection by classifying news propagation paths as a multivariate time series. Characteristics of each user involved in spreading news   were represented by a numerical vector. Then a time series classifier was built by combining CNN and RNN. This classifier was used for fake news detection by capturing the local and global variations of observed characteristics along the propagation path. This model is considered as more robust, as it relies on common user characteristics which are more reliable and accessible in the early stage of news propagation. The experiments were performed on two real-world data sets based on Weibo~\cite{ma2016detecting} and Twitter~\cite{ma2017detect}. The proposed model detected fake news within~5 minutes of its spread with~92 percent accuracy for Weibo and~85 percent accuracy for Twitter data sets. %In addition, the authors intend to study the semi-supervised learning techniques to deal with extensive unlabeled articles on social media.

%One of the challenges is to early detect fake news while maintaining strict confidentiality. Different models have been proposed to achieve this goal, such as: user flagging activity~\cite{kim2018leveraging}, building predictive models~\cite{volkova2017separating}, etc. 
Sebastian \textit{et al.}~\cite{tschiatschek2018fake} examined the ways to minimize the spread of fake news at an early stage by stopping its propagation in the network. They aggregated user flagging, a feature introduced by Facebook that allows users to flag fake news. In order to utilize this feature efficiently, the authors developed a technique called `DETECTIVE' which uses Bayesian Inference to learn flagging accuracy. Extensive experiments were performed by using a publicly available data set~\cite{leskovec2012learning} from Facebook. The results indicated that even with minimum user engagement DETECTIVE can leverage crowd signals to detect fake news. It delivered better results in comparison to existing algorithms, i.e. NO-Learn and RANDOM. 

The dissemination of misinformation on OSNs has a particularly undesirable effect when it comes to public emergencies. Dynamic Linear Threshold (DLT) model~\cite{litou2016real} was developed to attempt and limit this type of information. It analyzes the user's probability, based on an analysis of competing beliefs, of propagating either credible or non-credible news. Moreover, an optimization problem based on DLT was formulated to identify a certain set of users that could be responsible for limiting the spread of misinformation by initiating the propagation of credible information.  

A study by Garcia \textit{et al.}~\cite{garcia2017understanding} focused on examining reputation~\cite{Michalas:12:StR,Michalas:14:StRM,Michalas:14:Lord,Michalas:20:NordSec:FunctionalSift}, popularity and social influence on Twitter using digital traces from~2009 to~2016. They evaluated the global features and specific parameters that make users more popular, keep them more active and determine their social influence. Global measures of reputation were calculated by taking into account the network information for more than~40 million users. These new global features of reputation are based on the D-core decomposition method~\cite{giatsidis2013d} and The Twitter's bow-tie structure~\cite{broder2000graph} in~2009. 
%The analysis was performed on large-scale digital traces that consisted of~40 million users collected for~7 years. 
The results indicated that social influence is more related to popularity then reputation, and global network metrics such as social reputation are more accurate predictors for social influence than local metrics such as followers, etc. 

Soroush \textit{et al.}~\cite{vosoughi2018spread} collected and studied twitter data from~2006 to~2007 in order to classify it as true or false news. News is classified as true or false based on information collected from six independent fact-checking organizations. They generated a data set that consisted of approximately~126,000 tweets, tweeted by~3 million twitter users approximately~4.5 million times. They found that fake news was more novel and inspired surprise, fear, and disgust in replies, while true news inspired trust, sadness, anticipation, and joy. As people prefer to share novel information, false news spreads more rapidly, deeply and broadly than true news. According to Panos \textit{et al.}~\cite{constantinides2018introduction}, rapid dissemination of information on social media is due to information cascade. Liang Wu and Huan Liu~\cite{wu2018tracing} also classified twitter messages using diffusion network information. Instead of using content features, they focused on the propagation of Twitter messages. They proposed trace miner, a novel approach that uses diffusion network information to classify social media messages. Trace miner accepts message traces as inputs and outputs its category. Table~\ref{tab:networkandcontent} presents a detailed summary of the studies used in network as well as  content and context analysis.

\setlength\LTleft{-.7cm}
\afterpage{
	{\footnotesize
		\begin{longtable}{|p{0.18\textwidth}|p{0.2\textwidth}|p{0.13\textwidth}|p{0.2\textwidth}|p{0.17\textwidth}|p{0.07\textwidth}|}
			\hline
			\rowcolor{Gray}
			\textbf{Approach} & \textbf{Data set and Features} & \textbf{Evaluation Metrics} & \textbf{Finding or Outcomes} & \textbf{Weaknesses} & \textbf{Platform}\\
			\hline
			RNN and CNN for fake news detection~\cite{liu2018early}&Weibo~\cite{ma2016detecting}, Twitter15 and Twitter16~\cite{ma2017detect}, user profiles&Effectiveness, efficiency&Outperforms a state-of-the-art model in terms of both effectiveness and efficiency&Problem with computational efficiency and interpretability~\cite{zhou2019network}&Twitter, Weibo\\ \hline
			Bayesian inference for fake news detection~\cite{tschiatschek2018fake}&4,039 users, 88,234 edges, users and spammers&Utility, engagement, robustness&Outperforms NO-LEARN and RANDOM algorithms&Trustworthiness of news sources is ambiguous&Facebook\\ \hline
			Diffusion of network information for classification~\cite{vosoughi2018spread}& 126,000 stories, 3 million users, 4.5 million tweets, retweets, users&Diffusion dynamics&Fake news spreads rapidly and deeply and is more novel than true news&Information cascades~\cite{constantinides2018introduction}&Twitter\\ \hline			
			LSTM-RNN for fake news detection~\cite{wu2018tracing}&3,600 fake news, 68,892 real news, network information&Micro-F1, Macro-F1&Trace miner: classifying social media messages&Only considers network information&Twitter\\ \hline	
			Network flow model for fact-checking~\cite{shiralkar2017finding}&Synthetic corpora, real-world data set, Edge capacity&AUROC&Network flow techniques are promising for fact-checking&Limitation of content-based approach~\cite{pan2018content}& WSDM-Cup  2017~\cite{Hannah}\\ \hline
			Bow-Tie and D-core decomposition for user analysis~\cite{garcia2017understanding}&40 million users, 1.47 billion follower links&Reputation, social influence, popularity&Global metrics are more predictive than local&Theory-driven approach~\cite{hasani2018consensus}&Twitter\\ \hline
			Hybrid CNN for fake news detection~\cite{wang2017liar}&LIAR 12,836 short statements,Metadata and text features&Accuracy&LIAR data set, Integrate text and metadata&Justification and evidence are ignored in experiments \cmmnt{~\cite{thorne2018fever}}&\href{http://static.politifact.com/api/
				v2apidoc.html}{Politifact}\\ \hline
			RNN and user behavior for fake news detection~\cite{ruchansky2017csi}&Two real-world data sets (Twitter and Weibo)~\cite{ma2016detecting}, text &Classification accuracy&More accurate in fake news classification&No assumptions about user distribution behavior&Twitter, Weibo\\ \hline
			
			DNN for fake news detection~\cite{o2018language}&12,000 fake and 12,000 real news, language patterns&Accuracy&Observes subtle differences in language patterns of real and fake news&Only predicts truthfulness of claims&Different websites\\ \hline
			\caption{Detailed Summary of the Studies used in Network as well as Content and Context Analysis}
			\label{tab:networkandcontent}
\end{longtable}}}

After reviewing the studies discussed above, it became evident there is no `one size fits all' when it comes to fake news detection. Extensive research is still required to fully understand the dynamic nature of this problem.  %that diverse methods have been applied to the fake news detection problem.  
%For instance, some researchers have utilized feature-based methods and approaches based on the NN models. One limitation posed by this approach is that it requires human involvement. 
%Evidently, 

\section{Fact Checking}
\label{sec:factchecking}

The rapid spread of fraudulent information is a big problem for readers who fail to determine whether a piece of information is real or fake. Since fake news is a big threat to society and responsible for spreading confusion, it is necessary to have an efficient and accurate solution to verify information in order to secure the global content platform. To address the problem of fake news, the American media education agency \href{https://www.poynter.org/ifcn/}{Poynter} established the  \href{https://ifcncodeofprinciples.poynter.org/}{International Fact-Checking Network~(IFCN)} in~2015, which is responsible for observing trends in fact-checking as well as providing training to fact-checkers. A great deal of effort has already been devoted to providing a platform where fact-checking organizations around the world can use a uniform code of principles to prevent the spread of fake news. Two fact-checking organizations, \href{https://www.snopes.com}{Snopes} and \href{(https://www.politifact.com}{Politifact}, developed a fake news detection tool useful in classifying fake news levels in stages. However, this tool requires a lot of manual work. There is a profound need for a model that can automatically detect fake news. 
%First, we will look at different competitions that have been organized for detecting fake news. 
%Due to increasing use of the Internet, an enormous volume of information is generated. For this huge volume of information, the traditional fact-checking method where an expert journalist manually checks the credibility and reliability of information is not feasible. Manual fact-checking by a human being is, however, both important and incredibly daunting.
 
Giovanni \textit{et al.} reduced the complex manual fact-checking task to a simple network analysis problem~\cite{ciampaglia2015computational}, as such problems are easy to solve computationally. The proposed approach was evaluated by analyzing tens of thousands of statements related to culture, history, biographical and geographical information using a public knowledge graph extracted from Wikipedia. They found that true statements consistently receive higher support in comparison to false ones and concluded that applying network analytics to large-scale knowledge repositories provides new strategies for automatic fact-checking. Below, we examine two facets of fact checking problem. In~\autoref{subsec:automaticfactchecking} we look into computational approaches to automatic fact checking, whereas in~\autoref{subsec:trustandcredibility}, we concentrate on the issue of trust and credibility of the information and the source providing it. 

\subsection{Towards Automatic Fact Checking}
\label{subsec:automaticfactchecking}
Computational approaches to fact-checking are considered key to tackling the massive spread of misinformation. These approaches are scalable and effective in evaluating the accuracy of dubious claims. In addition, they improve the productivity of human fact-checkers.

One of the proposed approaches is an unsupervised network flow-based approach~\cite{shiralkar2017finding}, which helps to ascertain the credibility of a statement of fact. The statement of fact is available as a set of three elements that consist of the subject entity, the object entity, and the relation between them. First, the background information of any real-world entity is viewed on a knowledge graph as a flow of the network. Then, a knowledge stream is built by computational fact-checking which shows the connection between the subject and object of a set. The authors evaluated network flow model on actual and customized fact data sets and found it to be quite effective in separating true and false statements.

A study by Baly \textit{et al.}~\cite{baly2018predicting} examined on the factuality and bias of claims across various news media. They collected features from articles of the target news websites, their URL structures, the web traffic they attract, their twitter accounts (where applicable) as well as Wikipedia pages. These features were then used to train the SVM classifier for bias and factuality separately. 
%Estimated accuracy, macro-averaged $F_{1}$ score, Mean Average Error (MAE), and a variant of MAE (more robust to class variance) were used as an evaluation metrics. 
The evaluation, showed that the articles' features achieved the best performance on factuality and bias, Wikipedia features were somewhat useful for bias but not for factuality, and Twitter and URL features faired better in factuality than bias.

%\textcolor{red}{It is evident that inflammatory, hyperpartisan (extremely one-sided), hoax and emotional news spreads more successfully on social media. Martin \textit{et al.}~\cite{potthast2017stylometric} focused on the analysis of hyperpartisan news by assessing the writing style similarity between text categories. The researchers used hyperpartisan news involving a corpus of~1,627 articles manually reviewed by BuzzFeed professionals. They applied Unmasking -- a meta-learning approach for assessing style similarities between text groups. They managed to successfully distinguish hyperpartisan from mainstream news and satire news from both hyperpartisan and mainstream ones. However, the proposed approach didn't show promising results in detecting fake news.} 
 
A different approach for an automatic fake news detection~\cite{perez2017automatic} was based on several exploratory analyses to identify the linguistic differences between legitimate and fake news. It involved the introduction of two novel data sets, the first collected using both manual and crowdsource annotation, and the second generated directly from the web. Based on these, first several exploratory analyses were performed to identify the linguistic properties most common for fake news. Secondly, a fake news detector model based on these extracted linguistic features was built. They concluded that the proposed system performed better than humans in certain scenarios with respect to more serious and diverse news sources. However, human beings outperformed the proposed model in the celebrity domain.

OSNs are also used as a vector for the diffusion of hoaxes. Hoaxes spread uncontrollably as propagation of such news depends on very active users. At the same time, news organizations devote a great deal of time and effort to high-quality fact-checking of information online. Eugenio \textit{et al.}~\cite{tacchini2017some} used two classification algorithms: LR and Boolean Crowd Sourcing~(BCS) for classifying Facebook posts as hoaxes or non-hoaxes based on users who ``liked'' it. On a data set of~15,500 posts and~909,236 users, they obtained a classification accuracy of more than~99 percent. The proposed technique even worked for users who ``liked'' both hoax and non-hoax posts. Similarly, Kumar \textit{et al.}~\cite{kumar2016disinformation} studied the presence of hoaxes in Wikipedia articles  based on a data set consisting of~20K hoaxes explicitly and manually labeled by Wikipedia editors. According to their findings, hoaxes have very little impact and can be easily detected. A multi-modal hoax detection system that merges diverse modalities -- the source, text, and the image of a tweet was proposed by Maigrot \textit{et al.}~\cite{maigrot2016mediaeval}. Their findings suggested that using only source or text modality ensures high performance in comparison to using all the modalities. Marcella \textit{et al.}~\cite{tambuscio2015fact} focused on the diffusion of hoaxes on OSNs by considering hoaxes as viruses in which a normal user, once infected, behaves as a hoax-spreader. The proposed stochastic epidemic model can be interpreted as a Susceptible-Infected-Susceptible~(SIS) or Susceptible-Infected-Recovered~(SIS) model -- the infected user can either be a believer (someone who believes the fake news) or a fact-checker (checking the news before believing it). The model was implemented and tested on homogeneous, heterogeneous and real networks. Based on a wide range of values and topologies, the fact-checking activity was analysed and then a threshold was defined for fact-checking probability (verifying probability). This threshold was used to achieve the complete removal of fake news based on the number of fact-checkers considering the news as fake or real. A study by Shao \textit{et al.} focused on the temporal relation between the spread of misinformation and fact-checking, and the different ways in which both are shared by users. They proposed Hoaxy~\cite{shao2016hoaxy} -- a model useful in the collection, detection, and analysis of this type of misinformation. They generated a data set by collecting data from both fake news (71 sites,~1,287,768 tweets,~171,035 users and~96,400 URLs) and fact-checking (6 sites,~154,526 tweets,~78,624 users and~11,183 URLs) sources. According to their results, fact-checking data sharing lags behind misinformation by~10-20 hours. They suggested that social news observatories could play an important role by providing the dynamics of real and fake news distribution and the associated risks.

\subsection{Trust and credibility}
\label{subsec:trustandcredibility}
The ease of sharing and discovering information on social media results in a huge amount of content published for target audiences. Both participants (those who share and consume) must check the credibility of shared content. Social media also enables its users to act simultaneously as content producers and consumers. The content consumer has more flexibility in what content to follow. For the content producer, it is necessary to check and evaluate the source of information. If a user is interested in receiving information regarding a particular topic of interest from a specific source, his primary task is to check the credibility, relevance, and quality of that source. Different ways of checking credibility include:

\begin{itemize}
	\item Looking for other users who have subscribed to such information~\cite{canini2011finding}.
	\item Assessing both the expertise (support and recommendations from other professionals)~\cite{ericsson2018cambridge, wang2015trust} and user credibility.
	\item Assessing the credibility of the sources (examining the content and peer support)~\cite{rieh2007credibility}.
\end{itemize}

Researchers have proposed different techniques for identifying credible and reputable sources of information. Canini \textit{et al.}~\cite{canini2011finding} proposed an algorithm based on both the content and social status of the user. Weng \textit{et al.} merged the web page ranking technique and topic modelling to compute the rank of a Twitter user~\cite{weng2010twitterrank}. Cha \textit{et al.}~\cite{cha2010measuring} studied the factors that specify user influence. A random walk approach~\cite{perozzi2014deepwalk} was proposed for separating credible sources from malicious ones by performing network feature analysis. %This analysis traces the distinct dependencies and relations that form multiple networks. 

TweetCred, a real-time web-based system, was developed to evaluate the credibility of tweets~\cite{gupta2014tweetcred}. It assigns a credibility score to each tweet on a user time line rating from~1 (low credibility) to~7 (high credibility). The credibility score is then computed using a semi-supervised ranking algorithm trained on a data set consisting of an extensive set of features collected from previous work~\cite{perez2017automatic} and manually labelled by humans. The TweetCred evaluation was performed based on its usability, effectiveness, and response time. An~80 percent credibility score was calculated and displayed within~6 seconds. Additionally,~63 percent of users either agreed or disagreed with the generated score by~1-2 points. Irrespective of its effectiveness, the results were still influenced by user personalization and the context of tweets which did not involve factual information.

A different research model was developed -- based on perceptions related to news
authors, news sharers, and users -- to test verification
behaviours of users. ~\cite{torres2018combating}. The aim was to study the validation of content published by users on Social Networking Sites (SNSs). The results were assessed using a three-step analysis to evaluate the measurement model, structural model, and common method bias. It focused on the epistemology of declarations of interpersonal trust to examine factors that influence user trust in disseminated news on SNSs. To test the full model, the researchers used SmartPLS 2.0. The evaluation showed that the variety in social ties on SNSs increases trust among network participants and trust in the network reduces news verification behaviours. However, the evaluation disregards the importance of the nature of news connected with the recipient.

Trust is an important factor to be considered when engaging in social interaction on social media. When measuring trust between two unknown users, the challenging task is the discovery of a reliable trust path. In~\cite{ghavipour2018trust}, Ghavipour \textit{et al.} addressed the problem of reliable trust paths by utilizing a heuristic algorithm built on learning automata called DLATrust. They proposed a new approach for aggregating the trust values from multiple paths based on a standard collaborative filtering mechanism. The experiments performed on \href{https://web.archive.org/web/20170715120119/http://advogato.org/}{Advogato} -- a well-known network data set for trust, showed the efficiency and high accuracy in predicting the trust of reliable paths between two indirectly connected users.

Liu and Wu \textit{et al.}~\cite{shu2018understanding} studied the correlation between user profiles and the fake news shared on social media. A real-world data set comprising social context and news content was built for categorizing users based on measuring their trust in fake news. Representative groups of both experienced users~(able to differentiate between real and fake news) and naive users~(unable to differentiate between real and fake news) were selected. They proposed that the features relevant to these users could be useful in identifying fake news. The results for identified user groups showed that the distribution satisfies power-law distribution~\cite{clauset2009power} with high $R^{2}$ scores. This result indicated a significant difference between features of experienced and naive users. However, the paper left unexplored the credibility and political bias of experienced users before characterizing them for fake news detection.

The timely detection of misinformation and sharing of credible information during emergency situations are of utmost importance. The challenge of distinguishing  useful information from misinformation during these events is however still significant. Moreover, a lack of know-how about social networks makes it even more challenging to discern the credibility of shared information~\cite{Webwise}. Antoniadis \textit{et al.}~\cite{antoniadis2015model} developed a detection model to identify misinformation and suspicious behavioural patterns during emergency events on the Twitter platform. The model was based on a supervised learning technique using the user's profile and tweets. The experiments were performed on a data set consisting of 59,660~users and 80,294~tweets. The authors filtered 81 percent of the tweets and claimed that more than 23 percent were misrepresentations. Although the proposed technique makes no distinction between intentional and unintentional information~\cite{balestrucci2019identification}, it successfully achieved timely detection. %This model can be used in the future to limit the spread of misinformation. 

In Table~\ref{tab:trustandandcredibility}, we analyze trust and reputation models in terms of the mechanism used, data set as well as the outcomes and weaknesses of each model. 
In the existing literature, insufficient importance is given to the sources responsible for spreading the fake news. Evaluating the source is not straightforward process, as there are multiple variables to be considered in source verification, such as affiliation and reputation of the source, expertise in the domain, agreement or disapproval of other sources etc. Moreover, the absence of a source makes information unreliable, regardless of whether it is generated by an authentic source or not. Hence, fake news evaluation requires a model capable of performing source tracking, verification and validation. 
\setlength\LTleft{-.7cm}
\afterpage{
	{\footnotesize
		\begin{longtable}{|p{0.18\textwidth}|p{0.2\textwidth}|p{0.13\textwidth}|p{0.2\textwidth}|p{0.17\textwidth}|p{0.07\textwidth}|}
			\hline
			\rowcolor{Gray}
			\textbf{Approach} & \textbf{Data set and Features} & \textbf{Evaluation Metrics} & \textbf{Finding or Outcomes} & \textbf{Weaknesses} & \textbf{Platform}\\
			\hline
			Structural modeling~\cite{torres2018combating}&541 users, Age, gender, network size&Reliability, validity&Development of a research model based on perceptions&Ignores news connection with recipient&Social networking sites\\ \hline
			Measuring user trust for fake news detection~\cite{shu2018understanding}& Two data sets,Explicit and implicit profile features&Follower to following counts ratio&Expert and naive user features differ&Does not consider the bias and credibility of users &Twitter\\ \hline
			SVM-rank for credibility assessment~\cite{gupta2014tweetcred}&10,074,150 tweets, 4,996,448 users, features obtained from high impact crisis events & Response time, usability, effectiveness, &TweetCred browser extension&Results influenced by context of tweets and personalization&Twitter\\ \hline 
			Automated learning and Standard collaborative filtering~\cite{ghavipour2018trust}&Advogato, Observer, Apprentice, Journeyer, Master&Coverage, prediction accuracy&Efficient and accurate trust path discovery&Does not consider the dynamic nature of trust~\cite{ghavipour2018dynamic}&Advogato\\ \hline
			Spam and bot detection~\cite{cresci2019cashtag}&9 million tweets, 30,032 companies, Market capitalization, industrial classification&Cashtag&Uncovering malicious practices--cashtag piggybacking&--&Twitter\\ \hline
			Supervised learning for misinformation detection~\cite{antoniadis2015model}& 80,294 tweets, 59,660 users&Accuracy, precision, recall, F-Measure&Accuracy of timely identification of misinformation at 70\% &Undefined intentions~\cite{balestrucci2019identification}&Twitter\\ \hline
			Stochastic epidemic model  for fact-checking~\cite{tambuscio2015fact}&Network with 1,000 nodes, Spreading rate, forgetting probability&Probability&Define a fact-checking probability for hoaxes&Does not consider the heterogeneity of agents&Facebook\\ \hline
			Hoaxy for fact-checking~\cite{shao2016hoaxy}&Fake news and fact-checking sources, data volume, time series  &Keyword correlation& Propagation of fake news is dominated by active users &Fake news makes more of a contribution to data set generation&Twitter\\ \hline
			Random forest classifier for fake news detection~\cite{potthast2017stylometric}&1,627 articles, Writing style&Accuracy, precision, recall, F1&Distinguished hyperpartisan and mainstream&Not applicable for fake news detection&Facebook\\ \hline
			Linear SVM classifier for fake news detection~\cite{perez2017automatic}&100 fake and 100 legitimate articles&Accuracy, precision, recall, F1 measures&Two data sets, accuracy comparable to humans in detecting fake news&Humans perform better in celebrity domain&Web\\ \hline
			LR, BCS algorithm for classification~\cite{tacchini2017some}&15,500 posts, 909,236 users, likes&Accuracy&Classification accuracy 99\% for hoaxes and non-hoaxes&Limited conspiracy theories in data set~\cite{shu2018fakenewsnet}&Facebook\\ \hline
			SVM classifier for predicting factuality~\cite{baly2018predicting}&1,066 news websites, URL, article, account&Accuracy, $F_{1}$ score, MAE and its variant&Predicting the factuality of reports and bias of news media&Limiting sharing of false content is challenging~\cite{paschen2019investigating}&Entire news medium\\ \hline
			\caption{Detailed summary of the studies used in Fact-checking, Trust and Credibility}
			\label{tab:trustandandcredibility}
	\end{longtable}}
}

\section{Tools and Data Resources}
\label{sec:toolsanddataset}

%\subsection{Fake News Detectors}
%\label{subsec: fake news detector}

Social media popularity, the availability of the internet, the extreme growth of user-generated website content, the lack of quality control and poor governance all provide fertile ground for sharing and spreading false and unverified information. This has led to continuous deterioration of information veracity. As the significance of the fake news problem is growing, the research community is proposing increasingly robust and accurate solutions. Some of the proposed solutions are discussed below and their characteristics are provided in Table~\ref{tab:Web-browsing-tools}.

\begin{itemize}
	
	\item \texttt{BS-Detector\footnote{\url{https://gitlab.com/bs-detector/bs-detector}}:} Available as a browser extension for both Mozilla and Chrome. It searches for all the links available on a webpage that are linked to unreliable sources and checks these links against a manually compiled list of domains. It can classify the domains as fake news, conspiracy theory, clickbait, extremely biased, satire, proceed with caution, etc. The BS detector has been downloaded and installed around about~25,000 times~\cite{AlisonGriswold}.  
%	
%	\item Fake News Detector AI\footnote{\url{http://www.fakenewsai.com/}}: There is insufficient information available regarding the Fake News Detector apart from that it uses NN approach to analyze the website by checking if it matches a fake website.
	
	\item \texttt{FiB\footnote{\url{https://devpost.com/software/fib}}:} The distribution of content is as important as its creation, FiB takes both post creation as well as distribution into account. It verifies the authenticity of a post in real time using AI. The AI uses keyword extraction, image recognition and source verification to check the authenticity of posts and provide a trust score. In addition, FiB tries to provide true information for posts that are deemed false~\cite{figueira2017current}.
	
	\item \texttt{Trusted News add-on\footnote{\url{https://trusted-news.com/}}:} Built in conjunction with \href{https://metacertprotocol.com/}{MetaCertProtocol} powered by the Metacert organization to help users spot suspicious or fake news. It is used to measure the credibility of website content and flags content as good, questionable or harmful. It gives a wider set of outputs, including marking website contents as malicious, satirical, trustworthy, untrustworthy, biased, clickbait and unknown~\cite{paul2019}.  
	
	\item \texttt{SurfSafe\footnote{\url{https://chrome.google.com/webstore/detail/surfsafe-join-the-fight-a/hbpagabeiphkfhbboacggckhkkipgdmh?hl=en}}:} There are different ways to analyze fake news such as textual analysis, image analysis, etc. Ash~Bhat and Rohan~Phadte focused on the analysis of fake news using images and generated a data set which consists of images collected from~100 fact-checking and trusted new sites. They developed a plug-in that checks the images against a generated data set. The main idea is to check each new image against the generated image data set. If the image is used in a fake context or modified, the information as a whole is considered fake~\cite{BryanClark}.  
	
	\item \texttt{BotOrNot:} A publicly available service used to assign a classification score to a Twitter account. This score is assigned to an account on the basis of the similarity it exhibits to the known characteristics of social bots. This classification system leverages more than~1,000 features extracted from contents, interaction patterns and available metadata~\cite{davis2016botornot}. These features are further grouped into six sub-classes:
	
	\begin{itemize}
		\item Network features - built by extracting the statistical features for mentions, retweets, and hashtag co-occurrence. 
		\item User features - based on twitter metadata such as creation time of account, languages, locations. 
		\item Friend features - dependent on the statistics of social contacts such as number of followers, posts, followees and so on.
		\item Temporal features - recording the timing pattern for content generation and distribution.
		\item Content features - based on part-of-speech tagging.
		\item Sentiment features - built by using a sentiment analysis algorithm that takes into account happiness, emotion scores, etc.
	\end{itemize}
	
	\item \texttt{Decodex\footnote{\url{https://chrome.google.com/webstore/detail/decodex/kbpkclapffgmndlaifaaalgkaagkfdod?hl=fr}}:} An online fake news detection tool that alerts the user to the potential of fake news by labeling the information as `satire', `info' and `no information'~\cite{gielczyk2019evaluation}. 
	
	\item \texttt{TrustyTweet\footnote{\url{https://peasec.de/2019/trustytweet/}}:}  TrustyTweet is a browser plug-in, proposed for twitter users to assess and increase media literacy. It shifts the focus from fake news detection by labelling to supporting users to make their own assessment by providing transparent, neutral and intuitive hints when dealing with the fake news. TrustyTweet is based on gathering the potential indicators for fake news, already identified and proven to be promising in previous studies~\cite{hartwig2019trustytweet}. %Based on these indicators, TrustyTweet can assist Twitter users in tweet assessment by providing an intuitive warning without creating reactance.     
	
	\item \texttt{Fake News Detector\footnote{\url{https://github.com/fake-news-detector/fake-news-detector/tree/master/robinho}}:} The Fake News Detector is an open source project used for flagging news. A user can flag news as either fake news, extremely biased or clickbait. The user flagging activity is visible to other fake news detector users who may flag it again. Once the news is flagged, it is saved in the repository and  accessible to \href{https://github.com/fake-news-detector/fake-news-detector/tree/master/robinho}{Robhino} -- an ML robot trained on the inputs provided by humans that flags news automatically as clickbait, fake news or extremely biased news.
	
	\item Fake News Guard\footnote{\url{https://www.eu-startups.com/directory/fake-news-guard/}}: Available as a browser extension, it can verify the links displayed on Facebook or any page visited by the user. There is insufficient information about the way this tool works, however the key idea is that the ``Fake news guard uses the AI technique along with network analysis and fact-checking''. 
	
	\item \texttt{TweetCred\footnote{\url{http://twitdigest.iiitd.edu.in/TweetCred/}}:} A web browser tool used for assessing the credibility of tweets by using a supervised ranking algorithm trained on more than~45 features. TweetCred assigns a credibility score for each tweet on the user time line. Over the course of three months, TweetCred was installed~1,127 times and computed the credibility score for~5.4 million tweets~\cite{gupta2014tweetcred}.
	
	\item \texttt{LiT.RL News Verification\footnote{\url{https://victoriarubin.fims.uwo.ca/2018/12/19/release-for-the-lit-rl-news-verification-browser-detecting-clickbait-satire-and-falsified-news/}}:} A research tool that analyses the language used on web pages. The core functionality of the News Verification browser is textual data analysis using NLP and automatic classification using a SVM. It automatically detects and highlights website news as clickbait, satirical fake news and fabricated news~\cite{rubin2019news}.
\end{itemize}
\afterpage{
{\footnotesize
\begin{longtable}{|p{0.1\textwidth}|p{0.1\linewidth}|p{0.1\linewidth}|p{0.2\linewidth}|p{0.1\linewidth}|p{0.15\linewidth}|p{0.2\linewidth}|}
			\hline
			\rowcolor{Gray}
			\textbf{Tools} & \textbf{Availability} & \textbf{Proposed} & \textbf{Technique} & \textbf{Input} & \textbf{Output} & \textbf{Source}\\ \hline
			SurfSafe&Browser extension&\href{https://www.theatlantic.com/technology/archive/2018/06/robhat-labs-surfsafe-fake-news-images/564101/}{Robhat labs}&Comparison and textual analysis&Images and text&Safe, warning, unsafe&100 fact-checking, trusted organizations\\ \hline
%			Fake News Detector AI&Website&\href{https://www.karansinghal.com/}{Karan Singhal}&AI&URLs&Fake, real&Existing fake news site\\ \hline
			Trusted News&Browser extension&Trusted News&--&Website content&Trustworthy, biased, satire&MetaCert protocol\\ \hline
			Fake News Detector&Browser extension&Robhino&Crowd sourcing, ML&News content&Fake news, clickbait, extremely biased&Feedback by other tools\\ \hline
			Fake News Guard&Browser extension&Fake News Guard&AI, network analysis, fact-checking&Webpages, links&Fake or not&Fact checkers\\ \hline
			Decodex&Browser extension&\href{https://digiday.com/uk/le-monde-identifies-600-unreliable-websites-fake-news-crackdown/}{Laurent's team}&-&Pieces of information&Satire, info, no information&600 websites\\ \hline
			BS Detector&Browser extension&Daniel Sieradski,&Comparison model&URLs&Fake news, conspiracy theory, clickbait, extremely biased etc&Data set of unreliable domains\\ \hline
			TrustyTweet&Browser extension&Katrin Hartwig and Christian Reuter~\cite{hartwig2019trustytweet}&Media literacy&Tweets&Politically neutral, transparent and intuitive warnings&Potential indicators from previous studies\\ \hline
			TweetCred&Browser extension&-&Semi-supervised ranking model&Tweets&Credibility score&Twitter data\\ \hline
			FiB&Browser extension&\href{https://post.devpost.com/}{DEVPOST}&Text analysis, image analysis, web scraping&Facebook posts&Trust score&Verification using keyword extraction, image recognition, source verification\\ \hline
			BotOrNot&Website, REST API&Clayton \textit{et al.}&Classification algorithm &Twitter screen name&Bot likelihood score&Accounts for recent history including tweet mentions\\ \hline
			LiT.RL News Verification&Web browser&~Rubin \textit{et al.}&NLP, support vector machine&Language used&Satirical news, clickbait, falsified news& Lexico-syntactic features in text \\ \hline
\caption{Detailed Summary of the Online Web Browsing Tools}
\label{tab:Web-browsing-tools}
\end{longtable}}}

%\subsection{Datasets for Fake News}
%\label{subsec:DFN}
Detecting fake news on social media poses many challenges as most fake news is intentionally written. Researcher are considering different information, such as user behaviour, the engagement content of news, etc. to tackle the problem. However, there is no data set available that could provide the information on how fake news propagates, how different users interact with fake news, how to extract temporal features which could help to detect it and what the impact of fake news truly is. 

%In this section, we will discuss the problem of fake news detection from an ML perspective. The aim of using ML is to automate the process of fake news detection and reduce the time and effort required for humans to detect such news. 
%% A survey was conducted using Natural Language Processing~(NLP)~\cite{oshikawa2018survey} for fake news detection. 
%One of the major issues with automated fake news detection from an ML perspective is the availability and quality of the data set. In order to facilitate research in this field, a comprehensive data set is required, consisting of all the necessary information related to fake news such as social context, content and spatiotemporal information~\cite{shu2018fakenewsnet}. Some of the dataset cover the problem of fake news detection while some of them do not consider this problem. We will discuss some of the data sets that are publicly available and then compared them in Table~\ref{dataset}. 

In the previous section, we discussed the automatic detection of fake new using ML models. ML models require high quality data set to be efficient. This continues to be a major challenge when it comes to social media data due to it's unstructured nature,  high dimensionality, etc. % As a consequence, the data set that is currently available is of low quality. 
In order to facilitate research in this field, a comprehensive guide to existing data sets is required. Below we present the details for some of the more widely used ones: 

\begin{itemize}
	
	\item \texttt{CredBank:} Collected by tracking more than~1 billion tweets between October~2014 and February~2015~\cite{MitraTanushree}. It consists of tweets, events, topics and an associated credibility judgment assigned by humans. The data set comprises~60 million tweets which are further categorized into~1,049 real-world events. Further, the data is spread into a streaming tweet file, topic file, credibility annotation file and searched tweet file~\cite{mitra2015credbank}. 
	
	\item \texttt{LIAR:} A publicly available fake news detection data set~\cite{thiagorainmaker} that can be used for fact-checking. It consists of~12,836 short statements labelled manually by humans. In order to verify their truthfulness, each statement is evaluated by the editor of \href{https://www.politifact.com/}{POLITIFACT.COM}. Each statement is labelled in any of the following six categories: true, mostly-true, half-true, barely-true, false, pants on fire~\cite{wang2017liar}.
	
%	\item \texttt{Twitter7\footnote{\url{http://snap.stanford.edu/data/twitter7.html}}:} This data set consists of~580 million tweets collected from~20 million users over 8~months. Among these tweets, approximately~6 million different hashtags were identified and a total of~144 million URLs were mentioned.  Yang \textit{et al.}~\cite{yang2011patterns} demonstrated their approach based on two massive data sets. The first was Twitter7 and the second consisted of~170 million news media articles and blog posts. These articles were collected over~1 year from~1 million online sources. 
	
	\item \texttt{Memetracker9:} This data set~\cite{leskovec2009meme} recorded social media activity and online mainstream content over a three-month period. They used a \href{https://www.datastreamer.io//}{Spinn3rAPI} and collected~90 million documents from~165 million different websites~\cite{JureLeskovec1}. The data set they generated is~350GB in size. First, they extracted~112 million quotes, which were further refined and from which~22 million distinct phrases were collected.
	
	\item \texttt{FakeNewsNet\footnote{\url{https://github.com/KaiDMML/FakeNewsNet}}:} A multi-dimensional data repository consisting of social context, content and spatiotemporal information~\cite{shu2018fakenewsnet}. The data set was constructed using \href{http://blogtrackers.fulton.asu.edu:3000/\#/dashboard}{FakeNewsTracker}, a tool used for collecting, analyzing as well as visualizing fake news. In the given data set, the content consists of news, articles and images while context consists of information related to the user, post, response and network. The spatiotemporal information consists of spatial (user profile with location, tweets with location) and temporal information (timestamp for news and responses).
	
	\item \texttt{BuzzFeedNews\footnote{\url{https://www.buzzfeednews.com/article/craigsilverman/partisan-fb-pages-analysis}}:} This data set, recorded all the news published by~9 news agencies on Facebook regarding the US election. The articles and news were fact-checked by journalists from BuzzFeed. It contains~1,627 articles and~826 streams from hyperpartisan Facebook pages which publish misleading and false information at an alarming rate~\cite{jsvine}. 
	
	\item \texttt{BS Detector~\cite{ThiagoVieira2017}:} This data set was collected by using \href{https://github.com/selfagency/bs-detector}{BS Detector}, a web browser extension for both Chrome and Mozilla. It is used to search all the links linked to unreliable sources on a given web page. These links are checked across a manually compiled list of domains.
	
	\item \texttt{BuzzFace:} This data set~\cite{santia2018buzzface} consists of~2,263 news articles and~1.6 million comments. \href{https://github.com/BuzzFeedNews/2016-10-facebook-fact-check}{Buzzface} is based on extending the BuzzFeed data set by adding comments related to Facebook news articles. The news articles were categorized as ``mostly true'',``mixture of true and false'', ``mostly false'' and ``no factual comments''. 
	
	\item \texttt{FacebookHoax:} In this data set~\cite{tacchini2017some}, \href{https://developers.facebook.com/docs/graph-api}{facebook graph API} is used for the collection of data. It consists of~15,500 posts, of which~8,923 are hoaxes and the rest are non-hoaxes. These posts were collected from~32 pages:~14 conspiracy and~18 scientific pages. In addition, the data set also includes the number of likes, which exceeds~2.3 millions. 
	
	\item \texttt{Higgs-Twitter:} This data set~\cite{de2013anatomy} consists of  Twitter posts related to the discovery of the new Higgs boson particle. The tweets were collected using the Twitter API. It consists of all the tweets that contain one of the following hashtags or keywords: cern, higgs, lhc, boson. The data set consists of~527,496 users and~985,590 analysed tweets of which~632,207 were geo-located tweets.
	
	\item \texttt{Trust and Believe:} This data set consists of~50000 Twitter users, all of whom were politicians~\cite{tanveer_khan_2021_4428240}. For each user, a unique profile is created containing~19
	features. A total of~1000 user was manually annotated, with the rest being classified using an active learning approach. 
	\end{itemize}

\setlength\LTleft{-.5cm}
\afterpage{
{\footnotesize
\begin{longtable}[hbt!]{|p{0.09\textwidth}|p{0.13\textwidth}|p{0.16\textwidth}|p{0.13\textwidth}|p{0.15\textwidth}|p{0.09\textwidth}|p{0.15\textwidth}|}
		\hline
		\rowcolor{Gray}
		\textbf{Dataset} & \textbf{Statistics} & \textbf{Observations} & \textbf{Goal}& \textbf{Approach} & \textbf{Sources} & \textbf{Limitations}\\ 
		\hline
		CRED-BANK&60 million tweets, 1,049 real events&Manually annotated&Credibility assessment&Media events are linked to a human credibility judgement&Twitter&Collected tweets are not related to fake news articles~\cite{shu2018fakenewsnet} \\ 
		\hline
		LIAR&12,836 statements&Manually annotated&Fact-checking&Assessment of truthfulness of claim&\href{https://www.politifact.com/truth-o-meter/statements/}{TruthO-Meter}, \href{https://www.channel4.com/news/factcheck/}{fact-checking}&Instead of entire article based on short statements~\cite{shu2018fakenewsnet}\\
		\hline
		FAKE NEWS Net&Social context, content, spatiotemporal information&Fake News Tracker&Analyzing and visualizing fake news&Fake news diffusion, user engagement&\href{https://www.politifact.com/}{PolitiFact}, Twitter&Social engagement of the articles \cmmnt{~\cite{oshikawa2018survey}} \\ 
		\hline
%		Twitter7&Tweets: 580 million, Users: 20 million&6 million hashtags, 144 million URLs &Temporal pattern of twitter&Temporal patterns &Twitter&No labelling of fake news\\
%		\hline
		Memetrac-ker9& 90 million documents&22 million district phrases are extracted&Temporal patterns&Tracking ideas, new topic memes&1.65 million sites&--\\
		\hline
		BuzzFeed& left-wing and right-wing articles&Rates post as ``true'',``mixture of true and false'',``false''&Fact-checked&Facebook engagement number&Facebook&Based on headlines and text only~\cite{shu2018fakenewsnet}\\ \hline
		BS Detector&--&BS Detector assigned labels&News veracity&Manually compiled list of domains&Web pages&Instead of a human expert, a tool is used for news veracity\\ \hline
		BuzzFace&2,263 news articles, 1.6 million comments&``true'',``mixture of true and false'', ``false'' and ``no factual comments''&Veracity assessment&Extension of BuzzFeed including comments&Facebook&Context and content information but no temporal information~\cite{shu2018fakenewsnet}\\ \hline
		Facebook HOAX&15,500 posts, 32 pages, 2,300,000 likes&Scientific pages are non-hoaxes, conspiracy pages are hoaxes&Post classification into hoaxes and non-hoaxes&Number of likes per post and per user, relation between pages&Facebook&Few instances of news and conspiracy theories~\cite{shu2018fakenewsnet}\\ \hline
		Higgs-Twitter&527,496 users, 985,590 tweets &632,207 geo-located tweets&User behavior accuracy&Analysis of spatial and temporal user activity&Twitter&No labelling of fake news\\ \hline
		Trust and Believe&50,000 users &Manually annotated, Active learning&Influence score&Active learning approach&Twitter&Small dataset\\ \hline
\caption{Detailed Summary of the Available Data Sets in the Existing Literature}
\label{tab:dataset}
\end{longtable}}}
Table~\ref{tab:dataset} presents a detailed summary of the available data sets used for fake news detection in existing literature. Most are either small in size or contain mainly uni-modal data. The existing multi-modal data sets, unfortunately, still can't be used as a benchmark for training and testing models for fake news detection~\cite{jindalnewsbag}. The next step is to generate large and comprehensive data sets that would include resources from which all relevant information could be extracted.

\section{Discussion and Challenges}
\label{sec:discussionandchallenges}
Solving the problem of fake news detection and minimizing their impact on society is one of the important issues being considered in the research community. In this review, we analysed different studies using varying methods for detecting fake news. With the aim of aiding future research we provide their classification based on the social media platform used in Table~\ref{tab:classificationofstudies}. %is a major issue that needs to be addressed urgently because of itimplications. Many researchers have invested numerous resources in finding ways to mitigate this problem. 
%In this review, we analysed different studies using varying methods for detecting fake news. We  n\ref{tab:classificationofstudies},  we have classified them according to the social media platform on which they based their research. %This categorization is by no means a comprehension guide but it will help prospective researchers to focus on the studies based on the platform they are interested in.} 
%In this section, we studied the existing literature for fake news detection and then classified it on the basis of the social media platform used and the different ways that are proposed for fake news detection. First, we assigned the existing literature to either the Twitter or Facebook group as seen in Table~\ref{Intro:TandF}.  	

{\footnotesize
\begin{longtable}[hbt!]{|p{0.09\linewidth}|p{0.91\linewidth}|}
		\hline
		\rowcolor{Gray}
		\textbf{Platform} & \textbf{Research Papers}\\ 
		\hline
		Twitter& Grier \textit{et al.}~\cite{grier2010spam},Ye \textit{et al.}~\cite{ye2010measuring}, Chengcheng \textit{et al.}~\cite{shao2016hoaxy}, Soroush \textit{et al.}~\cite{vosoughi2018spread}, Liang Wu and Huan liu~\cite{wu2018tracing}, Hartwig \textit{et al.}~\cite{hartwig2019trustytweet}, Davis \textit{et al.}~\cite{davis2016botornot}, Cha \textit{et al.}~\cite{cha2010measuring}, Weng \textit{et al.}~\cite{weng2010twitterrank}, Canini \textit{et al.}~\cite{canini2011finding}, Thomas Kurt~\cite{thomas2013role}, Holton \textit{et al.}~\cite{holton2011journalists}, Antoniadis \textit{et al.}~\cite{antoniadis2015model}, Alessandro \textit{et al.}~\cite{balestrucci2019identification}, Zhao \textit{et al.}~\cite{zhao2015enquiring}, Gupta \textit{et al.}~\cite{gupta2014tweetcred}, Khan and Michalas~\cite{Michalas:20:TrustCom:FakeNews} \\ 
		\hline
		Facebook& Tambuscio \textit{et al.}~\cite{tambuscio2015fact}, Joon Ian Wong~\cite{All:Fake:News:Traffic:Facebook}, Monther \textit{et al.}~\cite{aldwairi2018detecting}, Potthast \textit{et al.}~\cite{potthast2017stylometric}, Alexey Grigorev~\cite{grigorev2017identifying}, Fake News Guard\footnote{\url{https://www.eu-startups.com/directory/fake-news-guard/}}, BuzzFace~\cite{santia2018buzzface}, FacebookHoax~\cite{tacchini2017some, shu2018fakenewsnet}, Sebastian \textit{et al.}~\cite{tschiatschek2018fake}, Detective~\cite{leskovec2012learning}\\
\hline
\caption{Classification of the Studies Surveyed based on the Platform Used -- Facebook and Twitter}
\label{tab:classificationofstudies}
\end{longtable}}

\begin{figure}[hbt!]
\begin{forest}
	for tree={
		grow=east,
		parent anchor=south east,
		child anchor=south west,
		anchor=south,
		align=center,
		l sep+=2.5pt,
		s sep+=-5pt,
		inner sep=0pt,
		outer sep=0pt,
		edge path={
			\noexpand\path [draw, rounded corners=5pt, \forestoption{edge}] (!u.parent anchor) [out=0, in=180] to (.child anchor)\forestoption{edge label} -- (.south east);},
		for root={ellipse,draw,parent anchor=east,},}
	[Fake News\\Detection
	[Knowledge\\Based
	[Info\\Retrieval
	[\cite{ferreira2016emergent, baly2018predicting, holton2011journalists, gupta2014tweetcred}]]
	[Semantics
	[\cite{ciampaglia2015computational, tambuscio2015fact, shao2016hoaxy, shiralkar2017finding}]]]
	[, calign with current edge
	[Network\\Propagation
	[Link\\Based[\cite{lee2011seven, aldwairi2018detecting, shiralkar2017finding}]]
	[Neighbor\\Based[\cite{hu2013social,rayana2015collective, li2016robust}]]
	[Group\\Based[\cite{jindal2007review, gao2010community, akoglu2015graph, wu2017adaptive, wu2018tracing}]]]]
	[Feature\\Based
	[Text\\Categorization
	[\cite{potthast2017stylometric, perez2017automatic, tacchini2017some, wu2017adaptive, o2018language}]]
	[Context and \\ Content  Based
	[\cite{zhao2015enquiring, ruchansky2017csi, wang2017liar, wang2010don}]]
	[Account \\ Based
	[\cite{gilani2017classification, cresci2017paradigm, edwards2014bot, lee2011seven, wu2013detecting, chu2010tweeting, chu2012detecting, ercsahin2017twitter, STCSZafar, giachanou2019bot}]]]
    [Hybrid\\Approach
	[\cite{liu2018early, Sean, Andreas, pham2019transferring, liu2020fned, wang2017liar, yang2019fake, Michalas:20:TrustCom:FakeNews}]]]
	%	[great-grandchild\\node
	%	[great-great grandchild node]
	%	[great-great grandchild node]]]]]
\end{forest}
  \caption{Classification of the Existing Literature Based on Four Paradigms -- Hybrid Approach, Feature Based, Network Propagation, and Knowledge Based}\label{fig:class}
\end{figure}
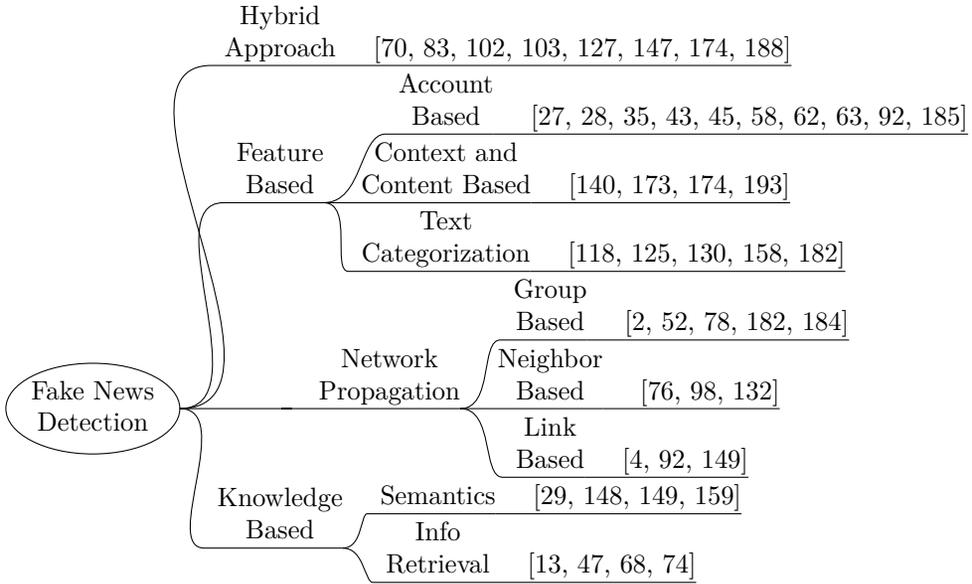

Similarly, a study of the current literature on false news identification can be divided into four paradigms: hybrid approach, feature-based, network propagation and knowledge-based. The hybrid approaches employ both human and ML approaches for the detection of fake news. In the feature-based method, multiple features associated with a specific social media account are used to detect fake news. This paradigm can further be divided into three sub-categories -- account-based, context and content-based and Text categorization. These methods are explicitly discussed in section~\ref{sec:fakenewsanalysis}. The third paradigm, network propagation, describes the potential methods for discovering, flagging and stopping the propagation of fake news in its infancy. The final paradigm entails supplementing AI models with human expert knowledge for decision-making (see section~\ref{sec:fakenewsanalysis}). An overview of these paradigms is given in Figure~\ref{fig:class}. 

%Secondly, we classified the existing literature for fake news detection into four paradigms, i.e. feature-based, network propagation, knowledge-based and hybrid approach as seen in Figure~\ref{fig:class}. The feature-based paradigm is further categorized into account-based classification, content- and context-based and text categorization. Details on this are provided in subsection~\ref{subsec:CA} and \ref{subsec: AA}. The second paradigm, network propagation, is further classified into group-based, neighbor-based and link-based, which are discussed in detail in subsection~\ref{subsec:PN}. The knowledge-based, which can be also referred to as fact-checking, is covered in detail in subsection~\ref{subsec:FC} and subsection~\ref{sec:TAC} while the last paradigm, which is the hybrid approach is mostly discussed in section~\ref{sec:OCFND}.

Identifying and mitigating the spread of fake news and its variants presents a set of unique challenges.  Fake news dissemination is a part of coordinated campaigns  targeting a specific audience with the aim of generating a plausible impact on either local or global level.
%\subsection{Governmental Approaches to Tackle the Problem of Fake News}
%\label{subsec: government approaches}
%Looking at the high level of social engagement generated by the distribution of fake news, it is obvious that the widespread of misleading information has a global impact. 
Many companies as well as entire countries were faced with the need to start building mechanisms to protect citizens from fake news. In September~2019, Facebook announced it was contributing~\$10 million to a fund to improve deepfake detection technologies while several governments have taken different initiatives to defeat this problem~\cite{FunkeDaniel, rusu2019legislative, TheLawLibrary}. 
Educational institutions and non-profit organizations have also tried to mitigate the problem through advocacy and literacy campaigns. Specifically, these institutions in collaboration with technology companies have designed various techniques for detecting, flagging, and reporting fake news~\cite{Eryn, Tora, Shayan, Danielle}.

Table~\ref{tab:Approaches} summarizes the actions that have been taken by governments around the world in order to battle the spread of fake news.

	{\footnotesize	
		\begin{longtable}{|p{0.135\textwidth}|p{0.18\textwidth}|p{0.7\textwidth}|}
			\hline
			\rowcolor{Gray}
			\textbf{Country} & \textbf{Focus} & \textbf{Approach/Action
			}\\ 
			\hline
			Argentina & Fact-checking resources for public & -- Commission created to verify fake news during national election campaign; \ \ \newline  -- Imposing sanctions for spreading fake news.
			\\ \hline
			Sweden &Foreign disinformation campaign& -- Media broadcasts and publications are governed by law; \ \ \newline  -- Educating citizens  \\
			\hline
			Canada&Foreign disinformation campaign& -- No specific law developed to prohibit the spread of fake news. Laws related to the criminal code or broadcasting distribution regulation may be relevant to spreading fake news.\\
			\hline
			China&Election misinformation& -- Spreading fake news is a crime under China's criminal law; \ \ \newline -- Imposition of a fine and imprisonment; \ \  \newline -- Reliable information is published to systematically rebut fake news \\
			\hline
			Egypt&Media regulation& -- Three domestic laws have been passed to regulate information distribution and its accuracy; \ \ \newline -- Imposing sanctions for spreading fake news \\
			\hline
			France &Election misinformation& -- No specific law but there is general legislation against fake news; \ \ \newline -- Imposing sanctions for spreading fake news\\
			\hline
			Germany &Hate speech& -- A number of civil and criminal laws exist for fake news; \ \ \newline -- Network enforcement act specific for fighting fake news\\
			\hline
			Israel&Foreign disinformation campaign& -- High-level committee appointed by the president to examine the current law for threats and find ways to address them; \ \ \newline -- Imposing sanctions for spreading fake news\\
			\hline
			Japan&Media regulation& -- A law exists to counter fake news; \ \ \newline -- Ministry of Communication and Internal Affairs work jointly to counter fake news\\
			\hline
			Kenya&Election misinformation& -- Computer misuse and cyber-crime act has been passed, not yet in force; \ \ \newline -- Educating citizens \\
			\hline
			Malaysia&Election misinformation& -- Malaysian anti-fake News Act 2018; \ \ \newline -- A fact-checking portal is operated by government agencies; \ \ \newline -- Imposing sanctions for spreading fake news      \\ 		\hline
			Nicaragua&Media regulation& -- No specific law available, however some provisions can be found within the penal code and election law\\
			\hline
			Russia &Election misinformation& -- Passed legislation that addresses the spread of fake news; \ \ \newline -- Imposing sanctions for spreading fake news\\
			\hline
			Brazil&Election misinformation& -- No law but the topic is under discussion in congress; \ \ \newline -- Fines and imprisonment\\
			\hline
			United Kingdom&Foreign disinformation campaign& -- No legislation to scrutinize or validate news on social media; \ \ \newline -- Reliable information is published to systematically rebut fake news\\
			\hline
			United Arab Emirates&Election misinformation& -- Sharing misinformation is a crime by law; \ \ \newline -- Imposition of a fine \\
			\hline
			United States&disinformation, misinformation& -- Proposed a federal law; \ \ \newline -- State media literacy initiatives\\
			\hline
			\caption{Approaches Taken by Governments to Tackle the Problem of  Fake News}
			\label{tab:Approaches}
\end{longtable}}

%\textcolor{red}{Information shared on social media during public emergencies has a rather significant impact on society.}\textcolor{red}{Often, people rely on social media for information and then share it with their friends. The current challenge is to identify whether the information shared is credible or not.} 

%Essentially, the detection of fake news has become the crux for numerous researchers in diverse fields. To some extent, the proposed methods have succeeded in achieving promising results. However, the dynamic and diverse nature of this problem has rendered detection algorithms ineffective in comprehensively addressing this challenging issue. In addition, fake news is intentionally spread-out to mislead the readers, thus, using only verbal contents to detect fake news is not enough. Furthermore, social users' engagement with false information produces data that is big, unstructured, incomplete and noisy. Consequently, using auxiliary information for fake news detection is complex and problematic. In the rest of this section, the critical challenges surrounding the detection of fake news are discussed. These challenges are also summarized in Figure~\ref{fig:futurechallenges}.

\begin{figure}
	\centering
	\includegraphics[width=0.9\textwidth]{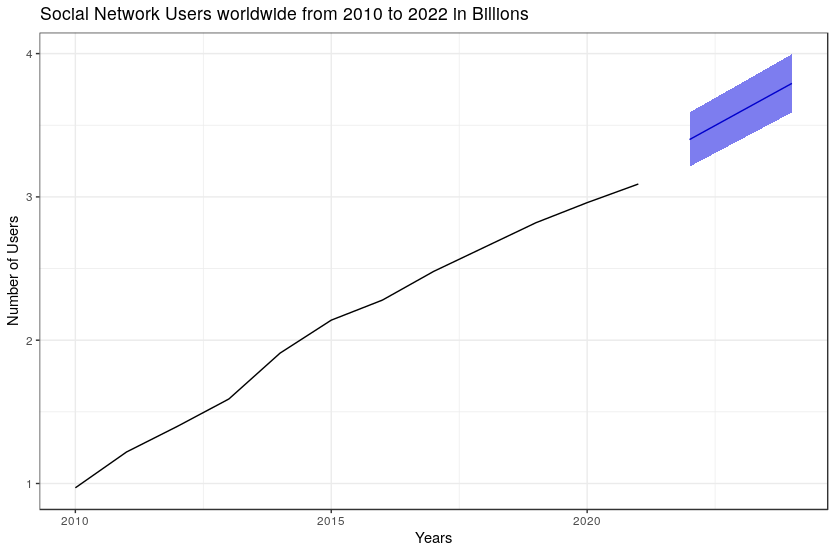}
	\caption{Number of Social Users}
	\label{fig:socialusers}
\end{figure}

The greatest obstacle in fake news detection is that the information spreads through social media platforms like forest fire (especially if it's polarizing) which when not addressed, becomes viral in a matter of milliseconds~\cite{stahl2018fake}. The implications of this instantaneous consumption of information, on the other hand, are long-lasting. As a result, fake news becomes indistinguishable from real information, and the ongoing trends are difficult to recognize. We believe that fake news propagation can only be successfully controlled through early detection (see section~\ref{subsec:networkpropagation}).  Another significant problem is that the rise in the influence of social media is closely connected to the increase in the number of users. According to Figure~\ref{fig:socialusers}, there are currently more than~3 billion users and by~2024 this number is expected to exceed~4 billion, a development that will eventually lead to an exponential rise in data~\cite{Tankovska}. This data is most likely to be potentially uncertain due to inconsistencies, incompleteness, noise and unstructured nature. This complexity increases the velocity, variety, and amount of data and will most probably jeopardize the legitimacy of the results of any standard analytic processes and decisions that would be based on them. Analysis of such data requires tailor-made advanced analytical mechanisms. Designing techniques that could efficiently predict or evaluate future courses of action with high precision thus remains very challenging.

To summarize, humans are susceptible to becoming victims of false information due to their intrinsic way of processing and interpreting  information being influenced by cognitive biases -- namely, by the Truth Bias, Naive Realism and Confirmation Bias~\cite{stahl2018fake}. Consequently, all fake information floating around can lead to false information which is capable of ruining the ``balance of news ecosystem''. The main challenge is that most users do not pay more attention to the manipulated information, while those who are manipulating it are systematically trying to create more confusion. The outcome of this process is that the people's ability to decipher real from false information is further impeded~\cite{shu2017fake, rubin2017deception}. 

\begin{figure}
	\centering
	\includegraphics[width=0.9\textwidth]{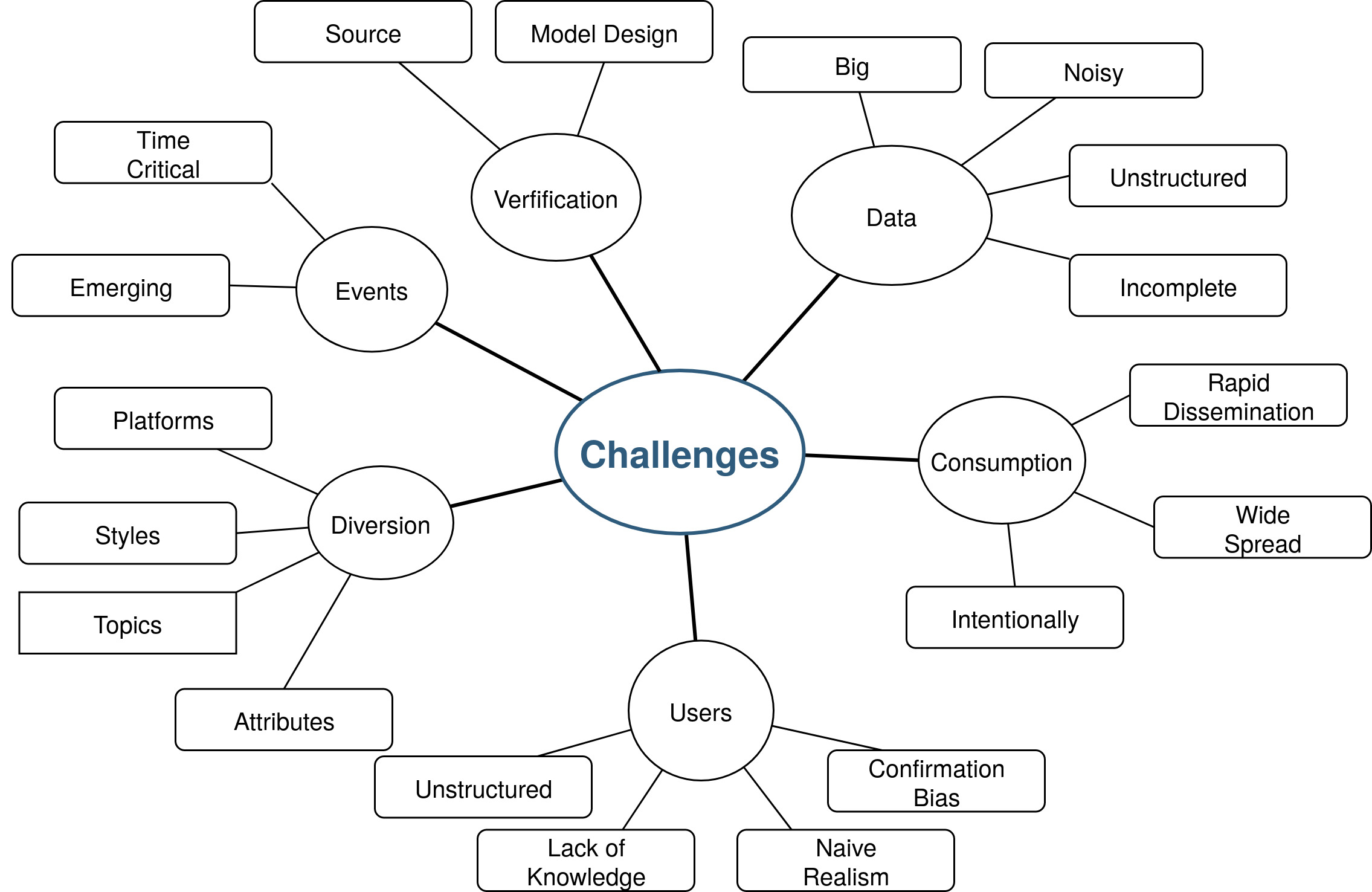}
	\caption{Future Challenges}
	\label{fig:futurechallenges}
\end{figure}

Can we stop the viral spread?, the answer obviously is \textit{Not yet} and it is because of the critical challenges surrounding the detection of fake news (see Figure~\ref{fig:futurechallenges}). Several efforts, however, have been put in place to help limit it such as media literacy. Media literacy comprised of practices that enable people to access and critically evaluate content across different media seems like the only valid solution. Although this is, and always was a challenging task, a coherent understanding, proper education, training, awareness and responsible media engagement could change this~\cite{bulger2018promises}. In the mean time, resisting disinformation and ``fake news'' culture should be promoted and encouraged. In addition, cross-disciplinary collaboration (i.e., social psychology, political science, sociology, communication studies etc.) can help and streamline findings across diverse disciplines to devise a holistic approach for understanding the media environment structure and how it operates.

\section{Conclusion}
\label{sec:Conclusion}
Today, OSNs can be seen as platforms where people from all over the world can instantly communicate with strangers and even influence people's actions. Social media has shaped the digital world to an extent that they now seem like an indispensable part of our daily lives.  However, social networks' ease of use has also revolutionized the generation and distribution of fake news. This prevailing trend has had a significant impact on our societies.

In this survey paper, we studied the problem of fake news detection from two different perspectives. Firstly, to assist users in identifying who they are interacting with, we looked at different approaches in existing literature used for the identification and classification of user accounts. To this end, we analysed in depth both the users' context~(anyone) and content~(anything). For the early identification and mitigation of fake news, we studied different approaches that focus on data network features. Recently proposed  approaches for measuring the relevance, credibility, and quality of sources were analysed in detail.

Secondly, we approached the problem of automating fake news detection by elaborating on the top three approaches used during fake news detection competitions and looked at the characteristics of more robust and accurate web-browsing tools. We also examined the statistical outputs, advantages, and disadvantages of some of the publicly available data sets. 
As the detection and prevention of fake news presents specific challenges, our conclusion identified potential challenges and  promising research directions.

\section*{\uppercase{Acknowledgement}}
%\noindent{This} research has received funding from the ASCLEPIOS: Advanced Secure Cloud Encrypted Platform for Internationally Orchestrated Solutions in Healthcare Project No. 826093 EU research project and the European Union’s Horizon 2020 research and innovation Programme under grant agreement No 825355 (CYBELE).
\noindent{This} research has received funding from the EU research projects ASCLEPIOS (No. 826093) and CYBELE (No 825355).

\bibliographystyle{ACM-Reference-Format}
\bibliography{SoK_Fake_News_Detection}

%%% -*-BibTeX-*-
%%% Do NOT edit. File created by BibTeX with style
%%% ACM-Reference-Format-Journals [18-Jan-2012].

\begin{thebibliography}{196}

%%% ====================================================================
%%% NOTE TO THE USER: you can override these defaults by providing
%%% customized versions of any of these macros before the \bibliography
%%% command.  Each of them MUST provide its own final punctuation,
%%% except for \shownote{}, \showDOI{}, and \showURL{}.  The latter two
%%% do not use final punctuation, in order to avoid confusing it with
%%% the Web address.
%%%
%%% To suppress output of a particular field, define its macro to expand
%%% to an empty string, or better, \unskip, like this:
%%%
%%% \newcommand{\showDOI}[1]{\unskip}   % LaTeX syntax
%%%
%%% \def \showDOI #1{\unskip}           % plain TeX syntax
%%%
%%% ====================================================================

\ifx \showCODEN    \undefined \def \showCODEN     #1{\unskip}     \fi
\ifx \showDOI      \undefined \def \showDOI       #1{#1}\fi
\ifx \showISBNx    \undefined \def \showISBNx     #1{\unskip}     \fi
\ifx \showISBNxiii \undefined \def \showISBNxiii  #1{\unskip}     \fi
\ifx \showISSN     \undefined \def \showISSN      #1{\unskip}     \fi
\ifx \showLCCN     \undefined \def \showLCCN      #1{\unskip}     \fi
\ifx \shownote     \undefined \def \shownote      #1{#1}          \fi
\ifx \showarticletitle \undefined \def \showarticletitle #1{#1}   \fi
\ifx \showURL      \undefined \def \showURL       {\relax}        \fi
% The following commands are used for tagged output and should be
% invisible to TeX
\providecommand\bibfield[2]{#2}
\providecommand\bibinfo[2]{#2}
\providecommand\natexlab[1]{#1}
\providecommand\showeprint[2][]{arXiv:#2}

\bibitem[\protect\citeauthoryear{Adewole, Anuar, Kamsin, Varathan, and
  Razak}{Adewole et~al\mbox{.}}{2017}]%
        {adewole2017malicious}
\bibfield{author}{\bibinfo{person}{Kayode~Sakariyah Adewole},
  \bibinfo{person}{Nor~Badrul Anuar}, \bibinfo{person}{Amirrudin Kamsin},
  \bibinfo{person}{Kasturi~Dewi Varathan}, {and} \bibinfo{person}{Syed~Abdul
  Razak}.} \bibinfo{year}{2017}\natexlab{}.
\newblock \showarticletitle{Malicious accounts: Dark of the social networks}.
\newblock \bibinfo{journal}{\emph{Journal of Network and Computer
  Applications}}  \bibinfo{volume}{79} (\bibinfo{year}{2017}),
  \bibinfo{pages}{41--67}.
\newblock


\bibitem[\protect\citeauthoryear{Akoglu, Tong, and Koutra}{Akoglu
  et~al\mbox{.}}{2015}]%
        {akoglu2015graph}
\bibfield{author}{\bibinfo{person}{Leman Akoglu}, \bibinfo{person}{Hanghang
  Tong}, {and} \bibinfo{person}{Danai Koutra}.}
  \bibinfo{year}{2015}\natexlab{}.
\newblock \showarticletitle{Graph based anomaly detection and description: a
  survey}.
\newblock \bibinfo{journal}{\emph{Data mining and knowledge discovery}}
  \bibinfo{volume}{29}, \bibinfo{number}{3} (\bibinfo{year}{2015}),
  \bibinfo{pages}{626--688}.
\newblock


\bibitem[\protect\citeauthoryear{Albright}{Albright}{2016}]%
        {Jonathan}
\bibfield{author}{\bibinfo{person}{Jonathan Albright}.}
  \bibinfo{year}{2016}\natexlab{}.
\newblock \bibinfo{title}{The \#Election2016 Micro-Propaganda Machine}.
\newblock
  \bibinfo{howpublished}{\url{https://medium.com/@d1gi/the-election2016-micro-propaganda-machine-383449cc1fba}}.
\newblock


\bibitem[\protect\citeauthoryear{Aldwairi and Alwahedi}{Aldwairi and
  Alwahedi}{2018}]%
        {aldwairi2018detecting}
\bibfield{author}{\bibinfo{person}{Monther Aldwairi} {and} \bibinfo{person}{Ali
  Alwahedi}.} \bibinfo{year}{2018}\natexlab{}.
\newblock \showarticletitle{Detecting fake news in social media networks}.
\newblock \bibinfo{journal}{\emph{Procedia Computer Science}}
  \bibinfo{volume}{141} (\bibinfo{year}{2018}), \bibinfo{pages}{215--222}.
\newblock


\bibitem[\protect\citeauthoryear{Allcott and Gentzkow}{Allcott and
  Gentzkow}{2017}]%
        {allcott2017social}
\bibfield{author}{\bibinfo{person}{Hunt Allcott} {and} \bibinfo{person}{Matthew
  Gentzkow}.} \bibinfo{year}{2017}\natexlab{}.
\newblock \showarticletitle{Social media and fake news in the 2016 election}.
\newblock \bibinfo{journal}{\emph{Journal of economic perspectives}}
  \bibinfo{volume}{31}, \bibinfo{number}{2} (\bibinfo{year}{2017}),
  \bibinfo{pages}{211--36}.
\newblock


\bibitem[\protect\citeauthoryear{Alothali, Zaki, Mohamed, and
  Alashwal}{Alothali et~al\mbox{.}}{2018}]%
        {alothali2018detecting}
\bibfield{author}{\bibinfo{person}{Eiman Alothali}, \bibinfo{person}{Nazar
  Zaki}, \bibinfo{person}{Elfadil~A Mohamed}, {and} \bibinfo{person}{Hany
  Alashwal}.} \bibinfo{year}{2018}\natexlab{}.
\newblock \showarticletitle{Detecting social bots on Twitter: a literature
  review}. In \bibinfo{booktitle}{\emph{2018 International Conference on
  Innovations in Information Technology (IIT)}}. IEEE,
  \bibinfo{pages}{175--180}.
\newblock


\bibitem[\protect\citeauthoryear{Anna~Escher}{Anna~Escher}{2016}]%
        {AnnaEscher}
\bibfield{author}{\bibinfo{person}{Anthony~Ha Anna~Escher}.}
  \bibinfo{year}{2016}\natexlab{}.
\newblock \bibinfo{title}{WTF is clickbait?}
\newblock
  \bibinfo{howpublished}{\url{https://techcrunch.com/2016/09/25/wtf-is-clickbait/}}.
\newblock


\bibitem[\protect\citeauthoryear{Antoniadis, Litou, and Kalogeraki}{Antoniadis
  et~al\mbox{.}}{2015}]%
        {antoniadis2015model}
\bibfield{author}{\bibinfo{person}{Sotirios Antoniadis},
  \bibinfo{person}{Iouliana Litou}, {and} \bibinfo{person}{Vana Kalogeraki}.}
  \bibinfo{year}{2015}\natexlab{}.
\newblock \showarticletitle{A model for identifying misinformation in online
  social networks}. In \bibinfo{booktitle}{\emph{OTM Confederated International
  Conferences" On the Move to Meaningful Internet Systems"}}. Springer,
  \bibinfo{pages}{473--482}.
\newblock


\bibitem[\protect\citeauthoryear{Asano}{Asano}{2017}]%
        {Evanasano}
\bibfield{author}{\bibinfo{person}{Evan Asano}.}
  \bibinfo{year}{2017}\natexlab{}.
\newblock \showarticletitle{How much time do people spend on social media?}
\newblock \bibinfo{journal}{\emph{Social Media Today}} (\bibinfo{year}{2017}),
  \bibinfo{pages}{290--306}.
\newblock


\bibitem[\protect\citeauthoryear{Aslam}{Aslam}{2021}]%
        {Aslam2021}
\bibfield{author}{\bibinfo{person}{Salman Aslam}.}
  \bibinfo{year}{2021}\natexlab{}.
\newblock \bibinfo{title}{{Twitter by the Numbers: Stats, Demographics \& Fun
  Facts}}.
\newblock
  \bibinfo{howpublished}{\url{https://www.omnicoreagency.com/twitter-statistics/}}.
\newblock


\bibitem[\protect\citeauthoryear{Bakas, Michalas, and Ullah}{Bakas
  et~al\mbox{.}}{2020}]%
        {Michalas:20:NordSec:FunctionalSift}
\bibfield{author}{\bibinfo{person}{Alexandros Bakas}, \bibinfo{person}{Antonis
  Michalas}, {and} \bibinfo{person}{Amjad Ullah}.}
  \bibinfo{year}{2020}\natexlab{}.
\newblock \showarticletitle{(F)unctional Sifting: {A} Privacy-Preserving
  Reputation System Through Multi-Input Functional Encryption}. In
  \bibinfo{booktitle}{\emph{Secure {IT} Systems - 25th Nordic Conference,
  NordSec 2020, Virtual Event, November 23-24, 2020, Proceedings}}
  \emph{(\bibinfo{series}{Lecture Notes in Computer Science})},
  \bibfield{editor}{\bibinfo{person}{Mikael Asplund} {and}
  \bibinfo{person}{Simin Nadjm{-}Tehrani}} (Eds.),
  Vol.~\bibinfo{volume}{12556}. \bibinfo{publisher}{Springer},
  \bibinfo{pages}{111--126}.
\newblock
\urldef\tempurl%
\url{https://doi.org/10.1007/978-3-030-70852-8\_7}
\showDOI{\tempurl}


\bibitem[\protect\citeauthoryear{Balestrucci, De~Nicola, Inverso, and
  Trubiani}{Balestrucci et~al\mbox{.}}{2019}]%
        {balestrucci2019identification}
\bibfield{author}{\bibinfo{person}{Alessandro Balestrucci},
  \bibinfo{person}{Rocco De~Nicola}, \bibinfo{person}{Omar Inverso}, {and}
  \bibinfo{person}{Catia Trubiani}.} \bibinfo{year}{2019}\natexlab{}.
\newblock \showarticletitle{Identification of credulous users on Twitter}. In
  \bibinfo{booktitle}{\emph{Proceedings of the 34th ACM/SIGAPP Symposium on
  Applied Computing}}. \bibinfo{pages}{2096--2103}.
\newblock


\bibitem[\protect\citeauthoryear{Baly, Karadzhov, Alexandrov, Glass, and
  Nakov}{Baly et~al\mbox{.}}{2018}]%
        {baly2018predicting}
\bibfield{author}{\bibinfo{person}{Ramy Baly}, \bibinfo{person}{Georgi
  Karadzhov}, \bibinfo{person}{Dimitar Alexandrov}, \bibinfo{person}{James
  Glass}, {and} \bibinfo{person}{Preslav Nakov}.}
  \bibinfo{year}{2018}\natexlab{}.
\newblock \showarticletitle{Predicting factuality of reporting and bias of news
  media sources}.
\newblock \bibinfo{journal}{\emph{arXiv preprint arXiv:1810.01765}}
  (\bibinfo{year}{2018}).
\newblock


\bibitem[\protect\citeauthoryear{Benevenuto, Magno, Rodrigues, and
  Almeida}{Benevenuto et~al\mbox{.}}{2010}]%
        {benevenuto2010detecting}
\bibfield{author}{\bibinfo{person}{Fabricio Benevenuto},
  \bibinfo{person}{Gabriel Magno}, \bibinfo{person}{Tiago Rodrigues}, {and}
  \bibinfo{person}{Virgilio Almeida}.} \bibinfo{year}{2010}\natexlab{}.
\newblock \showarticletitle{Detecting spammers on twitter}. In
  \bibinfo{booktitle}{\emph{Collaboration, electronic messaging, anti-abuse and
  spam conference (CEAS)}}, Vol.~\bibinfo{volume}{6}. \bibinfo{pages}{12}.
\newblock


\bibitem[\protect\citeauthoryear{Binham}{Binham}{2019}]%
        {CarolineBinham}
\bibfield{author}{\bibinfo{person}{Caroline Binham}.}
  \bibinfo{year}{2019}\natexlab{}.
\newblock \bibinfo{title}{Companies fear rise of fake news and social media
  rumours}.
\newblock
  \bibinfo{howpublished}{\url{https://www.ft.com/content/4241a2f6-e080-11e9-9743-db5a370481bc}}.
\newblock


\bibitem[\protect\citeauthoryear{Bollen, Mao, and Pepe}{Bollen
  et~al\mbox{.}}{2011}]%
        {bollen2011modeling}
\bibfield{author}{\bibinfo{person}{Johan Bollen}, \bibinfo{person}{Huina Mao},
  {and} \bibinfo{person}{Alberto Pepe}.} \bibinfo{year}{2011}\natexlab{}.
\newblock \showarticletitle{Modeling public mood and emotion: Twitter sentiment
  and socio-economic phenomena}. In \bibinfo{booktitle}{\emph{Fifth
  International AAAI Conference on Weblogs and Social Media}}.
\newblock


\bibitem[\protect\citeauthoryear{Broder, Kumar, Maghoul, Raghavan, Rajagopalan,
  Stata, Tomkins, and Wiener}{Broder et~al\mbox{.}}{2000}]%
        {broder2000graph}
\bibfield{author}{\bibinfo{person}{Andrei Broder}, \bibinfo{person}{Ravi
  Kumar}, \bibinfo{person}{Farzin Maghoul}, \bibinfo{person}{Prabhakar
  Raghavan}, \bibinfo{person}{Sridhar Rajagopalan}, \bibinfo{person}{Raymie
  Stata}, \bibinfo{person}{Andrew Tomkins}, {and} \bibinfo{person}{Janet
  Wiener}.} \bibinfo{year}{2000}\natexlab{}.
\newblock \showarticletitle{Graph structure in the web}.
\newblock \bibinfo{journal}{\emph{Computer networks}} \bibinfo{volume}{33},
  \bibinfo{number}{1-6} (\bibinfo{year}{2000}), \bibinfo{pages}{309--320}.
\newblock


\bibitem[\protect\citeauthoryear{Bulger and Davison}{Bulger and
  Davison}{2018}]%
        {bulger2018promises}
\bibfield{author}{\bibinfo{person}{Monica Bulger} {and}
  \bibinfo{person}{Patrick Davison}.} \bibinfo{year}{2018}\natexlab{}.
\newblock \showarticletitle{The promises, challenges, and futures of media
  literacy}.
\newblock  (\bibinfo{year}{2018}).
\newblock


\bibitem[\protect\citeauthoryear{Bytwerk}{Bytwerk}{2010}]%
        {bytwerk2010grassroots}
\bibfield{author}{\bibinfo{person}{Randall~L Bytwerk}.}
  \bibinfo{year}{2010}\natexlab{}.
\newblock \showarticletitle{Grassroots propaganda in the Third Reich: The Reich
  ring for National Socialist propaganda and public enlightenment}.
\newblock \bibinfo{journal}{\emph{German Studies Review}}
  (\bibinfo{year}{2010}), \bibinfo{pages}{93--118}.
\newblock


\bibitem[\protect\citeauthoryear{Canini, Suh, and Pirolli}{Canini
  et~al\mbox{.}}{2011}]%
        {canini2011finding}
\bibfield{author}{\bibinfo{person}{Kevin~R Canini}, \bibinfo{person}{Bongwon
  Suh}, {and} \bibinfo{person}{Peter~L Pirolli}.}
  \bibinfo{year}{2011}\natexlab{}.
\newblock \showarticletitle{Finding credible information sources in social
  networks based on content and social structure}. In
  \bibinfo{booktitle}{\emph{2011 IEEE Third International Conference on
  Privacy, Security, Risk and Trust and 2011 IEEE Third International
  Conference on Social Computing}}. IEEE, \bibinfo{pages}{1--8}.
\newblock


\bibitem[\protect\citeauthoryear{Carlson}{Carlson}{2017}]%
        {Eryn}
\bibfield{author}{\bibinfo{person}{Eryn Carlson}.}
  \bibinfo{year}{2017}\natexlab{}.
\newblock \bibinfo{title}{{Flagging Fake News}}.
\newblock
  \bibinfo{howpublished}{\url{https://niemanreports.org/articles/flagging-fake-news/}}.
\newblock


\bibitem[\protect\citeauthoryear{Carson}{Carson}{2019}]%
        {Thetelegraph}
\bibfield{author}{\bibinfo{person}{James Carson}.}
  \bibinfo{year}{2019}\natexlab{}.
\newblock \bibinfo{title}{{Fake news: What exactly is it and how can you spot
  it? }}.
\newblock
  \bibinfo{howpublished}{\url{https://www.telegraph.co.uk/technology/0/fake-news-exactly-has-really-had-influence/}}.
\newblock


\bibitem[\protect\citeauthoryear{Cha, Haddadi, Benevenuto, and Gummadi}{Cha
  et~al\mbox{.}}{2010}]%
        {cha2010measuring}
\bibfield{author}{\bibinfo{person}{Meeyoung Cha}, \bibinfo{person}{Hamed
  Haddadi}, \bibinfo{person}{Fabricio Benevenuto}, {and}
  \bibinfo{person}{Krishna~P Gummadi}.} \bibinfo{year}{2010}\natexlab{}.
\newblock \showarticletitle{Measuring user influence in twitter: The million
  follower fallacy}. In \bibinfo{booktitle}{\emph{fourth international AAAI
  conference on weblogs and social media}}.
\newblock


\bibitem[\protect\citeauthoryear{Chen, De, Hu, and Hwang}{Chen
  et~al\mbox{.}}{2014}]%
        {chen2014wisdom}
\bibfield{author}{\bibinfo{person}{Hailiang Chen}, \bibinfo{person}{Prabuddha
  De}, \bibinfo{person}{Yu~Jeffrey Hu}, {and} \bibinfo{person}{Byoung-Hyoun
  Hwang}.} \bibinfo{year}{2014}\natexlab{}.
\newblock \showarticletitle{Wisdom of crowds: The value of stock opinions
  transmitted through social media}.
\newblock \bibinfo{journal}{\emph{The Review of Financial Studies}}
  \bibinfo{volume}{27}, \bibinfo{number}{5} (\bibinfo{year}{2014}),
  \bibinfo{pages}{1367--1403}.
\newblock


\bibitem[\protect\citeauthoryear{Chen, Conroy, and Rubin}{Chen
  et~al\mbox{.}}{2015}]%
        {chen2015misleading}
\bibfield{author}{\bibinfo{person}{Yimin Chen}, \bibinfo{person}{Niall~J
  Conroy}, {and} \bibinfo{person}{Victoria~L Rubin}.}
  \bibinfo{year}{2015}\natexlab{}.
\newblock \showarticletitle{Misleading online content: recognizing clickbait
  as" false news"}. In \bibinfo{booktitle}{\emph{Proceedings of the 2015 ACM on
  workshop on multimodal deception detection}}. \bibinfo{pages}{15--19}.
\newblock


\bibitem[\protect\citeauthoryear{Christian~Reuter and
  Schlegel}{Christian~Reuter and Schlegel}{2019}]%
        {Fake:News:Perception:Germany}
\bibfield{author}{\bibinfo{person}{Jan~Kirchner Christian~Reuter,
  Katrin~Hartwig} {and} \bibinfo{person}{Noah Schlegel}.}
  \bibinfo{year}{2019}\natexlab{}.
\newblock \showarticletitle{Fake News Perception in Germany: A Representative
  Study of People's Attitudes and Approaches to Counteract Disinformation}. In
  \bibinfo{booktitle}{\emph{14th International Conference on
  Wirtschaftsinformatik}}.
\newblock


\bibitem[\protect\citeauthoryear{Chu, Gianvecchio, Wang, and Jajodia}{Chu
  et~al\mbox{.}}{2010}]%
        {chu2010tweeting}
\bibfield{author}{\bibinfo{person}{Zi Chu}, \bibinfo{person}{Steven
  Gianvecchio}, \bibinfo{person}{Haining Wang}, {and} \bibinfo{person}{Sushil
  Jajodia}.} \bibinfo{year}{2010}\natexlab{}.
\newblock \showarticletitle{Who is tweeting on Twitter: human, bot, or
  cyborg?}. In \bibinfo{booktitle}{\emph{Proceedings of the 26th annual
  computer security applications conference}}. ACM, \bibinfo{pages}{21--30}.
\newblock


\bibitem[\protect\citeauthoryear{Chu, Gianvecchio, Wang, and Jajodia}{Chu
  et~al\mbox{.}}{2012}]%
        {chu2012detecting}
\bibfield{author}{\bibinfo{person}{Zi Chu}, \bibinfo{person}{Steven
  Gianvecchio}, \bibinfo{person}{Haining Wang}, {and} \bibinfo{person}{Sushil
  Jajodia}.} \bibinfo{year}{2012}\natexlab{}.
\newblock \showarticletitle{Detecting automation of twitter accounts: Are you a
  human, bot, or cyborg?}
\newblock \bibinfo{journal}{\emph{IEEE Transactions on Dependable and Secure
  Computing}} \bibinfo{volume}{9}, \bibinfo{number}{6} (\bibinfo{year}{2012}),
  \bibinfo{pages}{811--824}.
\newblock


\bibitem[\protect\citeauthoryear{Ciampaglia, Shiralkar, Rocha, Bollen, Menczer,
  and Flammini}{Ciampaglia et~al\mbox{.}}{2015}]%
        {ciampaglia2015computational}
\bibfield{author}{\bibinfo{person}{Giovanni~Luca Ciampaglia},
  \bibinfo{person}{Prashant Shiralkar}, \bibinfo{person}{Luis~M Rocha},
  \bibinfo{person}{Johan Bollen}, \bibinfo{person}{Filippo Menczer}, {and}
  \bibinfo{person}{Alessandro Flammini}.} \bibinfo{year}{2015}\natexlab{}.
\newblock \showarticletitle{Computational fact checking from knowledge
  networks}.
\newblock \bibinfo{journal}{\emph{PloS one}} \bibinfo{volume}{10},
  \bibinfo{number}{6} (\bibinfo{year}{2015}), \bibinfo{pages}{e0128193}.
\newblock


\bibitem[\protect\citeauthoryear{Clark}{Clark}{201b}]%
        {BryanClark}
\bibfield{author}{\bibinfo{person}{Bryan Clark}.}
  \bibinfo{year}{201b}\natexlab{}.
\newblock \bibinfo{title}{{SurfSafe offers a browser-based solution to fake
  news}}.
\newblock
  \bibinfo{howpublished}{\url{https://thenextweb.com/insider/2018/08/21/surfsafe-offers-a-browser-based-solution-to-fake-news/}}.
\newblock


\bibitem[\protect\citeauthoryear{Clauset, Shalizi, and Newman}{Clauset
  et~al\mbox{.}}{2009}]%
        {clauset2009power}
\bibfield{author}{\bibinfo{person}{Aaron Clauset},
  \bibinfo{person}{Cosma~Rohilla Shalizi}, {and} \bibinfo{person}{Mark~EJ
  Newman}.} \bibinfo{year}{2009}\natexlab{}.
\newblock \showarticletitle{Power-law distributions in empirical data}.
\newblock \bibinfo{journal}{\emph{SIAM review}} \bibinfo{volume}{51},
  \bibinfo{number}{4} (\bibinfo{year}{2009}), \bibinfo{pages}{661--703}.
\newblock


\bibitem[\protect\citeauthoryear{Collins, Hoang, Nguyen, and Hwang}{Collins
  et~al\mbox{.}}{2020}]%
        {collins2020trends}
\bibfield{author}{\bibinfo{person}{Botambu Collins},
  \bibinfo{person}{Dinh~Tuyen Hoang}, \bibinfo{person}{Ngoc~Thanh Nguyen},
  {and} \bibinfo{person}{Dosam Hwang}.} \bibinfo{year}{2020}\natexlab{}.
\newblock \showarticletitle{Trends in combating fake news on social media--a
  survey}.
\newblock \bibinfo{journal}{\emph{Journal of Information and
  Telecommunication}} (\bibinfo{year}{2020}), \bibinfo{pages}{1--20}.
\newblock


\bibitem[\protect\citeauthoryear{Constantinides, Henfridsson, and
  Parker}{Constantinides et~al\mbox{.}}{2018}]%
        {constantinides2018introduction}
\bibfield{author}{\bibinfo{person}{Panos Constantinides}, \bibinfo{person}{Ola
  Henfridsson}, {and} \bibinfo{person}{Geoffrey~G Parker}.}
  \bibinfo{year}{2018}\natexlab{}.
\newblock \bibinfo{title}{Introduction-platforms and infrastructures in the
  digital age}.
\newblock
\newblock


\bibitem[\protect\citeauthoryear{Cresci, Di~Pietro, Petrocchi, Spognardi, and
  Tesconi}{Cresci et~al\mbox{.}}{2016}]%
        {cresci2016dna}
\bibfield{author}{\bibinfo{person}{Stefano Cresci}, \bibinfo{person}{Roberto
  Di~Pietro}, \bibinfo{person}{Marinella Petrocchi}, \bibinfo{person}{Angelo
  Spognardi}, {and} \bibinfo{person}{Maurizio Tesconi}.}
  \bibinfo{year}{2016}\natexlab{}.
\newblock \showarticletitle{DNA-inspired online behavioral modeling and its
  application to spambot detection}.
\newblock \bibinfo{journal}{\emph{IEEE Intelligent Systems}}
  \bibinfo{volume}{31}, \bibinfo{number}{5} (\bibinfo{year}{2016}),
  \bibinfo{pages}{58--64}.
\newblock


\bibitem[\protect\citeauthoryear{Cresci, Di~Pietro, Petrocchi, Spognardi, and
  Tesconi}{Cresci et~al\mbox{.}}{2017}]%
        {cresci2017paradigm}
\bibfield{author}{\bibinfo{person}{Stefano Cresci}, \bibinfo{person}{Roberto
  Di~Pietro}, \bibinfo{person}{Marinella Petrocchi}, \bibinfo{person}{Angelo
  Spognardi}, {and} \bibinfo{person}{Maurizio Tesconi}.}
  \bibinfo{year}{2017}\natexlab{}.
\newblock \showarticletitle{The paradigm-shift of social spambots: Evidence,
  theories, and tools for the arms race}. In
  \bibinfo{booktitle}{\emph{Proceedings of the 26th international conference on
  world wide web companion}}. \bibinfo{pages}{963--972}.
\newblock


\bibitem[\protect\citeauthoryear{Cresci, Lillo, Regoli, Tardelli, and
  Tesconi}{Cresci et~al\mbox{.}}{2019}]%
        {cresci2019cashtag}
\bibfield{author}{\bibinfo{person}{Stefano Cresci}, \bibinfo{person}{Fabrizio
  Lillo}, \bibinfo{person}{Daniele Regoli}, \bibinfo{person}{Serena Tardelli},
  {and} \bibinfo{person}{Maurizio Tesconi}.} \bibinfo{year}{2019}\natexlab{}.
\newblock \showarticletitle{Cashtag piggybacking: Uncovering spam and bot
  activity in stock microblogs on Twitter}.
\newblock \bibinfo{journal}{\emph{ACM Transactions on the Web (TWEB)}}
  \bibinfo{volume}{13}, \bibinfo{number}{2} (\bibinfo{year}{2019}),
  \bibinfo{pages}{11}.
\newblock


\bibitem[\protect\citeauthoryear{Dauphin, Fan, Auli, and Grangier}{Dauphin
  et~al\mbox{.}}{2017}]%
        {dauphin2017language}
\bibfield{author}{\bibinfo{person}{Yann~N Dauphin}, \bibinfo{person}{Angela
  Fan}, \bibinfo{person}{Michael Auli}, {and} \bibinfo{person}{David
  Grangier}.} \bibinfo{year}{2017}\natexlab{}.
\newblock \showarticletitle{Language modeling with gated convolutional
  networks}. In \bibinfo{booktitle}{\emph{Proceedings of the 34th International
  Conference on Machine Learning-Volume 70}}. JMLR. org,
  \bibinfo{pages}{933--941}.
\newblock


\bibitem[\protect\citeauthoryear{Davis, Varol, Ferrara, Flammini, and
  Menczer}{Davis et~al\mbox{.}}{2016}]%
        {davis2016botornot}
\bibfield{author}{\bibinfo{person}{Clayton~Allen Davis}, \bibinfo{person}{Onur
  Varol}, \bibinfo{person}{Emilio Ferrara}, \bibinfo{person}{Alessandro
  Flammini}, {and} \bibinfo{person}{Filippo Menczer}.}
  \bibinfo{year}{2016}\natexlab{}.
\newblock \showarticletitle{Botornot: A system to evaluate social bots}. In
  \bibinfo{booktitle}{\emph{Proceedings of the 25th International Conference
  Companion on World Wide Web}}. International World Wide Web Conferences
  Steering Committee, \bibinfo{pages}{273--274}.
\newblock


\bibitem[\protect\citeauthoryear{De~Domenico, Lima, Mougel, and
  Musolesi}{De~Domenico et~al\mbox{.}}{2013}]%
        {de2013anatomy}
\bibfield{author}{\bibinfo{person}{Manlio De~Domenico},
  \bibinfo{person}{Antonio Lima}, \bibinfo{person}{Paul Mougel}, {and}
  \bibinfo{person}{Mirco Musolesi}.} \bibinfo{year}{2013}\natexlab{}.
\newblock \showarticletitle{The anatomy of a scientific rumor}.
\newblock \bibinfo{journal}{\emph{Scientific reports}}  \bibinfo{volume}{3}
  (\bibinfo{year}{2013}), \bibinfo{pages}{2980}.
\newblock


\bibitem[\protect\citeauthoryear{DFRLab}{DFRLab}{2016}]%
        {DFRLab}
\bibfield{author}{\bibinfo{person}{DFRLab}.} \bibinfo{year}{2016}\natexlab{}.
\newblock \bibinfo{title}{{Human, Bot or Cyborg?}}
\newblock
  \bibinfo{howpublished}{\url{https://medium.com/@DFRLab/human-bot-or-cyborg-41273cdb1e17}}.
\newblock


\bibitem[\protect\citeauthoryear{Dimitriou and Michalas}{Dimitriou and
  Michalas}{2012}]%
        {Michalas:12:StR}
\bibfield{author}{\bibinfo{person}{T. Dimitriou} {and} \bibinfo{person}{A.
  Michalas}.} \bibinfo{year}{2012}\natexlab{}.
\newblock \showarticletitle{Multi-Party Trust Computation in Decentralized
  Environments}. In \bibinfo{booktitle}{\emph{2012 5th International Conference
  on New Technologies, Mobility and Security (NTMS)}}. \bibinfo{pages}{1--5}.
\newblock
\showISSN{2157-4952}
\urldef\tempurl%
\url{https://doi.org/10.1109/NTMS.2012.6208686}
\showDOI{\tempurl}


\bibitem[\protect\citeauthoryear{Dimitriou and Michalas}{Dimitriou and
  Michalas}{2014}]%
        {Michalas:14:StRM}
\bibfield{author}{\bibinfo{person}{Tassos Dimitriou} {and}
  \bibinfo{person}{Antonis Michalas}.} \bibinfo{year}{2014}\natexlab{}.
\newblock \showarticletitle{Multi-party Trust Computation in Decentralized
  Environments in the Presence of Malicious Adversaries}.
\newblock \bibinfo{journal}{\emph{Ad Hoc Networks}}  \bibinfo{volume}{15}
  (\bibinfo{date}{April} \bibinfo{year}{2014}), \bibinfo{pages}{53--66}.
\newblock
\showISSN{1570-8705}
\urldef\tempurl%
\url{https://doi.org/10.1016/j.adhoc.2013.04.013}
\showDOI{\tempurl}


\bibitem[\protect\citeauthoryear{Edwards, Edwards, Spence, and Shelton}{Edwards
  et~al\mbox{.}}{2014}]%
        {edwards2014bot}
\bibfield{author}{\bibinfo{person}{Chad Edwards}, \bibinfo{person}{Autumn
  Edwards}, \bibinfo{person}{Patric~R Spence}, {and}
  \bibinfo{person}{Ashleigh~K Shelton}.} \bibinfo{year}{2014}\natexlab{}.
\newblock \showarticletitle{Is that a bot running the social media feed?
  Testing the differences in perceptions of communication quality for a human
  agent and a bot agent on Twitter}.
\newblock \bibinfo{journal}{\emph{Computers in Human Behavior}}
  \bibinfo{volume}{33} (\bibinfo{year}{2014}), \bibinfo{pages}{372--376}.
\newblock


\bibitem[\protect\citeauthoryear{Ericsson, Hoffman, and Kozbelt}{Ericsson
  et~al\mbox{.}}{2018}]%
        {ericsson2018cambridge}
\bibfield{author}{\bibinfo{person}{K~Anders Ericsson},
  \bibinfo{person}{Robert~R Hoffman}, {and} \bibinfo{person}{Aaron Kozbelt}.}
  \bibinfo{year}{2018}\natexlab{}.
\newblock \bibinfo{booktitle}{\emph{The Cambridge handbook of expertise and
  expert performance}}.
\newblock \bibinfo{publisher}{Cambridge University Press}.
\newblock


\bibitem[\protect\citeauthoryear{Er{\c{s}}ahin, Akta{\c{s}},
  K{\i}l{\i}n{\c{c}}, and Akyol}{Er{\c{s}}ahin et~al\mbox{.}}{2017}]%
        {ercsahin2017twitter}
\bibfield{author}{\bibinfo{person}{Buket Er{\c{s}}ahin},
  \bibinfo{person}{{\"O}zlem Akta{\c{s}}}, \bibinfo{person}{Deniz
  K{\i}l{\i}n{\c{c}}}, {and} \bibinfo{person}{Ceyhun Akyol}.}
  \bibinfo{year}{2017}\natexlab{}.
\newblock \showarticletitle{Twitter fake account detection}. In
  \bibinfo{booktitle}{\emph{2017 International Conference on Computer Science
  and Engineering (UBMK)}}. IEEE, \bibinfo{pages}{388--392}.
\newblock


\bibitem[\protect\citeauthoryear{Fernandes}{Fernandes}{2019}]%
        {thiagorainmaker}
\bibfield{author}{\bibinfo{person}{Thiago Fernandes}.}
  \bibinfo{year}{2019}\natexlab{}.
\newblock \bibinfo{title}{{liardataset}}.
\newblock
  \bibinfo{howpublished}{\url{https://github.com/thiagorainmaker77/liar_dataset}}.
\newblock


\bibitem[\protect\citeauthoryear{Ferreira and Vlachos}{Ferreira and
  Vlachos}{2016}]%
        {ferreira2016emergent}
\bibfield{author}{\bibinfo{person}{William Ferreira} {and}
  \bibinfo{person}{Andreas Vlachos}.} \bibinfo{year}{2016}\natexlab{}.
\newblock \showarticletitle{Emergent: a novel data-set for stance
  classification}. In \bibinfo{booktitle}{\emph{Proceedings of the 2016
  conference of the North American chapter of the association for computational
  linguistics: Human language technologies}}. \bibinfo{pages}{1163--1168}.
\newblock


\bibitem[\protect\citeauthoryear{Figueira and Oliveira}{Figueira and
  Oliveira}{2017}]%
        {figueira2017current}
\bibfield{author}{\bibinfo{person}{{\'A}lvaro Figueira} {and}
  \bibinfo{person}{Luciana Oliveira}.} \bibinfo{year}{2017}\natexlab{}.
\newblock \showarticletitle{The current state of fake news: challenges and
  opportunities}.
\newblock \bibinfo{journal}{\emph{Procedia Computer Science}}
  \bibinfo{volume}{121} (\bibinfo{year}{2017}), \bibinfo{pages}{817--825}.
\newblock


\bibitem[\protect\citeauthoryear{for minds}{for minds}{[n. d.]}]%
        {Madeforminds}
\bibfield{author}{\bibinfo{person}{Made for minds}.} \bibinfo{year}{[n.
  d.]}\natexlab{}.
\newblock \bibinfo{title}{Spread of coronavirus fake news causes hundreds of
  deaths}.
\newblock
  \bibinfo{howpublished}{\url{https://www.dw.com/en/coronavirus-misinformation/a-54529310}}.
\newblock


\bibitem[\protect\citeauthoryear{Funke~Daniel}{Funke~Daniel}{2019}]%
        {FunkeDaniel}
\bibfield{author}{\bibinfo{person}{Flamini~Daniela Funke~Daniel}.}
  \bibinfo{year}{2019}\natexlab{}.
\newblock \bibinfo{title}{{A guide to anti-misinformation actions around the
  world}}.
\newblock
  \bibinfo{howpublished}{\url{https://www.poynter.org/ifcn/anti-misinformation-actions/}}.
\newblock


\bibitem[\protect\citeauthoryear{Gabrov{\v{s}}ek, Aleksovski, Mozeti{\v{c}},
  and Gr{\v{c}}ar}{Gabrov{\v{s}}ek et~al\mbox{.}}{2017}]%
        {gabrovvsek2017twitter}
\bibfield{author}{\bibinfo{person}{Peter Gabrov{\v{s}}ek},
  \bibinfo{person}{Darko Aleksovski}, \bibinfo{person}{Igor Mozeti{\v{c}}},
  {and} \bibinfo{person}{Miha Gr{\v{c}}ar}.} \bibinfo{year}{2017}\natexlab{}.
\newblock \showarticletitle{Twitter sentiment around the Earnings Announcement
  events}.
\newblock \bibinfo{journal}{\emph{PloS one}} \bibinfo{volume}{12},
  \bibinfo{number}{2} (\bibinfo{year}{2017}).
\newblock


\bibitem[\protect\citeauthoryear{Gao, Liang, Fan, Wang, Sun, and Han}{Gao
  et~al\mbox{.}}{2010}]%
        {gao2010community}
\bibfield{author}{\bibinfo{person}{Jing Gao}, \bibinfo{person}{Feng Liang},
  \bibinfo{person}{Wei Fan}, \bibinfo{person}{Chi Wang},
  \bibinfo{person}{Yizhou Sun}, {and} \bibinfo{person}{Jiawei Han}.}
  \bibinfo{year}{2010}\natexlab{}.
\newblock \showarticletitle{On community outliers and their efficient detection
  in information networks}. In \bibinfo{booktitle}{\emph{Proceedings of the
  16th ACM SIGKDD international conference on Knowledge discovery and data
  mining}}. ACM, \bibinfo{pages}{813--822}.
\newblock


\bibitem[\protect\citeauthoryear{Garcia, Mavrodiev, Casati, and
  Schweitzer}{Garcia et~al\mbox{.}}{2017}]%
        {garcia2017understanding}
\bibfield{author}{\bibinfo{person}{David Garcia}, \bibinfo{person}{Pavlin
  Mavrodiev}, \bibinfo{person}{Daniele Casati}, {and} \bibinfo{person}{Frank
  Schweitzer}.} \bibinfo{year}{2017}\natexlab{}.
\newblock \showarticletitle{Understanding popularity, reputation, and social
  influence in the twitter society}.
\newblock \bibinfo{journal}{\emph{Policy \& Internet}} \bibinfo{volume}{9},
  \bibinfo{number}{3} (\bibinfo{year}{2017}), \bibinfo{pages}{343--364}.
\newblock


\bibitem[\protect\citeauthoryear{Gazi and {\c{C}}etin}{Gazi and
  {\c{C}}etin}{2017}]%
        {gazi2017research}
\bibfield{author}{\bibinfo{person}{Mehmet~Ali Gazi} {and}
  \bibinfo{person}{Muharrem {\c{C}}etin}.} \bibinfo{year}{2017}\natexlab{}.
\newblock \showarticletitle{The research of the level of social media addiction
  of university students}.
\newblock \bibinfo{journal}{\emph{International Journal of Social Sciences and
  Education Research}} \bibinfo{volume}{3}, \bibinfo{number}{2}
  (\bibinfo{year}{2017}), \bibinfo{pages}{549--559}.
\newblock


\bibitem[\protect\citeauthoryear{Ghavipour and Meybodi}{Ghavipour and
  Meybodi}{2018a}]%
        {ghavipour2018dynamic}
\bibfield{author}{\bibinfo{person}{Mina Ghavipour} {and}
  \bibinfo{person}{Mohammad~Reza Meybodi}.} \bibinfo{year}{2018}\natexlab{a}.
\newblock \showarticletitle{A dynamic algorithm for stochastic trust
  propagation in online social networks: Learning automata approach}.
\newblock \bibinfo{journal}{\emph{Computer Communications}}
  \bibinfo{volume}{123} (\bibinfo{year}{2018}), \bibinfo{pages}{11--23}.
\newblock


\bibitem[\protect\citeauthoryear{Ghavipour and Meybodi}{Ghavipour and
  Meybodi}{2018b}]%
        {ghavipour2018trust}
\bibfield{author}{\bibinfo{person}{Mina Ghavipour} {and}
  \bibinfo{person}{Mohammad~Reza Meybodi}.} \bibinfo{year}{2018}\natexlab{b}.
\newblock \showarticletitle{Trust propagation algorithm based on learning
  automata for inferring local trust in online social networks}.
\newblock \bibinfo{journal}{\emph{Knowledge-Based Systems}}
  \bibinfo{volume}{143} (\bibinfo{year}{2018}), \bibinfo{pages}{307--316}.
\newblock


\bibitem[\protect\citeauthoryear{Ghosh, Viswanath, Kooti, Sharma, Korlam,
  Benevenuto, Ganguly, and Gummadi}{Ghosh et~al\mbox{.}}{2012}]%
        {ghosh2012understanding}
\bibfield{author}{\bibinfo{person}{Saptarshi Ghosh}, \bibinfo{person}{Bimal
  Viswanath}, \bibinfo{person}{Farshad Kooti}, \bibinfo{person}{Naveen~Kumar
  Sharma}, \bibinfo{person}{Gautam Korlam}, \bibinfo{person}{Fabricio
  Benevenuto}, \bibinfo{person}{Niloy Ganguly}, {and}
  \bibinfo{person}{Krishna~Phani Gummadi}.} \bibinfo{year}{2012}\natexlab{}.
\newblock \showarticletitle{Understanding and combating link farming in the
  twitter social network}. In \bibinfo{booktitle}{\emph{Proceedings of the 21st
  international conference on World Wide Web}}. \bibinfo{pages}{61--70}.
\newblock


\bibitem[\protect\citeauthoryear{Giachanou and Ghanem}{Giachanou and
  Ghanem}{2019}]%
        {giachanou2019bot}
\bibfield{author}{\bibinfo{person}{Anastasia Giachanou} {and}
  \bibinfo{person}{Bilal Ghanem}.} \bibinfo{year}{2019}\natexlab{}.
\newblock \showarticletitle{Bot and Gender Detection using Textual and
  Stylistic Information}.
\newblock \bibinfo{journal}{\emph{PAN}}  \bibinfo{volume}{16}
  (\bibinfo{year}{2019}), \bibinfo{pages}{5}.
\newblock


\bibitem[\protect\citeauthoryear{Giatsidis, Thilikos, and
  Vazirgiannis}{Giatsidis et~al\mbox{.}}{2013}]%
        {giatsidis2013d}
\bibfield{author}{\bibinfo{person}{Christos Giatsidis},
  \bibinfo{person}{Dimitrios~M Thilikos}, {and} \bibinfo{person}{Michalis
  Vazirgiannis}.} \bibinfo{year}{2013}\natexlab{}.
\newblock \showarticletitle{D-cores: measuring collaboration of directed graphs
  based on degeneracy}.
\newblock \bibinfo{journal}{\emph{Knowledge and information systems}}
  \bibinfo{volume}{35}, \bibinfo{number}{2} (\bibinfo{year}{2013}),
  \bibinfo{pages}{311--343}.
\newblock


\bibitem[\protect\citeauthoryear{Gibert, Mateu, and Planes}{Gibert
  et~al\mbox{.}}{2020}]%
        {gibert2020rise}
\bibfield{author}{\bibinfo{person}{Daniel Gibert}, \bibinfo{person}{Carles
  Mateu}, {and} \bibinfo{person}{Jordi Planes}.}
  \bibinfo{year}{2020}\natexlab{}.
\newblock \showarticletitle{The rise of machine learning for detection and
  classification of malware: Research developments, trends and challenges}.
\newblock \bibinfo{journal}{\emph{Journal of Network and Computer
  Applications}}  \bibinfo{volume}{153} (\bibinfo{year}{2020}),
  \bibinfo{pages}{102526}.
\newblock


\bibitem[\protect\citeauthoryear{Gie{\l}czyk, Wawrzyniak, and
  Chora{\'s}}{Gie{\l}czyk et~al\mbox{.}}{2019}]%
        {gielczyk2019evaluation}
\bibfield{author}{\bibinfo{person}{Agata Gie{\l}czyk},
  \bibinfo{person}{Rafa{\l} Wawrzyniak}, {and} \bibinfo{person}{Micha{\l}
  Chora{\'s}}.} \bibinfo{year}{2019}\natexlab{}.
\newblock \showarticletitle{Evaluation of the existing tools for fake news
  detection}. In \bibinfo{booktitle}{\emph{IFIP International Conference on
  Computer Information Systems and Industrial Management}}. Springer,
  \bibinfo{pages}{144--151}.
\newblock


\bibitem[\protect\citeauthoryear{gilani}{gilani}{2018}]%
        {STCSZafar}
\bibfield{author}{\bibinfo{person}{Zafar gilani}.}
  \bibinfo{year}{2018}\natexlab{}.
\newblock \bibinfo{title}{{STCS - Streaming Twitter Computation System}}.
\newblock \bibinfo{howpublished}{\url{https://github.com/zafargilani/stcs}}.
\newblock


\bibitem[\protect\citeauthoryear{Gilani, Kochmar, and Crowcroft}{Gilani
  et~al\mbox{.}}{2017}]%
        {gilani2017classification}
\bibfield{author}{\bibinfo{person}{Zafar Gilani}, \bibinfo{person}{Ekaterina
  Kochmar}, {and} \bibinfo{person}{Jon Crowcroft}.}
  \bibinfo{year}{2017}\natexlab{}.
\newblock \showarticletitle{Classification of twitter accounts into automated
  agents and human users}. In \bibinfo{booktitle}{\emph{Proceedings of the 2017
  IEEE/ACM International Conference on Advances in Social Networks Analysis and
  Mining 2017}}. ACM, \bibinfo{pages}{489--496}.
\newblock


\bibitem[\protect\citeauthoryear{Grice}{Grice}{2017}]%
        {Independent}
\bibfield{author}{\bibinfo{person}{Andrew Grice}.}
  \bibinfo{year}{2017}\natexlab{}.
\newblock \bibinfo{title}{{Fake news handed Brexiteers the referendum and now
  they have no idea what they're doing}}.
\newblock
  \bibinfo{howpublished}{\url{https://www.independent.co.uk/voices/michael-gove-boris-johnson-brexit-eurosceptic-press-theresa-may-a7533806.html}}.
\newblock


\bibitem[\protect\citeauthoryear{Grier, Thomas, Paxson, and Zhang}{Grier
  et~al\mbox{.}}{2010}]%
        {grier2010spam}
\bibfield{author}{\bibinfo{person}{Chris Grier}, \bibinfo{person}{Kurt Thomas},
  \bibinfo{person}{Vern Paxson}, {and} \bibinfo{person}{Michael Zhang}.}
  \bibinfo{year}{2010}\natexlab{}.
\newblock \showarticletitle{@ spam: the underground on 140 characters or less}.
  In \bibinfo{booktitle}{\emph{Proceedings of the 17th ACM conference on
  Computer and communications security}}. ACM, \bibinfo{pages}{27--37}.
\newblock


\bibitem[\protect\citeauthoryear{Grigorev}{Grigorev}{2017}]%
        {grigorev2017identifying}
\bibfield{author}{\bibinfo{person}{Alexey Grigorev}.}
  \bibinfo{year}{2017}\natexlab{}.
\newblock \showarticletitle{Identifying clickbait posts on social media with an
  ensemble of linear models}.
\newblock \bibinfo{journal}{\emph{arXiv preprint arXiv:1710.00399}}
  (\bibinfo{year}{2017}).
\newblock


\bibitem[\protect\citeauthoryear{Griswold}{Griswold}{2016}]%
        {AlisonGriswold}
\bibfield{author}{\bibinfo{person}{Alison Griswold}.}
  \bibinfo{year}{2016}\natexlab{}.
\newblock \bibinfo{title}{Facebook warned people that a popular fake news
  detector might be ``unsafe''}.
\newblock
  \bibinfo{howpublished}{\url{https://qz.com/851894/facebook-said-bs-detector-a-plugin-to-detect-fake-news-might-be-unsafe/}}.
\newblock


\bibitem[\protect\citeauthoryear{Gupta, Kumaraguru, Castillo, and Meier}{Gupta
  et~al\mbox{.}}{2014}]%
        {gupta2014tweetcred}
\bibfield{author}{\bibinfo{person}{Aditi Gupta}, \bibinfo{person}{Ponnurangam
  Kumaraguru}, \bibinfo{person}{Carlos Castillo}, {and}
  \bibinfo{person}{Patrick Meier}.} \bibinfo{year}{2014}\natexlab{}.
\newblock \showarticletitle{Tweetcred: Real-time credibility assessment of
  content on twitter}. In \bibinfo{booktitle}{\emph{International Conference on
  Social Informatics}}. Springer, \bibinfo{pages}{228--243}.
\newblock


\bibitem[\protect\citeauthoryear{Hannah~Bastl}{Hannah~Bastl}{2017}]%
        {Hannah}
\bibfield{author}{\bibinfo{person}{Elmar~Haussmann Hannah~Bastl,
  Bjorn~Buchhold}.} \bibinfo{year}{2017}\natexlab{}.
\newblock \bibinfo{title}{{Triple Scoring}}.
\newblock
  \bibinfo{howpublished}{\url{https://www.wsdm-cup-2017.org/triple-scoring.html}}.
\newblock


\bibitem[\protect\citeauthoryear{Hanselowski}{Hanselowski}{2017}]%
        {Andreas}
\bibfield{author}{\bibinfo{person}{Andreas Hanselowski}.}
  \bibinfo{year}{2017}\natexlab{}.
\newblock \bibinfo{title}{{Team Athene on the Fake News Challenge}}.
\newblock
  \bibinfo{howpublished}{\url{https://medium.com/@andre134679/team-athene-on-the-fake-news-challenge-28a5cf5e017b}}.
\newblock


\bibitem[\protect\citeauthoryear{Haralabopoulos, Anagnostopoulos, and
  Zeadally}{Haralabopoulos et~al\mbox{.}}{2015}]%
        {haralabopoulos2015lifespan}
\bibfield{author}{\bibinfo{person}{Giannis Haralabopoulos},
  \bibinfo{person}{Ioannis Anagnostopoulos}, {and} \bibinfo{person}{Sherali
  Zeadally}.} \bibinfo{year}{2015}\natexlab{}.
\newblock \showarticletitle{Lifespan and propagation of information in On-line
  Social Networks: A case study based on Reddit}.
\newblock \bibinfo{journal}{\emph{Journal of network and computer
  applications}}  \bibinfo{volume}{56} (\bibinfo{year}{2015}),
  \bibinfo{pages}{88--100}.
\newblock


\bibitem[\protect\citeauthoryear{Hartwig and Reuter}{Hartwig and
  Reuter}{2019}]%
        {hartwig2019trustytweet}
\bibfield{author}{\bibinfo{person}{Katrin Hartwig} {and}
  \bibinfo{person}{Christian Reuter}.} \bibinfo{year}{2019}\natexlab{}.
\newblock \showarticletitle{TrustyTweet: An Indicator-based Browser-Plugin to
  Assist Users in Dealing with Fake News on Twitter}. In
  \bibinfo{booktitle}{\emph{Proceedings of the International Conference on
  Wirtschaftsinformatik (WI)}}.
\newblock


\bibitem[\protect\citeauthoryear{Hasani-Mavriqi, Kowald, Helic, and
  Lex}{Hasani-Mavriqi et~al\mbox{.}}{2018}]%
        {hasani2018consensus}
\bibfield{author}{\bibinfo{person}{Ilire Hasani-Mavriqi},
  \bibinfo{person}{Dominik Kowald}, \bibinfo{person}{Denis Helic}, {and}
  \bibinfo{person}{Elisabeth Lex}.} \bibinfo{year}{2018}\natexlab{}.
\newblock \showarticletitle{Consensus dynamics in online collaboration
  systems}.
\newblock \bibinfo{journal}{\emph{Computational social networks}}
  \bibinfo{volume}{5}, \bibinfo{number}{1} (\bibinfo{year}{2018}),
  \bibinfo{pages}{2}.
\newblock


\bibitem[\protect\citeauthoryear{Holton and Lewis}{Holton and Lewis}{2011}]%
        {holton2011journalists}
\bibfield{author}{\bibinfo{person}{Avery~E Holton} {and}
  \bibinfo{person}{Seth~C Lewis}.} \bibinfo{year}{2011}\natexlab{}.
\newblock \showarticletitle{Journalists, social media, and the use of humor on
  Twitter}.
\newblock \bibinfo{journal}{\emph{Electronic Journal of Communication}}
  \bibinfo{volume}{21}, \bibinfo{number}{1/2} (\bibinfo{year}{2011}),
  \bibinfo{pages}{1--22}.
\newblock


\bibitem[\protect\citeauthoryear{Hong, Dan, and Davison}{Hong
  et~al\mbox{.}}{2011}]%
        {hong2011predicting}
\bibfield{author}{\bibinfo{person}{Liangjie Hong}, \bibinfo{person}{Ovidiu
  Dan}, {and} \bibinfo{person}{Brian~D Davison}.}
  \bibinfo{year}{2011}\natexlab{}.
\newblock \showarticletitle{Predicting popular messages in twitter}. In
  \bibinfo{booktitle}{\emph{Proceedings of the 20th international conference
  companion on World wide web}}. \bibinfo{pages}{57--58}.
\newblock


\bibitem[\protect\citeauthoryear{Hu, Tang, Zhang, and Liu}{Hu
  et~al\mbox{.}}{2013}]%
        {hu2013social}
\bibfield{author}{\bibinfo{person}{Xia Hu}, \bibinfo{person}{Jiliang Tang},
  \bibinfo{person}{Yanchao Zhang}, {and} \bibinfo{person}{Huan Liu}.}
  \bibinfo{year}{2013}\natexlab{}.
\newblock \showarticletitle{Social spammer detection in microblogging}. In
  \bibinfo{booktitle}{\emph{Twenty-Third International Joint Conference on
  Artificial Intelligence}}.
\newblock


\bibitem[\protect\citeauthoryear{IONOS}{IONOS}{2018}]%
        {Digital}
\bibfield{author}{\bibinfo{person}{Digital~Guide IONOS}.}
  \bibinfo{year}{2018}\natexlab{}.
\newblock \bibinfo{title}{{Social bots -- the technology behind fake news}}.
\newblock
  \bibinfo{howpublished}{\url{https://www.ionos.com/digitalguide/online-marketing/social-media/social-bots/}}.
\newblock


\bibitem[\protect\citeauthoryear{Jindal and Liu}{Jindal and Liu}{2007}]%
        {jindal2007review}
\bibfield{author}{\bibinfo{person}{Nitin Jindal} {and} \bibinfo{person}{Bing
  Liu}.} \bibinfo{year}{2007}\natexlab{}.
\newblock \showarticletitle{Review spam detection}. In
  \bibinfo{booktitle}{\emph{Proceedings of the 16th international conference on
  World Wide Web}}. ACM, \bibinfo{pages}{1189--1190}.
\newblock


\bibitem[\protect\citeauthoryear{Jindal, Sood, Singh, Vatsa, and
  Chakraborty}{Jindal et~al\mbox{.}}{2019}]%
        {jindalnewsbag}
\bibfield{author}{\bibinfo{person}{Sarthak Jindal}, \bibinfo{person}{Raghav
  Sood}, \bibinfo{person}{Richa Singh}, \bibinfo{person}{Mayank Vatsa}, {and}
  \bibinfo{person}{Tanmoy Chakraborty}.} \bibinfo{year}{2019}\natexlab{}.
\newblock \showarticletitle{NewsBag: A Multimodal Benchmark Dataset for Fake
  News Detection}.
\newblock  (\bibinfo{year}{2019}).
\newblock


\bibitem[\protect\citeauthoryear{Kaplan}{Kaplan}{2015}]%
        {kaplan2015social}
\bibfield{author}{\bibinfo{person}{Andreas~M Kaplan}.}
  \bibinfo{year}{2015}\natexlab{}.
\newblock \showarticletitle{Social media, the digital revolution, and the
  business of media}.
\newblock \bibinfo{journal}{\emph{International Journal on Media Management}}
  \bibinfo{volume}{17}, \bibinfo{number}{4} (\bibinfo{year}{2015}),
  \bibinfo{pages}{197--199}.
\newblock


\bibitem[\protect\citeauthoryear{Kaur, Singh, and Kumar}{Kaur
  et~al\mbox{.}}{2018}]%
        {kaur2018rise}
\bibfield{author}{\bibinfo{person}{Ravneet Kaur}, \bibinfo{person}{Sarbjeet
  Singh}, {and} \bibinfo{person}{Harish Kumar}.}
  \bibinfo{year}{2018}\natexlab{}.
\newblock \showarticletitle{Rise of spam and compromised accounts in online
  social networks: A state-of-the-art review of different combating
  approaches}.
\newblock \bibinfo{journal}{\emph{Journal of Network and Computer
  Applications}}  \bibinfo{volume}{112} (\bibinfo{year}{2018}),
  \bibinfo{pages}{53--88}.
\newblock


\bibitem[\protect\citeauthoryear{Khan}{Khan}{2021}]%
        {tanveer_khan_2021_4428240}
\bibfield{author}{\bibinfo{person}{Tanveer Khan}.}
  \bibinfo{year}{2021}\natexlab{}.
\newblock \bibinfo{title}{{Trust and Believe -- Should We? Evaluating the
  Trustworthiness of Twitter Users}}.
\newblock
\newblock
\urldef\tempurl%
\url{https://doi.org/10.5281/zenodo.4428240}
\showDOI{\tempurl}


\bibitem[\protect\citeauthoryear{Khan and Michalas}{Khan and Michalas}{2020}]%
        {Michalas:20:TrustCom:FakeNews}
\bibfield{author}{\bibinfo{person}{T. Khan} {and} \bibinfo{person}{A.
  Michalas}.} \bibinfo{year}{2020}\natexlab{}.
\newblock \showarticletitle{Trust and Believe - Should We? Evaluating the
  Trustworthiness of Twitter Users}. In \bibinfo{booktitle}{\emph{2020 IEEE
  19th International Conference on Trust, Security and Privacy in Computing and
  Communications (TrustCom)}}. \bibinfo{publisher}{IEEE Computer Society},
  \bibinfo{address}{Los Alamitos, CA, USA}, \bibinfo{pages}{1791--1800}.
\newblock
\urldef\tempurl%
\url{https://doi.org/10.1109/TrustCom50675.2020.00246}
\showDOI{\tempurl}


\bibitem[\protect\citeauthoryear{Kharratzadeh and Coates}{Kharratzadeh and
  Coates}{2012}]%
        {kharratzadeh2012weblog}
\bibfield{author}{\bibinfo{person}{Milad Kharratzadeh} {and}
  \bibinfo{person}{Mark Coates}.} \bibinfo{year}{2012}\natexlab{}.
\newblock \showarticletitle{Weblog analysis for predicting correlations in
  stock price evolutions}. In \bibinfo{booktitle}{\emph{Sixth International
  AAAI Conference on Weblogs and Social Media}}.
\newblock


\bibitem[\protect\citeauthoryear{Klyuev}{Klyuev}{2018}]%
        {klyuev2018fake}
\bibfield{author}{\bibinfo{person}{Vitaly Klyuev}.}
  \bibinfo{year}{2018}\natexlab{}.
\newblock \showarticletitle{Fake news filtering: Semantic approaches}. In
  \bibinfo{booktitle}{\emph{2018 7th International Conference on Reliability,
  Infocom Technologies and Optimization (Trends and Future
  Directions)(ICRITO)}}. IEEE, \bibinfo{pages}{9--15}.
\newblock


\bibitem[\protect\citeauthoryear{Kong, Shi, Yu, Liu, and Xia}{Kong
  et~al\mbox{.}}{2019}]%
        {kong2019academic}
\bibfield{author}{\bibinfo{person}{Xiangjie Kong}, \bibinfo{person}{Yajie Shi},
  \bibinfo{person}{Shuo Yu}, \bibinfo{person}{Jiaying Liu}, {and}
  \bibinfo{person}{Feng Xia}.} \bibinfo{year}{2019}\natexlab{}.
\newblock \showarticletitle{Academic social networks: Modeling, analysis,
  mining and applications}.
\newblock \bibinfo{journal}{\emph{Journal of Network and Computer
  Applications}}  \bibinfo{volume}{132} (\bibinfo{year}{2019}),
  \bibinfo{pages}{86--103}.
\newblock


\bibitem[\protect\citeauthoryear{Kshetri and Voas}{Kshetri and Voas}{2017}]%
        {kshetri2017economics}
\bibfield{author}{\bibinfo{person}{Nir Kshetri} {and} \bibinfo{person}{Jeffrey
  Voas}.} \bibinfo{year}{2017}\natexlab{}.
\newblock \showarticletitle{The economics of ``fake news''}.
\newblock \bibinfo{journal}{\emph{IT Professional}} \bibinfo{volume}{19},
  \bibinfo{number}{6} (\bibinfo{year}{2017}), \bibinfo{pages}{8--12}.
\newblock


\bibitem[\protect\citeauthoryear{Kucharski}{Kucharski}{2016}]%
        {kucharski2016study}
\bibfield{author}{\bibinfo{person}{Adam Kucharski}.}
  \bibinfo{year}{2016}\natexlab{}.
\newblock \showarticletitle{Study epidemiology of fake news}.
\newblock \bibinfo{journal}{\emph{Nature}} \bibinfo{volume}{540},
  \bibinfo{number}{7634} (\bibinfo{year}{2016}), \bibinfo{pages}{525--525}.
\newblock


\bibitem[\protect\citeauthoryear{Kumar, West, and Leskovec}{Kumar
  et~al\mbox{.}}{2016}]%
        {kumar2016disinformation}
\bibfield{author}{\bibinfo{person}{Srijan Kumar}, \bibinfo{person}{Robert
  West}, {and} \bibinfo{person}{Jure Leskovec}.}
  \bibinfo{year}{2016}\natexlab{}.
\newblock \showarticletitle{Disinformation on the web: Impact, characteristics,
  and detection of wikipedia hoaxes}. In \bibinfo{booktitle}{\emph{Proceedings
  of the 25th international conference on World Wide Web}}. International World
  Wide Web Conferences Steering Committee, \bibinfo{pages}{591--602}.
\newblock


\bibitem[\protect\citeauthoryear{Lazer, Baum, Benkler, Berinsky, Greenhill,
  Menczer, Metzger, Nyhan, Pennycook, Rothschild, et~al\mbox{.}}{Lazer
  et~al\mbox{.}}{2018}]%
        {lazer2018science}
\bibfield{author}{\bibinfo{person}{David~MJ Lazer}, \bibinfo{person}{Matthew~A
  Baum}, \bibinfo{person}{Yochai Benkler}, \bibinfo{person}{Adam~J Berinsky},
  \bibinfo{person}{Kelly~M Greenhill}, \bibinfo{person}{Filippo Menczer},
  \bibinfo{person}{Miriam~J Metzger}, \bibinfo{person}{Brendan Nyhan},
  \bibinfo{person}{Gordon Pennycook}, \bibinfo{person}{David Rothschild},
  {et~al\mbox{.}}} \bibinfo{year}{2018}\natexlab{}.
\newblock \showarticletitle{The science of fake news}.
\newblock \bibinfo{journal}{\emph{Science}} \bibinfo{volume}{359},
  \bibinfo{number}{6380} (\bibinfo{year}{2018}), \bibinfo{pages}{1094--1096}.
\newblock


\bibitem[\protect\citeauthoryear{Lee, Caverlee, and Webb}{Lee
  et~al\mbox{.}}{2010}]%
        {lee2010uncovering}
\bibfield{author}{\bibinfo{person}{Kyumin Lee}, \bibinfo{person}{James
  Caverlee}, {and} \bibinfo{person}{Steve Webb}.}
  \bibinfo{year}{2010}\natexlab{}.
\newblock \showarticletitle{Uncovering social spammers: social honeypots+
  machine learning}. In \bibinfo{booktitle}{\emph{Proceedings of the 33rd
  international ACM SIGIR conference on Research and development in information
  retrieval}}. ACM, \bibinfo{pages}{435--442}.
\newblock


\bibitem[\protect\citeauthoryear{Lee, Eoff, and Caverlee}{Lee
  et~al\mbox{.}}{2011}]%
        {lee2011seven}
\bibfield{author}{\bibinfo{person}{Kyumin Lee}, \bibinfo{person}{Brian~David
  Eoff}, {and} \bibinfo{person}{James Caverlee}.}
  \bibinfo{year}{2011}\natexlab{}.
\newblock \showarticletitle{Seven months with the devils: A long-term study of
  content polluters on twitter}. In \bibinfo{booktitle}{\emph{Fifth
  International AAAI Conference on Weblogs and Social Media}}.
\newblock


\bibitem[\protect\citeauthoryear{Lee and Kim}{Lee and Kim}{2013}]%
        {lee2013warningbird}
\bibfield{author}{\bibinfo{person}{Sangho Lee} {and} \bibinfo{person}{Jong
  Kim}.} \bibinfo{year}{2013}\natexlab{}.
\newblock \showarticletitle{Warningbird: A near real-time detection system for
  suspicious urls in twitter stream}.
\newblock \bibinfo{journal}{\emph{IEEE transactions on dependable and secure
  computing}} \bibinfo{volume}{10}, \bibinfo{number}{3} (\bibinfo{year}{2013}),
  \bibinfo{pages}{183--195}.
\newblock


\bibitem[\protect\citeauthoryear{Leskovec, Backstrom, and Kleinberg}{Leskovec
  et~al\mbox{.}}{2009a}]%
        {leskovec2009meme}
\bibfield{author}{\bibinfo{person}{Jure Leskovec}, \bibinfo{person}{Lars
  Backstrom}, {and} \bibinfo{person}{Jon Kleinberg}.}
  \bibinfo{year}{2009}\natexlab{a}.
\newblock \showarticletitle{Meme-tracking and the dynamics of the news cycle}.
  In \bibinfo{booktitle}{\emph{Proceedings of the 15th ACM SIGKDD international
  conference on Knowledge discovery and data mining}}. ACM,
  \bibinfo{pages}{497--506}.
\newblock


\bibitem[\protect\citeauthoryear{Leskovec, Backstrom, and Kleinberg}{Leskovec
  et~al\mbox{.}}{2009b}]%
        {JureLeskovec1}
\bibfield{author}{\bibinfo{person}{Jure Leskovec}, \bibinfo{person}{Lars
  Backstrom}, {and} \bibinfo{person}{Jon Kleinberg}.}
  \bibinfo{year}{2009}\natexlab{b}.
\newblock \showarticletitle{Meme-tracking and the dynamics of the news cycle}.
  In \bibinfo{booktitle}{\emph{Proceedings of the 15th ACM SIGKDD international
  conference on Knowledge discovery and data mining}}.
  \bibinfo{pages}{497--506}.
\newblock


\bibitem[\protect\citeauthoryear{Leskovec and Mcauley}{Leskovec and
  Mcauley}{2012}]%
        {leskovec2012learning}
\bibfield{author}{\bibinfo{person}{Jure Leskovec} {and}
  \bibinfo{person}{Julian~J Mcauley}.} \bibinfo{year}{2012}\natexlab{}.
\newblock \showarticletitle{Learning to discover social circles in ego
  networks}. In \bibinfo{booktitle}{\emph{Advances in neural information
  processing systems}}. \bibinfo{pages}{539--547}.
\newblock


\bibitem[\protect\citeauthoryear{Li, Chen, Mukherjee, Liu, and Shao}{Li
  et~al\mbox{.}}{2015}]%
        {li2015analyzing}
\bibfield{author}{\bibinfo{person}{Huayi Li}, \bibinfo{person}{Zhiyuan Chen},
  \bibinfo{person}{Arjun Mukherjee}, \bibinfo{person}{Bing Liu}, {and}
  \bibinfo{person}{Jidong Shao}.} \bibinfo{year}{2015}\natexlab{}.
\newblock \showarticletitle{Analyzing and detecting opinion spam on a
  large-scale dataset via temporal and spatial patterns}. In
  \bibinfo{booktitle}{\emph{ninth international AAAI conference on web and
  social Media}}.
\newblock


\bibitem[\protect\citeauthoryear{Li, Hu, Wu, and Liu}{Li et~al\mbox{.}}{2016}]%
        {li2016robust}
\bibfield{author}{\bibinfo{person}{Jundong Li}, \bibinfo{person}{Xia Hu},
  \bibinfo{person}{Liang Wu}, {and} \bibinfo{person}{Huan Liu}.}
  \bibinfo{year}{2016}\natexlab{}.
\newblock \showarticletitle{Robust unsupervised feature selection on networked
  data}. In \bibinfo{booktitle}{\emph{Proceedings of the 2016 SIAM
  International Conference on Data Mining}}. SIAM, \bibinfo{pages}{387--395}.
\newblock


\bibitem[\protect\citeauthoryear{Lim, Nguyen, Jindal, Liu, and Lauw}{Lim
  et~al\mbox{.}}{2010}]%
        {lim2010detecting}
\bibfield{author}{\bibinfo{person}{Ee-Peng Lim}, \bibinfo{person}{Viet-An
  Nguyen}, \bibinfo{person}{Nitin Jindal}, \bibinfo{person}{Bing Liu}, {and}
  \bibinfo{person}{Hady~Wirawan Lauw}.} \bibinfo{year}{2010}\natexlab{}.
\newblock \showarticletitle{Detecting product review spammers using rating
  behaviors}. In \bibinfo{booktitle}{\emph{Proceedings of the 19th ACM
  international conference on Information and knowledge management}}. ACM,
  \bibinfo{pages}{939--948}.
\newblock


\bibitem[\protect\citeauthoryear{Litou, Kalogeraki, Katakis, and
  Gunopulos}{Litou et~al\mbox{.}}{2016}]%
        {litou2016real}
\bibfield{author}{\bibinfo{person}{Iouliana Litou}, \bibinfo{person}{Vana
  Kalogeraki}, \bibinfo{person}{Ioannis Katakis}, {and}
  \bibinfo{person}{Dimitrios Gunopulos}.} \bibinfo{year}{2016}\natexlab{}.
\newblock \showarticletitle{Real-time and cost-effective limitation of
  misinformation propagation}. In \bibinfo{booktitle}{\emph{2016 17th IEEE
  International Conference on Mobile Data Management (MDM)}},
  Vol.~\bibinfo{volume}{1}. IEEE, \bibinfo{pages}{158--163}.
\newblock


\bibitem[\protect\citeauthoryear{Liu, Liu, and Ren}{Liu et~al\mbox{.}}{2019}]%
        {liu2019trust}
\bibfield{author}{\bibinfo{person}{Shuaipeng Liu}, \bibinfo{person}{Shuo Liu},
  {and} \bibinfo{person}{Lei Ren}.} \bibinfo{year}{2019}\natexlab{}.
\newblock \showarticletitle{Trust or Suspect? An Empirical Ensemble Framework
  for Fake News Classification}. In \bibinfo{booktitle}{\emph{Proceedings of
  the 12th ACM International Conference on Web Search and Data Mining,
  Melbourne, Australia}}. \bibinfo{pages}{11--15}.
\newblock


\bibitem[\protect\citeauthoryear{Liu and Wu}{Liu and Wu}{2018}]%
        {liu2018early}
\bibfield{author}{\bibinfo{person}{Yang Liu} {and}
  \bibinfo{person}{Yi-Fang~Brook Wu}.} \bibinfo{year}{2018}\natexlab{}.
\newblock \showarticletitle{Early detection of fake news on social media
  through propagation path classification with recurrent and convolutional
  networks}. In \bibinfo{booktitle}{\emph{Thirty-Second AAAI Conference on
  Artificial Intelligence}}.
\newblock


\bibitem[\protect\citeauthoryear{Liu and Wu}{Liu and Wu}{2020}]%
        {liu2020fned}
\bibfield{author}{\bibinfo{person}{Yang Liu} {and}
  \bibinfo{person}{Yi-Fang~Brook Wu}.} \bibinfo{year}{2020}\natexlab{}.
\newblock \showarticletitle{FNED: A Deep Network for Fake News Early Detection
  on Social Media}.
\newblock \bibinfo{journal}{\emph{ACM Transactions on Information Systems
  (TOIS)}} \bibinfo{volume}{38}, \bibinfo{number}{3} (\bibinfo{year}{2020}),
  \bibinfo{pages}{1--33}.
\newblock


\bibitem[\protect\citeauthoryear{Ma, Gao, Mitra, Kwon, Jansen, Wong, and
  Cha}{Ma et~al\mbox{.}}{2016}]%
        {ma2016detecting}
\bibfield{author}{\bibinfo{person}{Jing Ma}, \bibinfo{person}{Wei Gao},
  \bibinfo{person}{Prasenjit Mitra}, \bibinfo{person}{Sejeong Kwon},
  \bibinfo{person}{Bernard~J Jansen}, \bibinfo{person}{Kam-Fai Wong}, {and}
  \bibinfo{person}{Meeyoung Cha}.} \bibinfo{year}{2016}\natexlab{}.
\newblock \showarticletitle{Detecting rumors from microblogs with recurrent
  neural networks}.
\newblock  (\bibinfo{year}{2016}).
\newblock


\bibitem[\protect\citeauthoryear{Ma, Gao, and Wong}{Ma et~al\mbox{.}}{2017}]%
        {ma2017detect}
\bibfield{author}{\bibinfo{person}{Jing Ma}, \bibinfo{person}{Wei Gao}, {and}
  \bibinfo{person}{Kam-Fai Wong}.} \bibinfo{year}{2017}\natexlab{}.
\newblock \showarticletitle{Detect rumors in microblog posts using propagation
  structure via kernel learning}. Association for Computational Linguistics.
\newblock


\bibitem[\protect\citeauthoryear{Maigrot, Claveau, Kijak, and Sicre}{Maigrot
  et~al\mbox{.}}{2016}]%
        {maigrot2016mediaeval}
\bibfield{author}{\bibinfo{person}{C{\'e}dric Maigrot},
  \bibinfo{person}{Vincent Claveau}, \bibinfo{person}{Ewa Kijak}, {and}
  \bibinfo{person}{Ronan Sicre}.} \bibinfo{year}{2016}\natexlab{}.
\newblock \showarticletitle{Mediaeval 2016: A multimodal system for the
  verifying multimedia use task}.
\newblock


\bibitem[\protect\citeauthoryear{Mao, Wei, Wang, and Liu}{Mao
  et~al\mbox{.}}{2012}]%
        {mao2012correlating}
\bibfield{author}{\bibinfo{person}{Yuexin Mao}, \bibinfo{person}{Wei Wei},
  \bibinfo{person}{Bing Wang}, {and} \bibinfo{person}{Benyuan Liu}.}
  \bibinfo{year}{2012}\natexlab{}.
\newblock \showarticletitle{Correlating S\&P 500 stocks with Twitter data}. In
  \bibinfo{booktitle}{\emph{Proceedings of the first ACM international workshop
  on hot topics on interdisciplinary social networks research}}.
  \bibinfo{pages}{69--72}.
\newblock


\bibitem[\protect\citeauthoryear{Matsa and Shearer}{Matsa and Shearer}{2018}]%
        {ElisaShearer}
\bibfield{author}{\bibinfo{person}{Katerina~Eva Matsa} {and}
  \bibinfo{person}{Elisa Shearer}.} \bibinfo{year}{2018}\natexlab{}.
\newblock \showarticletitle{News use across social media platforms 2018}.
\newblock \bibinfo{journal}{\emph{Pew Research Center}}  \bibinfo{volume}{10}
  (\bibinfo{year}{2018}).
\newblock


\bibitem[\protect\citeauthoryear{Michalas and Komninos}{Michalas and
  Komninos}{2014}]%
        {Michalas:14:Lord}
\bibfield{author}{\bibinfo{person}{Antonis Michalas} {and}
  \bibinfo{person}{Nikos Komninos}.} \bibinfo{year}{2014}\natexlab{}.
\newblock \showarticletitle{The lord of the sense: A privacy preserving
  reputation system for participatory sensing applications}. In
  \bibinfo{booktitle}{\emph{Computers and Communication (ISCC), 2014 IEEE
  Symposium}}. IEEE, \bibinfo{pages}{1--6}.
\newblock


\bibitem[\protect\citeauthoryear{Michalas and Murray}{Michalas and
  Murray}{2017}]%
        {Michalas:17:Pies:CCS}
\bibfield{author}{\bibinfo{person}{Antonis Michalas} {and}
  \bibinfo{person}{Ryan Murray}.} \bibinfo{year}{2017}\natexlab{}.
\newblock \showarticletitle{Keep Pies Away from Kids: A Raspberry Pi Attacking
  Tool}. In \bibinfo{booktitle}{\emph{Proceedings of the 2017 Workshop on
  Internet of Things Security and Privacy}}
  \emph{(\bibinfo{series}{IoTS\&\#38;P '17})}. \bibinfo{publisher}{ACM},
  \bibinfo{address}{New York, NY, USA}, \bibinfo{pages}{61--62}.
\newblock
\showISBNx{978-1-4503-5396-0}
\urldef\tempurl%
\url{https://doi.org/10.1145/3139937.3139953}
\showDOI{\tempurl}


\bibitem[\protect\citeauthoryear{Mitra}{Mitra}{2016}]%
        {MitraTanushree}
\bibfield{author}{\bibinfo{person}{Eric~Gilbert Mitra, Tanushree}.}
  \bibinfo{year}{2016}\natexlab{}.
\newblock \bibinfo{title}{{CREDBANK-data}}.
\newblock
  \bibinfo{howpublished}{\url{https://github.com/compsocial/CREDBANK-data}}.
\newblock


\bibitem[\protect\citeauthoryear{Mitra and Gilbert}{Mitra and Gilbert}{2015}]%
        {mitra2015credbank}
\bibfield{author}{\bibinfo{person}{Tanushree Mitra} {and} \bibinfo{person}{Eric
  Gilbert}.} \bibinfo{year}{2015}\natexlab{}.
\newblock \showarticletitle{Credbank: A large-scale social media corpus with
  associated credibility annotations}. In \bibinfo{booktitle}{\emph{Ninth
  International AAAI Conference on Web and Social Media}}.
\newblock


\bibitem[\protect\citeauthoryear{Morris, Counts, Roseway, Hoff, and
  Schwarz}{Morris et~al\mbox{.}}{2012}]%
        {morris2012tweeting}
\bibfield{author}{\bibinfo{person}{Meredith~Ringel Morris},
  \bibinfo{person}{Scott Counts}, \bibinfo{person}{Asta Roseway},
  \bibinfo{person}{Aaron Hoff}, {and} \bibinfo{person}{Julia Schwarz}.}
  \bibinfo{year}{2012}\natexlab{}.
\newblock \showarticletitle{Tweeting is believing?: understanding microblog
  credibility perceptions}. In \bibinfo{booktitle}{\emph{Proceedings of the ACM
  2012 conference on computer supported cooperative work}}. ACM,
  \bibinfo{pages}{441--450}.
\newblock


\bibitem[\protect\citeauthoryear{Nasim, Nguyen, Lothian, Cope, and
  Mitchell}{Nasim et~al\mbox{.}}{2018}]%
        {nasim2018real}
\bibfield{author}{\bibinfo{person}{Mehwish Nasim}, \bibinfo{person}{Andrew
  Nguyen}, \bibinfo{person}{Nick Lothian}, \bibinfo{person}{Robert Cope}, {and}
  \bibinfo{person}{Lewis Mitchell}.} \bibinfo{year}{2018}\natexlab{}.
\newblock \showarticletitle{Real-time detection of content polluters in
  partially observable Twitter networks}. In
  \bibinfo{booktitle}{\emph{Companion Proceedings of the The Web Conference
  2018}}. \bibinfo{pages}{1331--1339}.
\newblock


\bibitem[\protect\citeauthoryear{Neander and Marlin}{Neander and
  Marlin}{2010}]%
        {neander2010media}
\bibfield{author}{\bibinfo{person}{Joachim Neander} {and}
  \bibinfo{person}{Randal Marlin}.} \bibinfo{year}{2010}\natexlab{}.
\newblock \showarticletitle{Media and Propaganda: The Northcliffe Press and the
  Corpse Factory Story of World War I.}
\newblock \bibinfo{journal}{\emph{Global Media Journal: Canadian Edition}}
  \bibinfo{volume}{3}, \bibinfo{number}{2} (\bibinfo{year}{2010}).
\newblock


\bibitem[\protect\citeauthoryear{News}{News}{2016}]%
        {jsvine}
\bibfield{author}{\bibinfo{person}{BuzzFeed News}.}
  \bibinfo{year}{2016}\natexlab{}.
\newblock \bibinfo{title}{{Fact-Checking Facebook Politics Pages}}.
\newblock
  \bibinfo{howpublished}{\url{https://github.com/BuzzFeedNews/2016-10-facebook-fact-check}}.
\newblock


\bibitem[\protect\citeauthoryear{Northman}{Northman}{2019}]%
        {Tora}
\bibfield{author}{\bibinfo{person}{Tora Northman}.}
  \bibinfo{year}{2019}\natexlab{}.
\newblock \bibinfo{title}{{Instagram Is Removing "Fake News" From the
  Platform}}.
\newblock
  \bibinfo{howpublished}{\url{https://hypebae.com/2019/8/instagram-fake-news-removing-tool-flagging-misinformation}}.
\newblock


\bibitem[\protect\citeauthoryear{O'Brien, Latessa, Evangelopoulos, and
  Boix}{O'Brien et~al\mbox{.}}{2018}]%
        {o2018language}
\bibfield{author}{\bibinfo{person}{Nicole O'Brien}, \bibinfo{person}{Sophia
  Latessa}, \bibinfo{person}{Georgios Evangelopoulos}, {and}
  \bibinfo{person}{Xavier Boix}.} \bibinfo{year}{2018}\natexlab{}.
\newblock \showarticletitle{The language of fake news: Opening the black-box of
  deep learning based detectors}.
\newblock  (\bibinfo{year}{2018}).
\newblock


\bibitem[\protect\citeauthoryear{of~Eastern~Finland}{of~Eastern~Finland}{2019}]%
        {UEFL}
\bibfield{author}{\bibinfo{person}{University of Eastern~Finland}.}
  \bibinfo{year}{2019}\natexlab{}.
\newblock \bibinfo{title}{{New application can detect Twitter bots in any
  language}}.
\newblock
  \bibinfo{howpublished}{\url{https://phys.org/news/2019-06-application-twitter-bots-language.html}}.
\newblock


\bibitem[\protect\citeauthoryear{Omidvar, Jiang, and An}{Omidvar
  et~al\mbox{.}}{2018}]%
        {omidvar2018using}
\bibfield{author}{\bibinfo{person}{Amin Omidvar}, \bibinfo{person}{Hui Jiang},
  {and} \bibinfo{person}{Aijun An}.} \bibinfo{year}{2018}\natexlab{}.
\newblock \showarticletitle{Using Neural Network for Identifying Clickbaits in
  Online News Media}. In \bibinfo{booktitle}{\emph{Annual International
  Symposium on Information Management and Big Data}}. Springer,
  \bibinfo{pages}{220--232}.
\newblock


\bibitem[\protect\citeauthoryear{Oshikawa, Qian, and Wang}{Oshikawa
  et~al\mbox{.}}{2018}]%
        {oshikawa2018survey}
\bibfield{author}{\bibinfo{person}{Ray Oshikawa}, \bibinfo{person}{Jing Qian},
  {and} \bibinfo{person}{William~Yang Wang}.} \bibinfo{year}{2018}\natexlab{}.
\newblock \showarticletitle{A survey on natural language processing for fake
  news detection}.
\newblock \bibinfo{journal}{\emph{arXiv preprint arXiv:1811.00770}}
  (\bibinfo{year}{2018}).
\newblock


\bibitem[\protect\citeauthoryear{Pan, Pavlova, Li, Li, Li, and Liu}{Pan
  et~al\mbox{.}}{2018}]%
        {pan2018content}
\bibfield{author}{\bibinfo{person}{Jeff~Z Pan}, \bibinfo{person}{Siyana
  Pavlova}, \bibinfo{person}{Chenxi Li}, \bibinfo{person}{Ningxi Li},
  \bibinfo{person}{Yangmei Li}, {and} \bibinfo{person}{Jinshuo Liu}.}
  \bibinfo{year}{2018}\natexlab{}.
\newblock \showarticletitle{Content based fake news detection using knowledge
  graphs}. In \bibinfo{booktitle}{\emph{International Semantic Web
  Conference}}. Springer, \bibinfo{pages}{669--683}.
\newblock


\bibitem[\protect\citeauthoryear{Paschen}{Paschen}{2019}]%
        {paschen2019investigating}
\bibfield{author}{\bibinfo{person}{Jeannette Paschen}.}
  \bibinfo{year}{2019}\natexlab{}.
\newblock \showarticletitle{Investigating the emotional appeal of fake news
  using artificial intelligence and human contributions}.
\newblock \bibinfo{journal}{\emph{Journal of Product \& Brand Management}}
  (\bibinfo{year}{2019}).
\newblock


\bibitem[\protect\citeauthoryear{Perez}{Perez}{2018}]%
        {Sarah}
\bibfield{author}{\bibinfo{person}{Sarah Perez}.}
  \bibinfo{year}{2018}\natexlab{}.
\newblock \bibinfo{title}{{Twitter's spam reporting tool now lets you specify
  type, including if it's a fake account}}.
\newblock
  \bibinfo{howpublished}{\url{https://techcrunch.com/2018/10/31/twitters-spam-reporting-tool-now-lets-you-specify-type-including-if-its-a-fake-account/}}.
\newblock


\bibitem[\protect\citeauthoryear{P{\'e}rez-Rosas, Kleinberg, Lefevre, and
  Mihalcea}{P{\'e}rez-Rosas et~al\mbox{.}}{2017}]%
        {perez2017automatic}
\bibfield{author}{\bibinfo{person}{Ver{\'o}nica P{\'e}rez-Rosas},
  \bibinfo{person}{Bennett Kleinberg}, \bibinfo{person}{Alexandra Lefevre},
  {and} \bibinfo{person}{Rada Mihalcea}.} \bibinfo{year}{2017}\natexlab{}.
\newblock \showarticletitle{Automatic detection of fake news}.
\newblock \bibinfo{journal}{\emph{arXiv preprint arXiv:1708.07104}}
  (\bibinfo{year}{2017}).
\newblock


\bibitem[\protect\citeauthoryear{Perozzi, Al-Rfou, and Skiena}{Perozzi
  et~al\mbox{.}}{2014}]%
        {perozzi2014deepwalk}
\bibfield{author}{\bibinfo{person}{Bryan Perozzi}, \bibinfo{person}{Rami
  Al-Rfou}, {and} \bibinfo{person}{Steven Skiena}.}
  \bibinfo{year}{2014}\natexlab{}.
\newblock \showarticletitle{Deepwalk: Online learning of social
  representations}. In \bibinfo{booktitle}{\emph{Proceedings of the 20th ACM
  SIGKDD international conference on Knowledge discovery and data mining}}.
  \bibinfo{pages}{701--710}.
\newblock


\bibitem[\protect\citeauthoryear{Pham}{Pham}{2019}]%
        {pham2019transferring}
\bibfield{author}{\bibinfo{person}{Lam Pham}.} \bibinfo{year}{2019}\natexlab{}.
\newblock \showarticletitle{Transferring, Transforming, Ensembling: The Novel
  Formula of Identifying Fake News}.
\newblock  (\bibinfo{year}{2019}).
\newblock


\bibitem[\protect\citeauthoryear{Posetti and Matthews}{Posetti and
  Matthews}{2018}]%
        {posetti2018short}
\bibfield{author}{\bibinfo{person}{Julie Posetti} {and} \bibinfo{person}{Alice
  Matthews}.} \bibinfo{year}{2018}\natexlab{}.
\newblock \showarticletitle{A short guide to the history of 'fake news' and
  disinformation}.
\newblock \bibinfo{journal}{\emph{International Center for Journalists}}
  \bibinfo{volume}{7} (\bibinfo{year}{2018}), \bibinfo{pages}{2018--07}.
\newblock


\bibitem[\protect\citeauthoryear{Potthast, Gollub, Wiegmann, Stein, Hagen,
  Komlossy, Schuster, and Fernandez}{Potthast et~al\mbox{.}}{2018}]%
        {PotthastMartin}
\bibfield{author}{\bibinfo{person}{Martin Potthast}, \bibinfo{person}{Tim
  Gollub}, \bibinfo{person}{Matti Wiegmann}, \bibinfo{person}{Benno Stein},
  \bibinfo{person}{Matthias Hagen}, \bibinfo{person}{Kristof Komlossy},
  \bibinfo{person}{Sebstian Schuster}, {and} \bibinfo{person}{Erika P.~Garces
  Fernandez}.} \bibinfo{year}{2018}\natexlab{}.
\newblock \bibinfo{title}{Webis Clickbait Corpus 2017 (Webis-Clickbait-17)}.
\newblock
\newblock
\urldef\tempurl%
\url{https://doi.org/10.5281/zenodo.3346491}
\showDOI{\tempurl}


\bibitem[\protect\citeauthoryear{Potthast, Kiesel, Reinartz, Bevendorff, and
  Stein}{Potthast et~al\mbox{.}}{2017}]%
        {potthast2017stylometric}
\bibfield{author}{\bibinfo{person}{Martin Potthast}, \bibinfo{person}{Johannes
  Kiesel}, \bibinfo{person}{Kevin Reinartz}, \bibinfo{person}{Janek
  Bevendorff}, {and} \bibinfo{person}{Benno Stein}.}
  \bibinfo{year}{2017}\natexlab{}.
\newblock \showarticletitle{A stylometric inquiry into hyperpartisan and fake
  news}.
\newblock \bibinfo{journal}{\emph{arXiv preprint arXiv:1702.05638}}
  (\bibinfo{year}{2017}).
\newblock


\bibitem[\protect\citeauthoryear{Rannard}{Rannard}{2020}]%
        {rannard2020australia}
\bibfield{author}{\bibinfo{person}{G Rannard}.}
  \bibinfo{year}{2020}\natexlab{}.
\newblock \bibinfo{title}{Australia fires: Misleading maps and pictures go
  viral}.
\newblock
\newblock


\bibitem[\protect\citeauthoryear{Rayana and Akoglu}{Rayana and Akoglu}{2015}]%
        {rayana2015collective}
\bibfield{author}{\bibinfo{person}{Shebuti Rayana} {and} \bibinfo{person}{Leman
  Akoglu}.} \bibinfo{year}{2015}\natexlab{}.
\newblock \showarticletitle{Collective opinion spam detection: Bridging review
  networks and metadata}. In \bibinfo{booktitle}{\emph{Proceedings of the 21th
  acm sigkdd international conference on knowledge discovery and data mining}}.
  ACM, \bibinfo{pages}{985--994}.
\newblock


\bibitem[\protect\citeauthoryear{Read}{Read}{2019}]%
        {Danielle}
\bibfield{author}{\bibinfo{person}{Danielle Read}.}
  \bibinfo{year}{2019}\natexlab{}.
\newblock \bibinfo{title}{{Social Media News: Fake News Flagging Tool, Clear
  Facebook History and More}}.
\newblock
  \bibinfo{howpublished}{\url{https://skedsocial.com/blog/fake-news-flagging-tool/}}.
\newblock


\bibitem[\protect\citeauthoryear{Riedel, Augenstein, Spithourakis, and
  Riedel}{Riedel et~al\mbox{.}}{2017}]%
        {riedel2017simple}
\bibfield{author}{\bibinfo{person}{Benjamin Riedel}, \bibinfo{person}{Isabelle
  Augenstein}, \bibinfo{person}{George Spithourakis}, {and}
  \bibinfo{person}{Sebastian Riedel}.} \bibinfo{year}{2017}\natexlab{}.
\newblock \bibinfo{title}{A simple but tough-to-beat baseline for the Fake News
  Challenge stance detection task. CoRR abs/1707.03264}.
\newblock
\newblock


\bibitem[\protect\citeauthoryear{Rieh and Danielson}{Rieh and
  Danielson}{2007}]%
        {rieh2007credibility}
\bibfield{author}{\bibinfo{person}{Soo~Young Rieh} {and}
  \bibinfo{person}{David~R Danielson}.} \bibinfo{year}{2007}\natexlab{}.
\newblock \showarticletitle{Credibility: A multidisciplinary framework}.
\newblock \bibinfo{journal}{\emph{Annual review of information science and
  technology}} \bibinfo{volume}{41}, \bibinfo{number}{1}
  (\bibinfo{year}{2007}), \bibinfo{pages}{307--364}.
\newblock


\bibitem[\protect\citeauthoryear{Risdal}{Risdal}{2017}]%
        {Kaggle}
\bibfield{author}{\bibinfo{person}{Megan Risdal}.}
  \bibinfo{year}{2017}\natexlab{}.
\newblock \bibinfo{title}{{Getting Real about Fake News}}.
\newblock
  \bibinfo{howpublished}{\url{https://www.kaggle.com/mrisdal/fake-news}}.
\newblock


\bibitem[\protect\citeauthoryear{Rubin, Brogly, Conroy, Chen, Cornwell, and
  Asubiaro}{Rubin et~al\mbox{.}}{2019}]%
        {rubin2019news}
\bibfield{author}{\bibinfo{person}{Victoria Rubin}, \bibinfo{person}{Chris
  Brogly}, \bibinfo{person}{Nadia Conroy}, \bibinfo{person}{Yimin Chen},
  \bibinfo{person}{Sarah~E Cornwell}, {and} \bibinfo{person}{Toluwase~V
  Asubiaro}.} \bibinfo{year}{2019}\natexlab{}.
\newblock \showarticletitle{A News Verification Browser for the Detection of
  Clickbait, Satire, and Falsified News}.
\newblock \bibinfo{journal}{\emph{The Journal of Open Source Software}}
  \bibinfo{volume}{4}, \bibinfo{number}{35} (\bibinfo{year}{2019}),
  \bibinfo{pages}{1}.
\newblock


\bibitem[\protect\citeauthoryear{Rubin}{Rubin}{2017}]%
        {rubin2017deception}
\bibfield{author}{\bibinfo{person}{Victoria~L Rubin}.}
  \bibinfo{year}{2017}\natexlab{}.
\newblock \showarticletitle{Deception detection and rumor debunking for social
  media}.
\newblock In \bibinfo{booktitle}{\emph{The SAGE Handbook of Social Media
  Research Methods}}. \bibinfo{publisher}{Sage}, \bibinfo{pages}{342}.
\newblock


\bibitem[\protect\citeauthoryear{Rubin, Chen, and Conroy}{Rubin
  et~al\mbox{.}}{2015}]%
        {rubin2015deception}
\bibfield{author}{\bibinfo{person}{Victoria~L Rubin}, \bibinfo{person}{Yimin
  Chen}, {and} \bibinfo{person}{Niall~J Conroy}.}
  \bibinfo{year}{2015}\natexlab{}.
\newblock \showarticletitle{Deception detection for news: three types of
  fakes}. In \bibinfo{booktitle}{\emph{Proceedings of the 78th ASIS\&T Annual
  Meeting: Information Science with Impact: Research in and for the
  Community}}. American Society for Information Science, \bibinfo{pages}{83}.
\newblock


\bibitem[\protect\citeauthoryear{Ruchansky, Seo, and Liu}{Ruchansky
  et~al\mbox{.}}{2017}]%
        {ruchansky2017csi}
\bibfield{author}{\bibinfo{person}{Natali Ruchansky}, \bibinfo{person}{Sungyong
  Seo}, {and} \bibinfo{person}{Yan Liu}.} \bibinfo{year}{2017}\natexlab{}.
\newblock \showarticletitle{Csi: A hybrid deep model for fake news detection}.
  In \bibinfo{booktitle}{\emph{Proceedings of the 2017 ACM on Conference on
  Information and Knowledge Management}}. ACM, \bibinfo{pages}{797--806}.
\newblock


\bibitem[\protect\citeauthoryear{Ruiz, Hristidis, Castillo, Gionis, and
  Jaimes}{Ruiz et~al\mbox{.}}{2012}]%
        {ruiz2012correlating}
\bibfield{author}{\bibinfo{person}{Eduardo~J Ruiz}, \bibinfo{person}{Vagelis
  Hristidis}, \bibinfo{person}{Carlos Castillo}, \bibinfo{person}{Aristides
  Gionis}, {and} \bibinfo{person}{Alejandro Jaimes}.}
  \bibinfo{year}{2012}\natexlab{}.
\newblock \showarticletitle{Correlating financial time series with
  micro-blogging activity}. In \bibinfo{booktitle}{\emph{Proceedings of the
  fifth ACM international conference on Web search and data mining}}.
  \bibinfo{pages}{513--522}.
\newblock


\bibitem[\protect\citeauthoryear{Rusu and Herman}{Rusu and Herman}{2019}]%
        {rusu2019legislative}
\bibfield{author}{\bibinfo{person}{Maria-Lucia Rusu} {and}
  \bibinfo{person}{Ramona-Elena Herman}.} \bibinfo{year}{2019}\natexlab{}.
\newblock \showarticletitle{Legislative Measures Adopted at the International
  Level Against Fake News}. In \bibinfo{booktitle}{\emph{International
  conference KNOWLEDGE-BASED ORGANIZATION}}, Vol.~\bibinfo{volume}{25}.
  Sciendo, \bibinfo{pages}{324--330}.
\newblock


\bibitem[\protect\citeauthoryear{Santia and Williams}{Santia and
  Williams}{2018}]%
        {santia2018buzzface}
\bibfield{author}{\bibinfo{person}{Giovanni~C Santia} {and}
  \bibinfo{person}{Jake~Ryland Williams}.} \bibinfo{year}{2018}\natexlab{}.
\newblock \showarticletitle{Buzzface: A news veracity dataset with facebook
  user commentary and egos}. In \bibinfo{booktitle}{\emph{Twelfth International
  AAAI Conference on Web and Social Media}}.
\newblock


\bibitem[\protect\citeauthoryear{Sardarizadeh}{Sardarizadeh}{2019}]%
        {Shayan}
\bibfield{author}{\bibinfo{person}{Shayan Sardarizadeh}.}
  \bibinfo{year}{2019}\natexlab{}.
\newblock \bibinfo{title}{{Instagram fact-check: Can a new flagging tool stop
  fake news?}}
\newblock
  \bibinfo{howpublished}{\url{https://www.bbc.com/news/blogs-trending-49449005}}.
\newblock


\bibitem[\protect\citeauthoryear{Saxena, Iyengar, and Gupta}{Saxena
  et~al\mbox{.}}{2015}]%
        {saxena2015understanding}
\bibfield{author}{\bibinfo{person}{Akrati Saxena}, \bibinfo{person}{SRS
  Iyengar}, {and} \bibinfo{person}{Yayati Gupta}.}
  \bibinfo{year}{2015}\natexlab{}.
\newblock \showarticletitle{Understanding spreading patterns on social networks
  based on network topology}. In \bibinfo{booktitle}{\emph{Proceedings of the
  2015 IEEE/ACM International Conference on Advances in Social Networks
  Analysis and Mining 2015}}. \bibinfo{pages}{1616--1617}.
\newblock


\bibitem[\protect\citeauthoryear{Schwartz}{Schwartz}{2018}]%
        {Oscar}
\bibfield{author}{\bibinfo{person}{Oscar Schwartz}.}
  \bibinfo{year}{2018}\natexlab{}.
\newblock \bibinfo{title}{{Your favorite Twitter bots are about die, thanks to
  upcoming rule changes}}.
\newblock
  \bibinfo{howpublished}{\url{https://qz.com/1422765/your-favorite-twitter-bots-are-about-die-thanks-to-upcoming-rule-changes/}}.
\newblock


\bibitem[\protect\citeauthoryear{Sean~Baird}{Sean~Baird}{2017}]%
        {Sean}
\bibfield{author}{\bibinfo{person}{Yuxi~Pan Sean~Baird, Doug~Sibley}.}
  \bibinfo{year}{2017}\natexlab{}.
\newblock \bibinfo{title}{{ Talos Targets Disinformation with Fake News
  Challenge Victory }}.
\newblock
  \bibinfo{howpublished}{\url{https://blog.talosintelligence.com/2017/06/talos-fake-news-challenge.html}}.
\newblock


\bibitem[\protect\citeauthoryear{Shao, Ciampaglia, Flammini, and Menczer}{Shao
  et~al\mbox{.}}{2016}]%
        {shao2016hoaxy}
\bibfield{author}{\bibinfo{person}{Chengcheng Shao},
  \bibinfo{person}{Giovanni~Luca Ciampaglia}, \bibinfo{person}{Alessandro
  Flammini}, {and} \bibinfo{person}{Filippo Menczer}.}
  \bibinfo{year}{2016}\natexlab{}.
\newblock \showarticletitle{Hoaxy: A platform for tracking online
  misinformation}. In \bibinfo{booktitle}{\emph{Proceedings of the 25th
  international conference companion on world wide web}}. International World
  Wide Web Conferences Steering Committee, \bibinfo{pages}{745--750}.
\newblock


\bibitem[\protect\citeauthoryear{Shiralkar, Flammini, Menczer, and
  Ciampaglia}{Shiralkar et~al\mbox{.}}{2017}]%
        {shiralkar2017finding}
\bibfield{author}{\bibinfo{person}{Prashant Shiralkar},
  \bibinfo{person}{Alessandro Flammini}, \bibinfo{person}{Filippo Menczer},
  {and} \bibinfo{person}{Giovanni~Luca Ciampaglia}.}
  \bibinfo{year}{2017}\natexlab{}.
\newblock \showarticletitle{Finding streams in knowledge graphs to support fact
  checking}. In \bibinfo{booktitle}{\emph{2017 IEEE International Conference on
  Data Mining (ICDM)}}. IEEE, \bibinfo{pages}{859--864}.
\newblock


\bibitem[\protect\citeauthoryear{Shu, Bhattacharjee, Alatawi, Nazer, Ding,
  Karami, and Liu}{Shu et~al\mbox{.}}{2020a}]%
        {shu2020combating}
\bibfield{author}{\bibinfo{person}{Kai Shu}, \bibinfo{person}{Amrita
  Bhattacharjee}, \bibinfo{person}{Faisal Alatawi}, \bibinfo{person}{Tahora~H
  Nazer}, \bibinfo{person}{Kaize Ding}, \bibinfo{person}{Mansooreh Karami},
  {and} \bibinfo{person}{Huan Liu}.} \bibinfo{year}{2020}\natexlab{a}.
\newblock \showarticletitle{Combating disinformation in a social media age}.
\newblock \bibinfo{journal}{\emph{Wiley Interdisciplinary Reviews: Data Mining
  and Knowledge Discovery}} \bibinfo{volume}{10}, \bibinfo{number}{6}
  (\bibinfo{year}{2020}), \bibinfo{pages}{e1385}.
\newblock


\bibitem[\protect\citeauthoryear{Shu, Mahudeswaran, Wang, Lee, and Liu}{Shu
  et~al\mbox{.}}{2020b}]%
        {shu2018fakenewsnet}
\bibfield{author}{\bibinfo{person}{Kai Shu}, \bibinfo{person}{Deepak
  Mahudeswaran}, \bibinfo{person}{Suhang Wang}, \bibinfo{person}{Dongwon Lee},
  {and} \bibinfo{person}{Huan Liu}.} \bibinfo{year}{2020}\natexlab{b}.
\newblock \showarticletitle{FakeNewsNet: A Data Repository with News Content,
  Social Context, and Spatiotemporal Information for Studying Fake News on
  Social Media}.
\newblock \bibinfo{journal}{\emph{Big Data}} \bibinfo{volume}{8},
  \bibinfo{number}{3} (\bibinfo{year}{2020}), \bibinfo{pages}{171--188}.
\newblock


\bibitem[\protect\citeauthoryear{Shu, Sliva, Wang, Tang, and Liu}{Shu
  et~al\mbox{.}}{2017}]%
        {shu2017fake}
\bibfield{author}{\bibinfo{person}{Kai Shu}, \bibinfo{person}{Amy Sliva},
  \bibinfo{person}{Suhang Wang}, \bibinfo{person}{Jiliang Tang}, {and}
  \bibinfo{person}{Huan Liu}.} \bibinfo{year}{2017}\natexlab{}.
\newblock \showarticletitle{Fake news detection on social media: A data mining
  perspective}.
\newblock \bibinfo{journal}{\emph{ACM SIGKDD Explorations Newsletter}}
  \bibinfo{volume}{19}, \bibinfo{number}{1} (\bibinfo{year}{2017}),
  \bibinfo{pages}{22--36}.
\newblock


\bibitem[\protect\citeauthoryear{Shu, Wang, and Liu}{Shu et~al\mbox{.}}{2018}]%
        {shu2018understanding}
\bibfield{author}{\bibinfo{person}{Kai Shu}, \bibinfo{person}{Suhang Wang},
  {and} \bibinfo{person}{Huan Liu}.} \bibinfo{year}{2018}\natexlab{}.
\newblock \showarticletitle{Understanding user profiles on social media for
  fake news detection}. In \bibinfo{booktitle}{\emph{2018 IEEE Conference on
  Multimedia Information Processing and Retrieval (MIPR)}}. IEEE,
  \bibinfo{pages}{430--435}.
\newblock


\bibitem[\protect\citeauthoryear{Silverman}{Silverman}{2016}]%
        {Buzzfeednews}
\bibfield{author}{\bibinfo{person}{Craig Silverman}.}
  \bibinfo{year}{2016}\natexlab{}.
\newblock \bibinfo{title}{{This Analysis Shows How Viral Fake Election News
  Stories Outperformed Real News On Facebook}}.
\newblock
  \bibinfo{howpublished}{\url{https://www.buzzfeednews.com/article/craigsilverman/viral-fake-election-news-outperformed-real-news-on-facebook\#.hlVK8Br7G}}.
\newblock


\bibitem[\protect\citeauthoryear{Stahl}{Stahl}{2018}]%
        {stahl2018fake}
\bibfield{author}{\bibinfo{person}{Kelly Stahl}.}
  \bibinfo{year}{2018}\natexlab{}.
\newblock \showarticletitle{Fake news detection in social media}.
\newblock \bibinfo{journal}{\emph{California State University Stanislaus}}
  \bibinfo{volume}{6} (\bibinfo{year}{2018}).
\newblock


\bibitem[\protect\citeauthoryear{Stone-Gross, Cova, Cavallaro, Gilbert,
  Szydlowski, Kemmerer, Kruegel, and Vigna}{Stone-Gross et~al\mbox{.}}{2009}]%
        {stone2009your}
\bibfield{author}{\bibinfo{person}{Brett Stone-Gross}, \bibinfo{person}{Marco
  Cova}, \bibinfo{person}{Lorenzo Cavallaro}, \bibinfo{person}{Bob Gilbert},
  \bibinfo{person}{Martin Szydlowski}, \bibinfo{person}{Richard Kemmerer},
  \bibinfo{person}{Christopher Kruegel}, {and} \bibinfo{person}{Giovanni
  Vigna}.} \bibinfo{year}{2009}\natexlab{}.
\newblock \showarticletitle{Your botnet is my botnet: analysis of a botnet
  takeover}. In \bibinfo{booktitle}{\emph{Proceedings of the 16th ACM
  conference on Computer and communications security}}. ACM,
  \bibinfo{pages}{635--647}.
\newblock


\bibitem[\protect\citeauthoryear{Suciu}{Suciu}{[n. d.]}]%
        {Peter2019}
\bibfield{author}{\bibinfo{person}{Peter Suciu}.} \bibinfo{year}{[n.
  d.]}\natexlab{}.
\newblock \bibinfo{title}{{More Americans Are Getting Their News From Social
  Media}}.
\newblock
\newblock


\bibitem[\protect\citeauthoryear{Tacchini, Ballarin, Della~Vedova, Moret, and
  de~Alfaro}{Tacchini et~al\mbox{.}}{2017}]%
        {tacchini2017some}
\bibfield{author}{\bibinfo{person}{Eugenio Tacchini}, \bibinfo{person}{Gabriele
  Ballarin}, \bibinfo{person}{Marco~L Della~Vedova}, \bibinfo{person}{Stefano
  Moret}, {and} \bibinfo{person}{Luca de Alfaro}.}
  \bibinfo{year}{2017}\natexlab{}.
\newblock \showarticletitle{Some like it hoax: Automated fake news detection in
  social networks}.
\newblock \bibinfo{journal}{\emph{arXiv preprint arXiv:1704.07506}}
  (\bibinfo{year}{2017}).
\newblock


\bibitem[\protect\citeauthoryear{Tambuscio, Ruffo, Flammini, and
  Menczer}{Tambuscio et~al\mbox{.}}{2015}]%
        {tambuscio2015fact}
\bibfield{author}{\bibinfo{person}{Marcella Tambuscio},
  \bibinfo{person}{Giancarlo Ruffo}, \bibinfo{person}{Alessandro Flammini},
  {and} \bibinfo{person}{Filippo Menczer}.} \bibinfo{year}{2015}\natexlab{}.
\newblock \showarticletitle{Fact-checking effect on viral hoaxes: A model of
  misinformation spread in social networks}. In
  \bibinfo{booktitle}{\emph{Proceedings of the 24th international conference on
  World Wide Web}}. ACM, \bibinfo{pages}{977--982}.
\newblock


\bibitem[\protect\citeauthoryear{Tandoc~Jr, Lim, and Ling}{Tandoc~Jr
  et~al\mbox{.}}{2018}]%
        {tandoc2018defining}
\bibfield{author}{\bibinfo{person}{Edson~C Tandoc~Jr},
  \bibinfo{person}{Zheng~Wei Lim}, {and} \bibinfo{person}{Richard Ling}.}
  \bibinfo{year}{2018}\natexlab{}.
\newblock \showarticletitle{Defining ``fake news'' A typology of scholarly
  definitions}.
\newblock \bibinfo{journal}{\emph{Digital journalism}} \bibinfo{volume}{6},
  \bibinfo{number}{2} (\bibinfo{year}{2018}), \bibinfo{pages}{137--153}.
\newblock


\bibitem[\protect\citeauthoryear{Tankovska}{Tankovska}{2021}]%
        {Tankovska}
\bibfield{author}{\bibinfo{person}{H. Tankovska}.}
  \bibinfo{year}{2021}\natexlab{}.
\newblock \bibinfo{title}{{ Number of social network users worldwide from 2017
  to 2025 }}.
\newblock
  \bibinfo{howpublished}{\url{https://www.statista.com/statistics/278414/number-of-worldwide-social-network-users/}}.
\newblock


\bibitem[\protect\citeauthoryear{Thakur}{Thakur}{2016}]%
        {AbhishekThakur}
\bibfield{author}{\bibinfo{person}{Abhishek Thakur}.}
  \bibinfo{year}{2016}\natexlab{}.
\newblock \bibinfo{title}{Identifying Clickbaits Using Machine Learning}.
\newblock
  \bibinfo{howpublished}{\url{https://www.linkedin.com/pulse/identifying-clickbaits-using-machine-learning-abhishek-thakur/}}.
\newblock


\bibitem[\protect\citeauthoryear{The Law Library~of Congress}{The Law
  Library~of Congress}{2019}]%
        {TheLawLibrary}
\bibfield{author}{\bibinfo{person}{Global Legal Research~Directorate The Law
  Library~of Congress}.} \bibinfo{year}{2019}\natexlab{}.
\newblock \bibinfo{title}{{53K rumors spread in Egypt in only 60 days, study
  reveals}}.
\newblock
  \bibinfo{howpublished}{\url{https://www.loc.gov/law/help/fake-news/counter-fake-news.pdf}}.
\newblock


\bibitem[\protect\citeauthoryear{Thomas}{Thomas}{2013}]%
        {thomas2013role}
\bibfield{author}{\bibinfo{person}{Kurt Thomas}.}
  \bibinfo{year}{2013}\natexlab{}.
\newblock \emph{\bibinfo{title}{The role of the underground economy in social
  network spam and abuse}}.
\newblock \bibinfo{thesistype}{Ph.D. Dissertation}. \bibinfo{school}{UC
  Berkeley}.
\newblock


\bibitem[\protect\citeauthoryear{Thomas, Grier, Ma, Paxson, and Song}{Thomas
  et~al\mbox{.}}{2011}]%
        {thomas2011design}
\bibfield{author}{\bibinfo{person}{Kurt Thomas}, \bibinfo{person}{Chris Grier},
  \bibinfo{person}{Justin Ma}, \bibinfo{person}{Vern Paxson}, {and}
  \bibinfo{person}{Dawn Song}.} \bibinfo{year}{2011}\natexlab{}.
\newblock \showarticletitle{Design and evaluation of a real-time url spam
  filtering service}. In \bibinfo{booktitle}{\emph{2011 IEEE symposium on
  security and privacy}}. IEEE, \bibinfo{pages}{447--462}.
\newblock


\bibitem[\protect\citeauthoryear{Torres, Gerhart, and Negahban}{Torres
  et~al\mbox{.}}{2018}]%
        {torres2018combating}
\bibfield{author}{\bibinfo{person}{Russell Torres}, \bibinfo{person}{Natalie
  Gerhart}, {and} \bibinfo{person}{Arash Negahban}.}
  \bibinfo{year}{2018}\natexlab{}.
\newblock \showarticletitle{Combating fake news: An investigation of
  information verification behaviors on social networking sites}. In
  \bibinfo{booktitle}{\emph{Proceedings of the 51st Hawaii International
  Conference on System Sciences}}.
\newblock


\bibitem[\protect\citeauthoryear{Tschiatschholek, Singla, Gomez~Rodriguez,
  Merchant, and Krause}{Tschiatschholek et~al\mbox{.}}{2018}]%
        {tschiatschek2018fake}
\bibfield{author}{\bibinfo{person}{Sebastian Tschiatschholek},
  \bibinfo{person}{Adish Singla}, \bibinfo{person}{Manuel Gomez~Rodriguez},
  \bibinfo{person}{Arpit Merchant}, {and} \bibinfo{person}{Andreas Krause}.}
  \bibinfo{year}{2018}\natexlab{}.
\newblock \showarticletitle{Fake news detection in social networks via crowd
  signals}. In \bibinfo{booktitle}{\emph{Companion Proceedings of the The Web
  Conference 2018}}. International World Wide Web Conferences Steering
  Committee, \bibinfo{pages}{517--524}.
\newblock


\bibitem[\protect\citeauthoryear{Vieira}{Vieira}{2017}]%
        {ThiagoVieira2017}
\bibfield{author}{\bibinfo{person}{Thiago Vieira}.}
  \bibinfo{year}{2017}\natexlab{}.
\newblock \bibinfo{title}{{Bs-detector-dataset}}.
\newblock
  \bibinfo{howpublished}{\url{https://github.com/thiagovas/bs-detector-dataset}}.
\newblock


\bibitem[\protect\citeauthoryear{Von~Ahn, Blum, and Langford}{Von~Ahn
  et~al\mbox{.}}{2004}]%
        {von2004telling}
\bibfield{author}{\bibinfo{person}{Luis Von~Ahn}, \bibinfo{person}{Manuel
  Blum}, {and} \bibinfo{person}{John Langford}.}
  \bibinfo{year}{2004}\natexlab{}.
\newblock \showarticletitle{Telling humans and computers apart automatically}.
\newblock \bibinfo{journal}{\emph{Commun. ACM}} \bibinfo{volume}{47},
  \bibinfo{number}{2} (\bibinfo{year}{2004}), \bibinfo{pages}{56--60}.
\newblock


\bibitem[\protect\citeauthoryear{Vosoughi, Roy, and Aral}{Vosoughi
  et~al\mbox{.}}{2018}]%
        {vosoughi2018spread}
\bibfield{author}{\bibinfo{person}{Soroush Vosoughi}, \bibinfo{person}{Deb
  Roy}, {and} \bibinfo{person}{Sinan Aral}.} \bibinfo{year}{2018}\natexlab{}.
\newblock \showarticletitle{The spread of true and false news online}.
\newblock \bibinfo{journal}{\emph{Science}} \bibinfo{volume}{359},
  \bibinfo{number}{6380} (\bibinfo{year}{2018}), \bibinfo{pages}{1146--1151}.
\newblock


\bibitem[\protect\citeauthoryear{Walsh}{Walsh}{2019}]%
        {paul2019}
\bibfield{author}{\bibinfo{person}{Paul Walsh}.}
  \bibinfo{year}{2019}\natexlab{}.
\newblock \bibinfo{title}{Factmata Trusted News Chrome Add-On Has Been Turned
  Off Until Further Notice}.
\newblock
  \bibinfo{howpublished}{\url{https://medium.com/@Paul__Walsh/factmata-trusted-news-chrome-add-on-has-been-turned-off-until-further-notice-7566f7312f86}}.
\newblock


\bibitem[\protect\citeauthoryear{Wanas, El-Saban, Ashour, and Ammar}{Wanas
  et~al\mbox{.}}{2008}]%
        {wanas2008automatic}
\bibfield{author}{\bibinfo{person}{Nayer Wanas}, \bibinfo{person}{Motaz
  El-Saban}, \bibinfo{person}{Heba Ashour}, {and} \bibinfo{person}{Waleed
  Ammar}.} \bibinfo{year}{2008}\natexlab{}.
\newblock \showarticletitle{Automatic scoring of online discussion posts}. In
  \bibinfo{booktitle}{\emph{Proceedings of the 2nd ACM workshop on Information
  credibility on the web}}. ACM, \bibinfo{pages}{19--26}.
\newblock


\bibitem[\protect\citeauthoryear{Wang}{Wang}{2010}]%
        {wang2010don}
\bibfield{author}{\bibinfo{person}{Alex~Hai Wang}.}
  \bibinfo{year}{2010}\natexlab{}.
\newblock \showarticletitle{Don't follow me: Spam detection in twitter}. In
  \bibinfo{booktitle}{\emph{2010 international conference on security and
  cryptography (SECRYPT)}}. IEEE, \bibinfo{pages}{1--10}.
\newblock


\bibitem[\protect\citeauthoryear{Wang}{Wang}{2017}]%
        {wang2017liar}
\bibfield{author}{\bibinfo{person}{William~Yang Wang}.}
  \bibinfo{year}{2017}\natexlab{}.
\newblock \showarticletitle{"liar, liar pants on fire": A new benchmark dataset
  for fake news detection}.
\newblock \bibinfo{journal}{\emph{arXiv preprint arXiv:1705.00648}}
  (\bibinfo{year}{2017}).
\newblock


\bibitem[\protect\citeauthoryear{Wang, Yin, Cai, Dong, and Dong}{Wang
  et~al\mbox{.}}{2015}]%
        {wang2015trust}
\bibfield{author}{\bibinfo{person}{Yingjie Wang}, \bibinfo{person}{Guisheng
  Yin}, \bibinfo{person}{Zhipeng Cai}, \bibinfo{person}{Yuxin Dong}, {and}
  \bibinfo{person}{Hongbin Dong}.} \bibinfo{year}{2015}\natexlab{}.
\newblock \showarticletitle{A trust-based probabilistic recommendation model
  for social networks}.
\newblock \bibinfo{journal}{\emph{Journal of Network and Computer
  Applications}}  \bibinfo{volume}{55} (\bibinfo{year}{2015}),
  \bibinfo{pages}{59--67}.
\newblock


\bibitem[\protect\citeauthoryear{Webwise}{Webwise}{2019}]%
        {Webwise}
\bibfield{author}{\bibinfo{person}{Webwise}.} \bibinfo{year}{2019}\natexlab{}.
\newblock \bibinfo{title}{{Explained: What is Fake News?}}
\newblock
  \bibinfo{howpublished}{\url{https://www.webwise.ie/teachers/what-is-fake-news/}}.
\newblock


\bibitem[\protect\citeauthoryear{Weerkamp and De~Rijke}{Weerkamp and
  De~Rijke}{2008}]%
        {weerkamp2008credibility}
\bibfield{author}{\bibinfo{person}{Wouter Weerkamp} {and}
  \bibinfo{person}{Maarten De~Rijke}.} \bibinfo{year}{2008}\natexlab{}.
\newblock \showarticletitle{Credibility improves topical blog post retrieval}.
  In \bibinfo{booktitle}{\emph{Proceedings of ACL-08: HLT}}.
  \bibinfo{pages}{923--931}.
\newblock


\bibitem[\protect\citeauthoryear{Weimer, Gurevych, and
  M{\"u}hlh{\"a}user}{Weimer et~al\mbox{.}}{2007}]%
        {weimer2007automatically}
\bibfield{author}{\bibinfo{person}{Markus Weimer}, \bibinfo{person}{Iryna
  Gurevych}, {and} \bibinfo{person}{Max M{\"u}hlh{\"a}user}.}
  \bibinfo{year}{2007}\natexlab{}.
\newblock \showarticletitle{Automatically assessing the post quality in online
  discussions on software}. In \bibinfo{booktitle}{\emph{Proceedings of the
  45th Annual Meeting of the ACL on Interactive Poster and Demonstration
  Sessions}}. Association for Computational Linguistics,
  \bibinfo{pages}{125--128}.
\newblock


\bibitem[\protect\citeauthoryear{Weng, Lim, Jiang, and He}{Weng
  et~al\mbox{.}}{2010}]%
        {weng2010twitterrank}
\bibfield{author}{\bibinfo{person}{Jianshu Weng}, \bibinfo{person}{Ee-Peng
  Lim}, \bibinfo{person}{Jing Jiang}, {and} \bibinfo{person}{Qi He}.}
  \bibinfo{year}{2010}\natexlab{}.
\newblock \showarticletitle{Twitterrank: finding topic-sensitive influential
  twitterers}. In \bibinfo{booktitle}{\emph{Proceedings of the third ACM
  international conference on Web search and data mining}}. ACM,
  \bibinfo{pages}{261--270}.
\newblock


\bibitem[\protect\citeauthoryear{Wong}{Wong}{2016}]%
        {All:Fake:News:Traffic:Facebook}
\bibfield{author}{\bibinfo{person}{Joon~Ian Wong}.}
  \bibinfo{year}{2016}\natexlab{}.
\newblock \bibinfo{title}{{Almost all the traffic to fake news sites is from
  Facebook, new data show}}.
\newblock
  \bibinfo{howpublished}{\url{https://qz.com/848917/facebook-fb-fake-news-data-from-jumpshot-its-the-biggest-traffic-referrer-to-fake-and-hyperpartisan-news-sites/}}.
\newblock
\newblock
\shownote{[Online; accessed 01-January-2020].}


\bibitem[\protect\citeauthoryear{Wong}{Wong}{2019}]%
        {Queenie}
\bibfield{author}{\bibinfo{person}{Queenie Wong}.}
  \bibinfo{year}{2019}\natexlab{}.
\newblock \bibinfo{title}{{Fake news is thriving thanks to social media users,
  study finds}}.
\newblock
  \bibinfo{howpublished}{\url{https://www.cnet.com/news/fake-news-more-likely-to-spread-on-social-media-study-finds/}}.
\newblock


\bibitem[\protect\citeauthoryear{Wu, Hu, Morstatter, and Liu}{Wu
  et~al\mbox{.}}{2017a}]%
        {wu2017adaptive}
\bibfield{author}{\bibinfo{person}{Liang Wu}, \bibinfo{person}{Xia Hu},
  \bibinfo{person}{Fred Morstatter}, {and} \bibinfo{person}{Huan Liu}.}
  \bibinfo{year}{2017}\natexlab{a}.
\newblock \showarticletitle{Adaptive spammer detection with sparse group
  modeling}. In \bibinfo{booktitle}{\emph{Eleventh International AAAI
  Conference on Web and Social Media}}.
\newblock


\bibitem[\protect\citeauthoryear{Wu, Hu, Morstatter, and Liu}{Wu
  et~al\mbox{.}}{2017b}]%
        {wu2017detecting}
\bibfield{author}{\bibinfo{person}{Liang Wu}, \bibinfo{person}{Xia Hu},
  \bibinfo{person}{Fred Morstatter}, {and} \bibinfo{person}{Huan Liu}.}
  \bibinfo{year}{2017}\natexlab{b}.
\newblock \showarticletitle{Detecting camouflaged content polluters}. In
  \bibinfo{booktitle}{\emph{Eleventh International AAAI Conference on Web and
  Social Media}}.
\newblock


\bibitem[\protect\citeauthoryear{Wu and Liu}{Wu and Liu}{2018}]%
        {wu2018tracing}
\bibfield{author}{\bibinfo{person}{Liang Wu} {and} \bibinfo{person}{Huan Liu}.}
  \bibinfo{year}{2018}\natexlab{}.
\newblock \showarticletitle{Tracing fake-news footprints: Characterizing social
  media messages by how they propagate}. In
  \bibinfo{booktitle}{\emph{Proceedings of the Eleventh ACM International
  Conference on Web Search and Data Mining}}. ACM, \bibinfo{pages}{637--645}.
\newblock


\bibitem[\protect\citeauthoryear{Wu, Feng, Fan, Gao, and Yu}{Wu
  et~al\mbox{.}}{2013}]%
        {wu2013detecting}
\bibfield{author}{\bibinfo{person}{Xian Wu}, \bibinfo{person}{Ziming Feng},
  \bibinfo{person}{Wei Fan}, \bibinfo{person}{Jing Gao}, {and}
  \bibinfo{person}{Yong Yu}.} \bibinfo{year}{2013}\natexlab{}.
\newblock \showarticletitle{Detecting marionette microblog users for improved
  information credibility}. In \bibinfo{booktitle}{\emph{Joint European
  Conference on Machine Learning and Knowledge Discovery in Databases}}.
  Springer, \bibinfo{pages}{483--498}.
\newblock


\bibitem[\protect\citeauthoryear{Xue, Yang, Yang, Wang, Chen, and Dai}{Xue
  et~al\mbox{.}}{2013}]%
        {xue2013votetrust}
\bibfield{author}{\bibinfo{person}{Jilong Xue}, \bibinfo{person}{Zhi Yang},
  \bibinfo{person}{Xiaoyong Yang}, \bibinfo{person}{Xiao Wang},
  \bibinfo{person}{Lijiang Chen}, {and} \bibinfo{person}{Yafei Dai}.}
  \bibinfo{year}{2013}\natexlab{}.
\newblock \showarticletitle{Votetrust: Leveraging friend invitation graph to
  defend against social network sybils}. In \bibinfo{booktitle}{\emph{2013
  Proceedings IEEE INFOCOM}}. IEEE, \bibinfo{pages}{2400--2408}.
\newblock


\bibitem[\protect\citeauthoryear{Yan}{Yan}{2006}]%
        {yan2006bot}
\bibfield{author}{\bibinfo{person}{Jeff Yan}.} \bibinfo{year}{2006}\natexlab{}.
\newblock \showarticletitle{Bot, cyborg and automated turing test}. In
  \bibinfo{booktitle}{\emph{International Workshop on Security Protocols}}.
  Springer, \bibinfo{pages}{190--197}.
\newblock


\bibitem[\protect\citeauthoryear{Yang, Niven, and Kao}{Yang
  et~al\mbox{.}}{2019}]%
        {yang2019fake}
\bibfield{author}{\bibinfo{person}{Kai-Chou Yang}, \bibinfo{person}{Timothy
  Niven}, {and} \bibinfo{person}{Hung-Yu Kao}.}
  \bibinfo{year}{2019}\natexlab{}.
\newblock \showarticletitle{Fake News Detection as Natural Language Inference}.
\newblock  (\bibinfo{year}{2019}).
\newblock


\bibitem[\protect\citeauthoryear{Yang, Wilson, Wang, Gao, Zhao, and Dai}{Yang
  et~al\mbox{.}}{2014}]%
        {yang2014uncovering}
\bibfield{author}{\bibinfo{person}{Zhi Yang}, \bibinfo{person}{Christo Wilson},
  \bibinfo{person}{Xiao Wang}, \bibinfo{person}{Tingting Gao},
  \bibinfo{person}{Ben~Y Zhao}, {and} \bibinfo{person}{Yafei Dai}.}
  \bibinfo{year}{2014}\natexlab{}.
\newblock \showarticletitle{Uncovering social network sybils in the wild}.
\newblock \bibinfo{journal}{\emph{ACM Transactions on Knowledge Discovery from
  Data (TKDD)}} \bibinfo{volume}{8}, \bibinfo{number}{1}
  (\bibinfo{year}{2014}), \bibinfo{pages}{2}.
\newblock


\bibitem[\protect\citeauthoryear{Yaraghi}{Yaraghi}{2019}]%
        {Niam}
\bibfield{author}{\bibinfo{person}{Niam Yaraghi}.}
  \bibinfo{year}{2019}\natexlab{}.
\newblock \bibinfo{title}{{How should social media platforms combat
  misinformation and hate speech?}}
\newblock
  \bibinfo{howpublished}{\url{https://www.brookings.edu/blog/techtank/2019/04/09/how-should-social-media-platforms-combat-misinformation-and-hate-speech/\#cancel}}.
\newblock


\bibitem[\protect\citeauthoryear{Ye, Kumar, and Akoglu}{Ye
  et~al\mbox{.}}{2016}]%
        {ye2016temporal}
\bibfield{author}{\bibinfo{person}{Junting Ye}, \bibinfo{person}{Santhosh
  Kumar}, {and} \bibinfo{person}{Leman Akoglu}.}
  \bibinfo{year}{2016}\natexlab{}.
\newblock \showarticletitle{Temporal opinion spam detection by multivariate
  indicative signals}. In \bibinfo{booktitle}{\emph{Tenth International AAAI
  Conference on Web and Social Media}}.
\newblock


\bibitem[\protect\citeauthoryear{Ye and Wu}{Ye and Wu}{2010}]%
        {ye2010measuring}
\bibfield{author}{\bibinfo{person}{Shaozhi Ye} {and} \bibinfo{person}{S~Felix
  Wu}.} \bibinfo{year}{2010}\natexlab{}.
\newblock \showarticletitle{Measuring message propagation and social influence
  on Twitter. com}. In \bibinfo{booktitle}{\emph{International conference on
  social informatics}}. Springer, \bibinfo{pages}{216--231}.
\newblock


\bibitem[\protect\citeauthoryear{Zhao, Resnick, and Mei}{Zhao
  et~al\mbox{.}}{2015}]%
        {zhao2015enquiring}
\bibfield{author}{\bibinfo{person}{Zhe Zhao}, \bibinfo{person}{Paul Resnick},
  {and} \bibinfo{person}{Qiaozhu Mei}.} \bibinfo{year}{2015}\natexlab{}.
\newblock \showarticletitle{Enquiring minds: Early detection of rumors in
  social media from enquiry posts}. In \bibinfo{booktitle}{\emph{Proceedings of
  the 24th International Conference on World Wide Web}}. International World
  Wide Web Conferences Steering Committee, \bibinfo{pages}{1395--1405}.
\newblock


\bibitem[\protect\citeauthoryear{Zhou and Zafarani}{Zhou and Zafarani}{2019}]%
        {zhou2019network}
\bibfield{author}{\bibinfo{person}{Xinyi Zhou} {and} \bibinfo{person}{Reza
  Zafarani}.} \bibinfo{year}{2019}\natexlab{}.
\newblock \showarticletitle{Network-based Fake News Detection: A Pattern-driven
  Approach}.
\newblock \bibinfo{journal}{\emph{ACM SIGKDD Explorations Newsletter}}
  \bibinfo{volume}{21}, \bibinfo{number}{2} (\bibinfo{year}{2019}),
  \bibinfo{pages}{48--60}.
\newblock


\bibitem[\protect\citeauthoryear{Zhou}{Zhou}{2017}]%
        {zhou2017clickbait}
\bibfield{author}{\bibinfo{person}{Yiwei Zhou}.}
  \bibinfo{year}{2017}\natexlab{}.
\newblock \showarticletitle{Clickbait detection in tweets using self-attentive
  network}.
\newblock \bibinfo{journal}{\emph{arXiv preprint arXiv:1710.05364}}
  (\bibinfo{year}{2017}).
\newblock


\bibitem[\protect\citeauthoryear{Zubiaga, Aker, Bontcheva, Liakata, and
  Procter}{Zubiaga et~al\mbox{.}}{2018}]%
        {zubiaga2018detection}
\bibfield{author}{\bibinfo{person}{Arkaitz Zubiaga}, \bibinfo{person}{Ahmet
  Aker}, \bibinfo{person}{Kalina Bontcheva}, \bibinfo{person}{Maria Liakata},
  {and} \bibinfo{person}{Rob Procter}.} \bibinfo{year}{2018}\natexlab{}.
\newblock \showarticletitle{Detection and resolution of rumours in social
  media: A survey}.
\newblock \bibinfo{journal}{\emph{ACM Computing Surveys (CSUR)}}
  \bibinfo{volume}{51}, \bibinfo{number}{2} (\bibinfo{year}{2018}),
  \bibinfo{pages}{32}.
\newblock


\end{thebibliography}

%%
%% If your work has an appendix, this is the place to put it.
%\appendix

\end{document}